\documentclass[twocolumn]{aastex631}

\usepackage{epsfig}
\usepackage{graphicx}
\usepackage{tabularx}
\usepackage{float}
\usepackage{placeins}
\usepackage{mathtools}
\usepackage{pgf}
\usepackage{ulem}
\usepackage{lipsum}
\usepackage{placeins}
\usepackage{enumitem}
\usepackage{booktabs}
\newcommand\doubleRule{\toprule\toprule}

\usepackage[T1]{fontenc}

\usepackage{amssymb}
\usepackage{hyperref}
\usepackage{tikz}
\usepackage{blindtext}
\def\gsim{\vcenter{\hbox{$>$}\offinterlineskip\hbox{$\sim$}}}

\shorttitle{Star Formation of NGC\,5128}
\shortauthors{Sima T. Aghdam et al.}

\begin{document}

\title{The Complex Star Formation History of the Halo of NGC\,5128 (Cen\,A)}

\email{atefeh@ipm.ir}

\author{Sima T. Aghdam}
\affil{School of Astronomy, Institute for Research in Fundamental Sciences (IPM), Tehran, 19568-36613, Iran}

\author{Atefeh Javadi}
\affiliation{School of Astronomy, Institute for Research in Fundamental Sciences (IPM), Tehran, 19568-36613, Iran}

\author{Seyedazim Hashemi}
\affiliation{Department of Physics and Astronomy, University of California Riverside, CA 92521, The USA}
\affiliation{Department of Physics, Sharif University of Technology, Tehran, 11155-9161, Iran}

\author{Mahdi Abdollahi}
\affil{School of Astronomy, Institute for Research in Fundamental Sciences (IPM), Tehran, 19568-36613, Iran}

\author{Jacco Th. van Loon}
\affiliation{Lennard-Jones Laboratories, Keele University, ST5 5BG, UK}

\author{Habib Khosroshahi}
\affiliation{School of Astronomy, Institute for Research in Fundamental Sciences (IPM), Tehran, 19568-36613, Iran}

\author{Roya H. Golshan}
\affiliation{I. Physikalisches Institut, Universität zu Köln, Zülpicher Straße 77, 50937 Cologne, Germany}

\author{Elham Saremi}
\affiliation{School of Astronomy, Institute for Research in Fundamental Sciences (IPM), Tehran, 19568-36613, Iran}
\affiliation{Instituto de Astrof{\`i}sica de Canarias, C/ V{\`i}a L{\`a}ctea s/n, 38205 La Laguna, Tenerife, Spain}
\affiliation{Departamento de Astrof{\`i}sica, Universidad de La Laguna, 38205 La Laguna, Tenerife, Spain}

\author{Maryam Saberi}
\affiliation{Rosseland Centre for Solar Physics, University of Oslo, P.O. Box 1029, Blindern, NO-0315, Oslo, Norway}

\begin{abstract}

NGC\,5128 (Cen\,A) is the nearest giant elliptical galaxy and one of the brightest extragalactic radio sources in the sky, boasting a prominent dust lane and jets emanating from its nuclear supermassive black hole. In this paper, we construct the star formation history (SFH) of two small fields in the halo of NGC\,5128: a northeastern field (Field\,1) at a projected distance of $\sim 18.8$ kpc from the center, and a southern field (Field\,2) $\sim 9.9$ kpc from the center. Our method is based on identifying long period variable (LPV) stars that trace their sibling stellar population and hence historical star formation due to their high luminosity and strong variability; we identified 395 LPVs in Field\,1 and 671 LPVs in Field\,2. Even though the two fields are $\sim 28$ kpc apart on opposite sides from the center, they show similar SFHs. In Field\,1, star formation rates (SFRs) increased significantly around $t\sim 800$ Myr and $t\sim 3.8$ Gyr; and in Field\,2, SFRs increased considerably  around $t\sim 800$ Myr, $t\sim 3.8$ Gyr, and $t\sim 6.3$ Gyr, where $t$ is look--back time. The increase in SFR $\sim 800$ Myr ago agrees with previous suggestions that the galaxy experienced a merger around that time. The SFH reconstructed from LPVs supports a scenario in which multiple episodes of nuclear activity lead to episodic jet-induced star formation. While there is no catalog of LPVs for the central part of NGC\,5128, applying our method to the outer regions (for the first time in a galaxy outside the Local Group) has enabled us to put constraints on the complex evolution of this cornerstone galaxy.

\end{abstract}



\section{Introduction} \label{sec:sec1}

Considering our location in the Local Group (LG), we have a great opportunity to study resolved populations of spiral galaxies and obtain more knowledge about their formation and evolution. However, there is no giant elliptical (GE) galaxy in the LG, therefore, our understanding is limited to the nearest elliptical galaxies in the other groups. NGC\,5128 (or Centaurus A), with a distance of 3.8 Mpc ($\mu = 27.87 \pm 0.16 $ mag; \citealp{Rejkuba2004a}, and $E(B-V) = 0.15 \pm 0.05$ mag; \citealp{Rejkuba2001}) offers us a unique opportunity to study the nearest GE galaxy up close (\citealp{harris1999}; \citealp{charmandaris2000}; \citealp{Rejkuba2004b}, \citeyear{Rejkuba2005}), located in the Centaurus group of galaxies (\citealp{karachentsev2005}). 

There is some doubt about the galaxy type of NGC\,5128. It has been classed as a giant elliptical galaxy based on its optical light distribution (\citealp{Baade1954}; \citealp{graham1979}). Other studies have suggested that it is an SO-type galaxy based on the dust lane and lenticular morphology (\citealp{morgan1958}; \citealp{vanden1990}). On balance, though, NGC\,5128 is generally considered as a giant elliptical galaxy (\citealp{harris2010}).

NGC\,5128's extended halo and radio lobes cover almost $2^\circ$ of the sky (\citealp{peng2002}) in optical maps. In recent years, the halo of NGC\,5128 has been subject of intense scrutiny (e.g., \citealp{hernandez2018}; \citealp{dsouza2018}; \citealp{crnojevic2016}; \citealp{salome2016a}, \citeyear{salome2016b}; \citealp{neff2015}; \citealp{crnojevic2014}; \citealp{Rejkuba2014}). NGC\,5128 is believed to be a post-merger galaxy (\citealp{peng2002}), one of only a handful of halos resolved into individual stars (\citealp{Rejkuba2011}). The stellar streams and field stars in the halo can be used to trace the signature of mergers and/or interactions. Moreover, an active galactic nucleus (AGN) at the center of NGC\,5128 produces the nearest example of powerful radio jets (\citealp{crockett2012}). AGN activity and its effect on the regulation of the star formation and evolution of the host galaxy is an important yet still open question in galaxy formation and evolution theory (e.g., \citealp{ciotti1997}; \citealp{silk1998}, \citeyear{silk2005}; \citealp{binney2004}; \citealp{springel2005}; \citealp{sijacki2007}; \citealp{schawinski2007}).

The star formation history (SFH) is a crucial component of galaxy formation and evolution. Recovering the SFH in resolved galaxies is typically based on the color-magnitude diagrams (CMDs) of individual stars where the signature of different stellar populations can be traced (e.g., \citealp{tolstoy1996}; \citealp{holtzman1999}; \citealp{olsen1999}; \citealp{dolphin2002}, \citealp{javadi2011b}). This type of analysis is confined to a few dozen galaxies that mostly lie within our LG because of the limited spatial resolution (within $\sim 2$ Mpc, \citealp{Ruiz2015}).

In this work, we aim to find the SFH of two small fields in the halo of NGC\,5128 using long period variable stars (LPVs) in order to understand the relation between the SFH of the halo and its merger history. Large amplitude ($>10$\% or so) regular or semi-regular LPVs are cool evolved stars spanning a wide age range from $\sim 10$ Myr to $\sim 10$ Gyr. These stars are very luminous, $\sim 1000$--$500,000$ $L_\odot$, in proportion to their birth mass, and of low temperature ($T \sim 2500$--$4500$ K), hence they are the most accessible tracers of stellar populations (e.g., \citealp{maraston2005}, \citeyear{maraston2006}; \citealp{javadi2011a}, \citeyear{javadi2011b}, \citeyear{javadi2013}, \citeyear{javadi2015}). LPV stars are mostly evolved asymptotic giant branch stars (AGBs) of $\gsim \ 30$ Myr (e.g., \citealp{fraser2005}, \citeyear{fraser2008}; \citealp{soszynski2009}) varying on timescales of $\approx 100$--$1300$ days making them easily identifiable (\citealp{javadi2011b}, \citeyear{javadi2011c}, \citeyear{javadi2015},  \citeyear{javadi2017}; \citealp{Rezaei2014}; \citealp{golshan2017}; \citealp{hashemi2019}; \citealp{navabi2021}; \citealp{saremi2020}, \citeyear{saremi2021}). Recent star formation ($\sim 10$--$30$ Myr) can be traced in a similar way by red supergiants (RSGs) with birth masses $\sim 8$--$30$ M$_\odot$. AGB stars inject up to $80\%$ of their mass into the interstellar medium (ISM) and play a significant role in chemical enrichment of the galaxy; mass loss can, in some cases, also be important for RSGs and their fate as supernova (\citealp{vassiliadis1993}; \citealp{javadi2013}; \citealp{vanloon1999}, \citeyear{vanloon2005}).

This paper is structured as follows. In Section \ref{sec:sec2}, the data used for the study are presented. There is a short explanation about detected LPV stars in section \ref{sec:sec3}. A discussion of removing the contaminated stars can be found in section \ref{sec:sec4}. The metallicity of the galaxy is discussed in section \ref{sec:sec5}. We provide a brief description of the method used for studying SFH in section \ref{sec:sec6}. Section \ref{sec:sec7} derives the SFH of two fields in NGC\,5128, and to obtain an accurate SFR, we identify a probability function to simulate the non-detected LPVs in section \ref{sec:sec8}. It is followed by discussion and a summary in sections \ref{sec:sec9} and \ref{sec:sec10}, respectively.


\section{Data} \label{sec:sec2}

We used near-IR photometry obtained with {\sc ISAAC} at the {\sc ESO} Paranal UT1 Antu 8.2\,m Very Large Telescope ({\sc VLT}) in two different fields in the halo of NGC\,5128. These data were published by \cite{Rejkuba2001} and analysed by \cite{Rejkuba2003a}. In addition, we also used near-IR photometry obtained with {\sc SOFI} at the {\sc ESO} La Silla 3.5\,m New Technology Telescope (NTT) for Field\,2. Fig.\,\ref{fig:fig1} presents an optical image of NGC\,5128 taken with the UK Schmidt Telescope (\citealp{ma1998}) overlaid by Field\,1 and Field\,2 that are studied in this work. It should be noted that Field\,1 and Field\,2 are the same as the fields studied in \cite{Rejkuba2003a} but smaller than the ones mentioned in \cite{Rejkuba2001}.

Field\,1 is centered at $\alpha = 13^{h}$ $26^{m}$ $23\rlap{.}^{s}5$, $\delta = -42^{\circ}$ $52^{\prime}$ $0^{\prime \prime}$ on the eminent north-eastern part of the halo, at a distance of $\sim 17^{\prime}$ ($\sim 18.8$ kpc) far from the center of the galaxy with a dimension of $2\rlap{.}^{\prime}28 \times 2\rlap{.}^{\prime}30$ ($5.7$ kpc$^2$).

Field\,1 is placed on the so-called inner filaments discovered by \cite{morganti1991}. The filaments, along the direction of the northern radio jet, are extended from $\sim 13$ to $\sim 22 $ kpc (\citealp{salome2016b}) and contain ionized gas, young star clusters, and ultraviolet emission (\cite{neff2015}). Young stars are $\la 10$ Myr old (e.g., \citealp{mould2000}; \citealp{crockett2012}), thus indicating recent and ongoing star formation (e.g., \citealp{Rejkuba2004a}; \citealp{neff2015}). The filaments and their star formation are suggested to be the result of interactions between jets and gas (e.g., \citealp{charmandaris2000}; \citealp{auld2012}; \citealp{crockett2012}; \citealp{santoro2015a}; \citeyear{santoro2015b}; \citealp{salome2016b}).

Field\,2 is centered at $\alpha = 13^{h}$ $25^{m}$ $26^{s}$, $\delta = -43^{\circ}$ $10^{\prime}$ $0^{\prime \prime}$ at a distance of $\sim 9^{\prime}$ ($\sim 9.9$ kpc) from the center with a dimension of $2\rlap{.}^{\prime}25 \times 2\rlap{.}^{\prime}31$ ($5.7$ kpc$^2$).


\section{LPVs in NGC\,5128} \label{sec:sec3}

To identify LPVs, \cite{Rejkuba2003a} performed  multi-epoch photometry in the $K_s$-band, along with single-epoch photometry in the $J_s$- and $H$-bands. The $50\%$ completeness limit in $K_s$ and $H$ is $22.5$ mag in Field\,1 and $21.5$ mag in Field\,2. As a result, $15574$ and $18098$ sources in Field\,1 and Field\,2 are detected, respectively, with at least 3 $K_s$-band observations among which more than $1500$ variable stars are identified. Based on Fourier analysis, \cite{Rejkuba2003a} could determine periods and amplitudes for $1046$ red variables with at least $10$ $K_s$-band measurements. Among them, $437$ and $709$ LPVs are detected in Field\,1 and Field\,2, respectively, most of which are brighter than the tip of the red giant branch (RGB) (\citealp{Rejkuba2003b}).

\begin{figure*}[ht!]
	\epsfig{figure=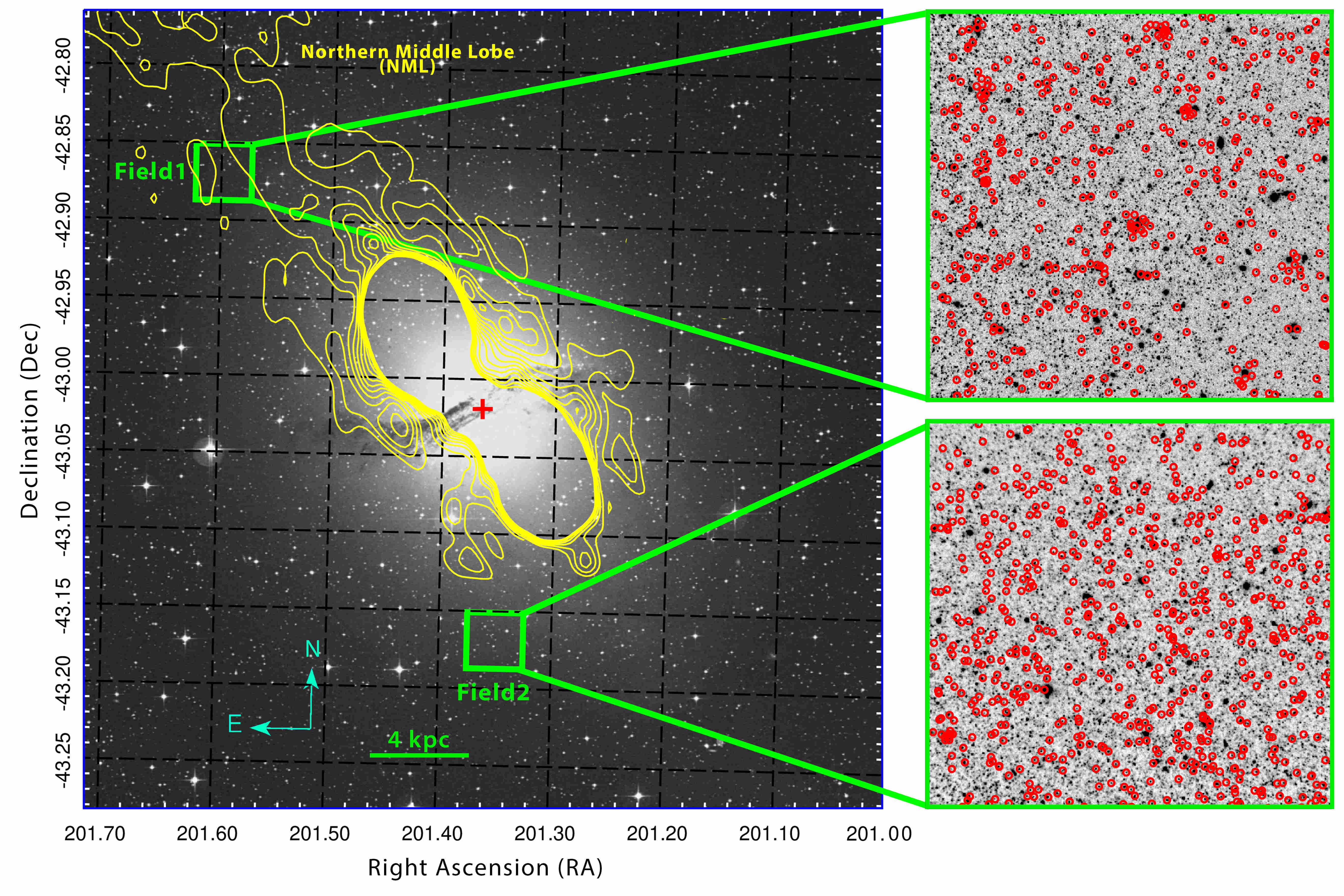, width=180mm, height=120mm}
	\caption{Archival optical image (at $468$ nm wavelength) of NGC\,5128 taken with the UK Schmidt Telescope overlain with the location of the fields studied here (each one is about $2\rlap{.}^{\prime}3 \times 2\rlap{.}^{\prime}3$) ({\it left}). The red cross indicates the center of NGC\,5128 (\citealp{ma1998}). Red circles represent the locations of the selected LPVs from the {\sc ISAAC} $K_s$-band data. The yellow overlain contours are the radio emission that show the structure of the lobes at $1392$ MHz and $128$ MHz presented in \cite{morganti1999}. The northern middle lobe (NML) is located in the upper left of the figure, whilst the north and south lobes are located in the north and south of the galaxy.}
	\label{fig:fig1}
\end{figure*}

Our selected sample to estimate the SFH includes all identified LPVs with periods longer than $70$ days (from \citealp{Rejkuba2003a}). Considering these criteria, we have $395$ and $671$ LPVs in Field\,1 and Field\,2, respectively. On the right side of Fig.\,\ref{fig:fig1} the distribution of selected LPVs (red circles) in their fields is depicted.

\begin{figure*}[ht!]
	\centering{\hbox{
			\epsfig{figure=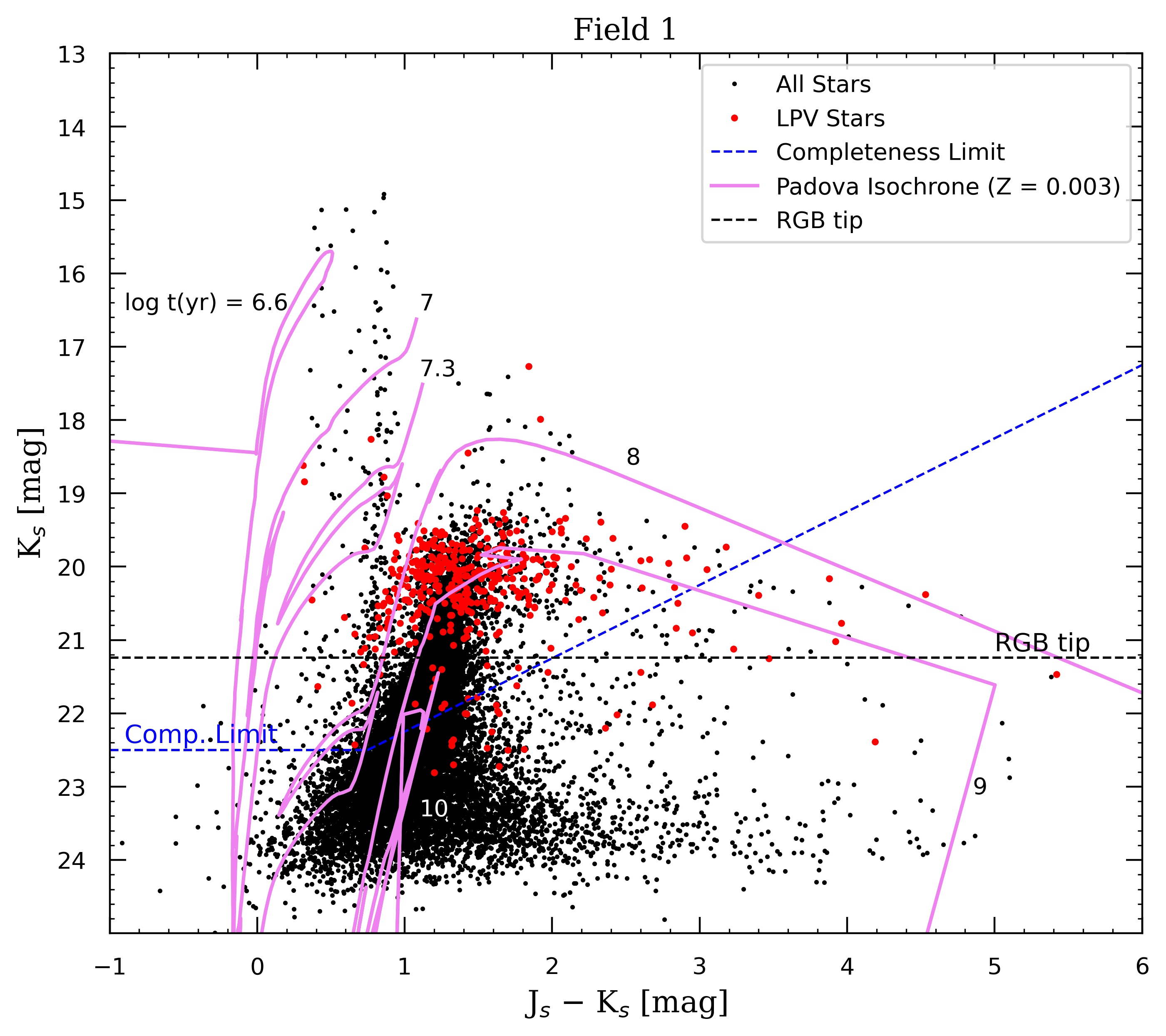,width=90mm,height=90mm}
			\epsfig{figure=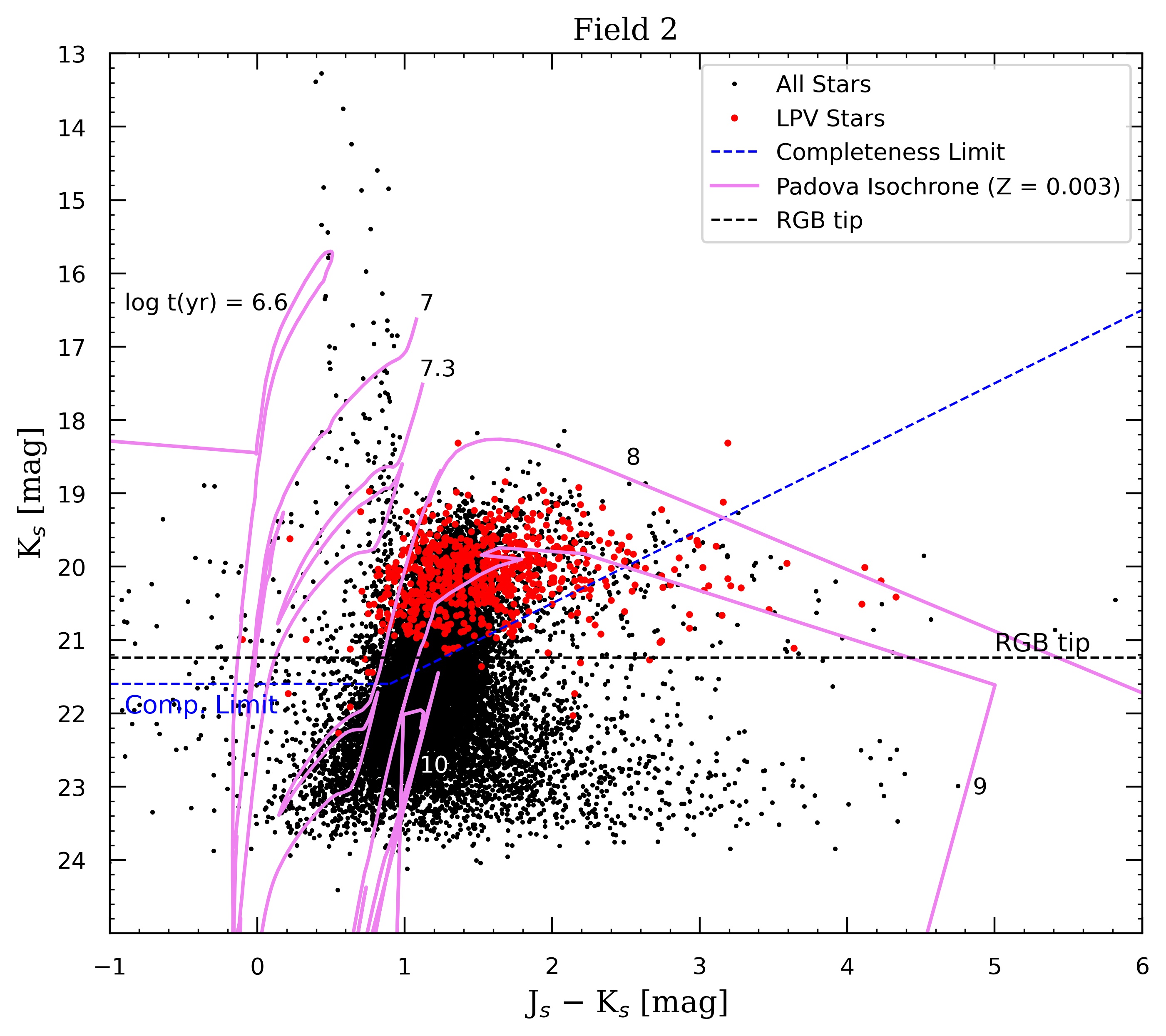,width=90mm,height=90mm}
	}}
	\caption{$K_s$ vs.\ $J_s$--$K_s$ CMD for the stars with at least 3 $K_s$-band detections (black dots) and LPVs (red dots) in Field\,1 ({\it left}) and Field\,2 ({\it right}). The black and blue dotted lines represent the RGB tip and completeness limit magnitudes in each field (\citealp{Rejkuba2003a}). The purple lines show theoretical stellar isochrones for a metallicity $Z = 0.003$ for $6$ different ages \cite{marigo2017}.}
	\label{fig:fig2}
\end{figure*}

Fig.\,\ref{fig:fig2} shows $K_s$ vs.\ $J_s$--$K_s$ CMDs for all those detected in at least 3 $K_s$-bands. The RGB tip is at $K_s = 21.24$ mag; the completeness limits are $K_s = 22.5$ mag and $K_s = 21.5$ mag in Field\,1 and Field\,2, respectively (\citealp{Rejkuba2003a}). \cite{Rejkuba2003b} found that it is necessary to subtract 0.1 mag from the originally published $K_s$-band magnitudes. A summary of the data used in this paper is presented in Table \ref{tab:tab1}.

\begin{table*}[ht]
	\centering
	\caption{Summary of the data taken from \cite{Rejkuba2001}, (\citeyear{Rejkuba2003a}), (\citeyear{Rejkuba2003b}) as used in this paper.}
	\label{tab:tab1}
	\begin{tabular}{ccccccccc}
		\hline
		Field    & RA                   & Dec                                                              & \llap{D}istance to Cente\rlap{r} & Coverage             & \multicolumn{2}{c}{Number} & Completeness Limit    & RGB Tip       \\
		---      & J2000                & J2000                                                            & arcmin (kpc)           & arcmin$^2$ (kpc$^2$) & Total        & LPV         & mag                   & mag           \\ \hline
		Field\,1 & $13^h$ $26^m$ $23\rlap{.}^s5$ & $-42^{\circ}$ $52^{\prime}$ $0^{\prime\prime}$ & $\sim$ 17 (18.8)       & $\sim 5.2$ (5.7)     & 15574        & 437         & 50 \% at K$_s = 22.5$ & K$_s = 21.24$ \\
		Field\,2 & $13^h$ $25^m$ $26\rlap{.}^s0$ & $-43^{\circ}$ $10^{\prime}$ $0^{\prime\prime}$ & $\sim$ 9 (9.9)         & $\sim 5.2$ (5.7)     & 18098        & 709         & 50 \% at K$_s = 21.5$ & K$_s = 21.24$ \\ \hline
	\end{tabular}
\end{table*}


\section{Contamination} \label{sec:sec4}

The observations of NGC\,5128 ($l = 309\rlap{.}^{\circ}515$, $b = +19\rlap{.}^{\circ}417$) are probably contaminated by foreground stars from the Milky Way. The contamination level can be determined by cross-matching our catalog with the recently published {\it Gaia} Early Data Release 3 (EDR3) (\citealp{brown2021}).
The completeness limit of {\it Gaia} is $G \approx 19$--$21$ mag, meaning that $50\%$ of stars are brighter than $G \sim 20$ mag (\citealp{brown2021}).

Stars were considered as foreground objects if they satisfied one of the following criteria as explained in detail by \cite{saremi2020}: (a) the star’s proper motion is consistent with the relation $\sqrt{\mu^{2}_{RA} + \mu^{2}_{DEC}} > 0.28 + 2error$ mas\,yr$^{-1}$ (\citealp{vandermarel2019}) or (b) the ratio of parallax to its error is larger than 2$\sigma$. We note that increasing the threshold from $2$ to $5$ did not affect the selection results. Thus, eight foreground stars were found, of which two in Field\,1 and six in Field\,2.

\begin{figure*}[ht!]
	\centering{\hbox{
			\epsfig{figure=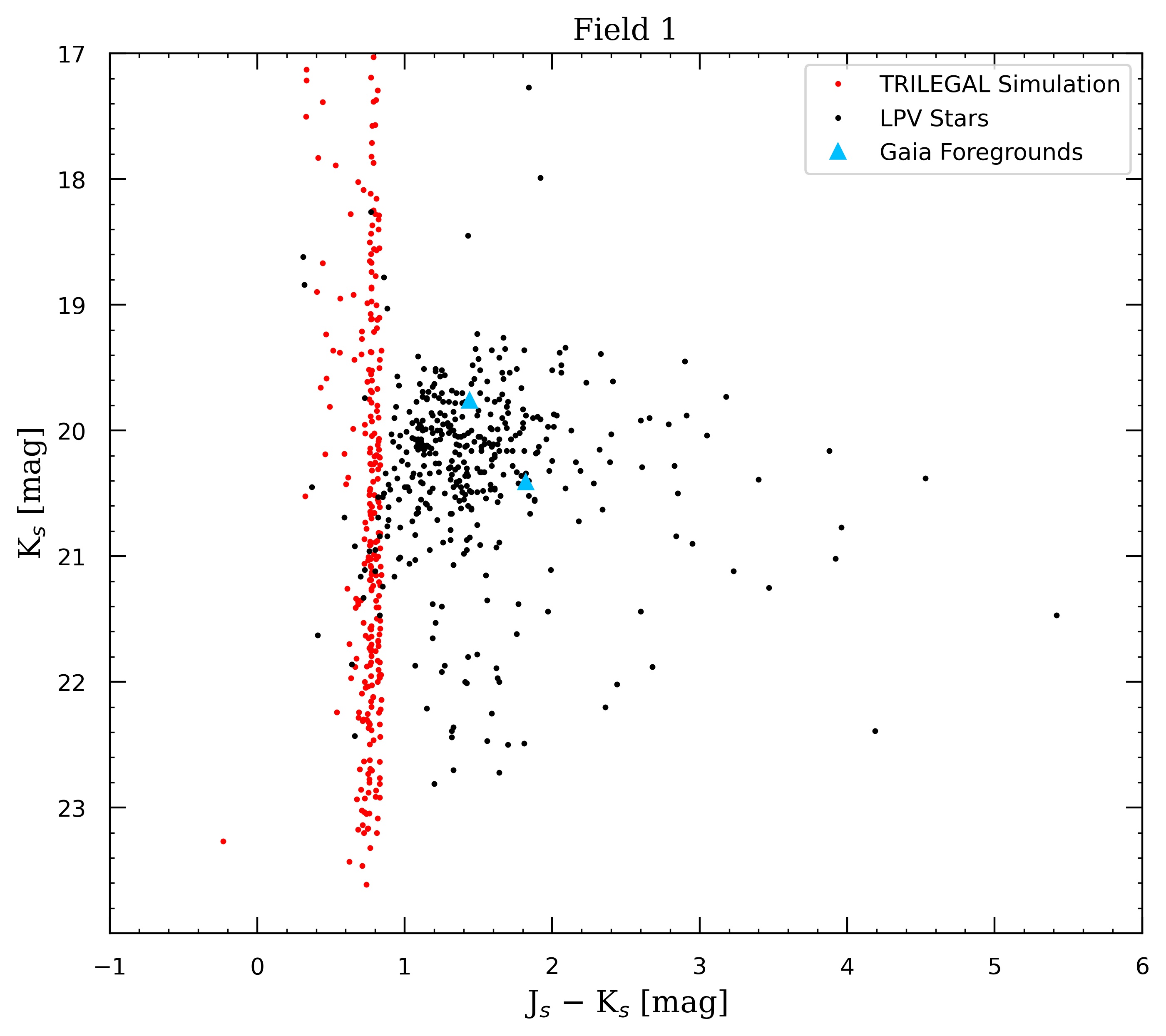,width=90mm}
			\epsfig{figure=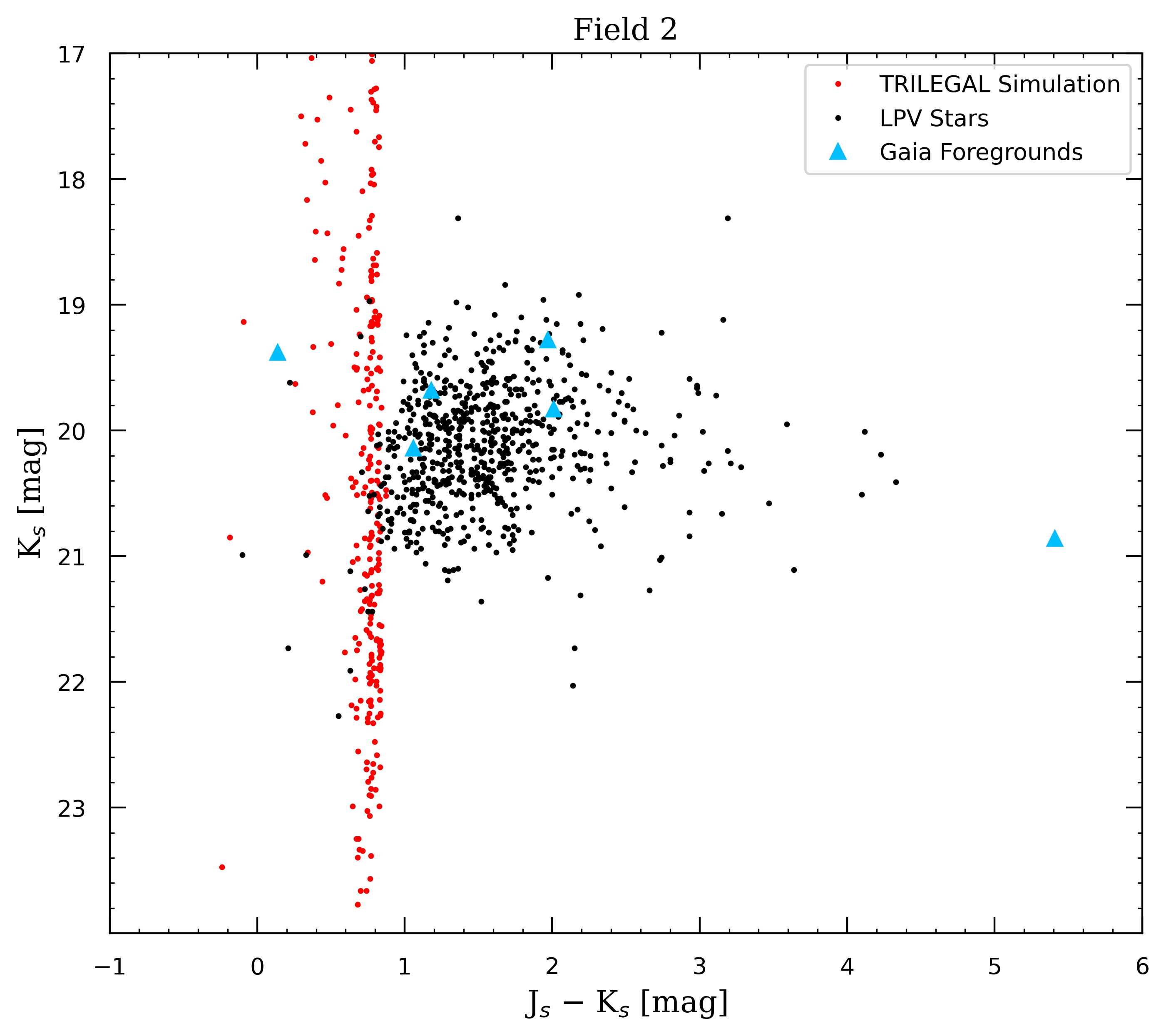,width=90mm}}}
	\caption{The obtained contamination of Field\,1, {\it left panel}, and Field\,2 {\it right panel}. Black dots represent detected LPVs. Red dots show the {\sc TRILEGAL} simulation in the $0.007$ deg$^2$ area, indicating low levels of contamination. Blue triangles are cross-matched with the {\it Gaia} catalog; we found only eight stars to be foreground, two of which are in Field\,1 and six in Field\,2.}
	\label{fig:fig3}
\end{figure*}

To verify the level of contamination, we simulated the foreground population using the {\sc TRILEGAL} stellar population synthesis code (\citealp{girardi2005}) via its web interface. We assumed two areas of $0.002$ deg$^2$ in the direction of Field\,1 and Field\,2. The results of these simulations along with detected LPVs in the two fields are shown in Fig.\,\ref{fig:fig3}, indicating that the color of simulated data is $J_s - K_s < 0.9$ mag, while the LPVs have a redder color. Therefore, there is no significant contamination between the Milky Way and the detected LPVs.

\section{Metallicity of the Halo of NGC\,5128} \label{sec:sec5}

With age, because of nucleosynthesis and chemical enrichment of the galaxy by old and dying stars, the metallicity at various locations within the galaxy changes. In general, it is expected that older stars formed in a metal-poor environment while younger stars formed in a metal-rich one (\citealp{javadi2011a}; \citealp{vanloon1999}; \citealp{vassiliadis1993}). The CMDs of Fig.\,\ref{fig:fig2}, specifically the evolved branch stars, reveal that the main chemical enrichment of the galaxy occurred $3$ Gyr ago.

The metallicity matters when reconstructing the SFH. A large number of globular clusters (GC) was studied spectroscopically \cite{woodley2010b}, yielding a metallicity range $-1.7 \la [M/H] \la 0.4$ dex ($0.0003 \la Z \la 0.05$). The majority of the halo was covered by their study, including Field\,1 and Field\,2. Their results are in agreement with \cite{Rejkuba2011}, who found $0.0001 < Z < 0.04$ by comparing an observed CMD obtained with the {\sc Hubble} Space Telescope of a field in the halo of NGC\,5128 (located at $38$ kpc south of the center) with simulated CMDs. While this region is far from the ones we present here, they exhibit a similar range of metallicity. Considering these results and other works by \cite{beasley2008}, \cite{peng2004b}, and \cite{Rejkuba2001}, we adopt the metallicity range $0.0003 < Z < 0.04$ for both fields.


\section{Method} \label{sec:sec6}

The method we used to calculate the SFH is based on LPVs and was developed by \cite{javadi2011b} and has since been applied in a variety of studies (\citealp{javadi2011b}, \citeyear{javadi2011c},  \citeyear{javadi2017}; \citealp{Rezaei2014}; \citealp{golshan2017}; \citealp{hashemi2019}; \citealp{navabi2021}; \citealp{saremi2021}). We estimated birth mass, age, and LPV phase duration (the duration that stars are in the LPV phase) of LPVs by applying Padova evolutionary models (\citealp{marigo2017}) and assuming a range of constant metallicities. The stellar mass is estimated by the mass--luminosity relation, and the mass--age relation gives the age of stars. The LPV phase duration can be investigated using the mass--LPV phase duration relation of the star as detailed in Appendix \ref{app:app}. Finally, LPVs are sorted according to their ages and divided into bins. SFRs for different bins with specified intervals in age and mass can be calculated using:

\begin{equation}
\xi(t) = \frac{dn^\prime(t)}{\delta t}\ \frac{\int_{\rm min}^{\rm max}f_{\rm IMF}(m)m\ dm}
{\int_{m(t)}^{m(t+dt)}f_{\rm IMF}(m)\ dm},
\label{eq:eq1}
\end{equation}
where $m$ is birth mass, $f_{\rm IMF}(m)$ is Kroupa initial mass function (IMF) defined as $f_{\rm IMF} = Am^{-\alpha}$ where $A$ is the normalization coefficient which is canceled out here and $\alpha$ depends on the mass
range, following Kroupa (2001):
\begin{equation}
\alpha=\left\{
\begin{array}{lll}
+0.3\pm0.7 & {\rm for} & {\rm min}\leq m/{\rm M}_\odot<0.08 \\
+1.3\pm0.5 & {\rm for} & 0.08\leq m/{\rm M}_\odot<     0.50 \\
+2.3\pm0.3 & {\rm for} & 0.50\leq m/{\rm M}_\odot<{\rm max} \\
\end{array}
\right.
\end{equation}
The minimum and maximum of the stellar mass range are adopted to be 0.02 and
200 M$_\odot$, respectively. Reasonable adjustments to these values are unlikely to cause the star formation rate to vary by more than a factor of two. $dn^\prime$ is the observed LPVs in each bin, and $\delta t$ is LPV phase duration.

The statistical error bars for each bin are calculated using the Poisson distribution:

\begin{equation}
\sigma_{\xi(t)}=\frac{\sqrt{N}}{N}\xi(t),
\label{eq:eq2}
\end{equation}
where $N$ is the number of stars in each age bin.

To estimate SFHs in this paper, we assumed $11$ metallicities to cover the metallicity of very old ($\gsim \ 12$ Gyr) to very young ($\sim 1$ Gyr) populations in the galaxy ($0.0003 < Z < 0.04$; \citealp{Rejkuba2005}, \citeyear{Rejkuba2011}; \citealp{woodley2010b}). The fitted lines/curves and corresponding equations for obtaining the mass, age, and LPV phase duration of LPVs are presented in Appendix \ref{app:app} for all metallicities with details of fitting and related plots. We aimed to use the latest version of the Padova evolutionary model \cite{marigo2017}; however, the final version fails to consider the variability of massive stars, but there are some stars in the halo of NGC\,5128 younger than $100$ Myr. Thus, we used the older release of \cite{marigo2008} for the LPV phase duration  of massive stars ($log(M/M_{\odot})>0.8$).

LPV stars in the AGB (or RSG) phase produce dust which attenuates their light. This effect depends on wavelength, resulting in reddening of the near-IR colors (\citealp{javadi2011b}; \citealp{vassiliadis1993}; \citealp{vanloon1999}, \citeyear{vanloon2005}). To obtain the intrinsic $K_s$-band magnitude, we need to apply a de-reddening process. To correct for circumstellar extinction, we plot the color--magnitude diagram of the LPV stars and theoretical isochrones by \cite{marigo2017}. The slope of the isochrones is related to dust regardless of whether it is oxygenous or carbonaceous in its composition. We expect the stars with $1.5 < M/M_\odot < 4$ birth mass to have become carbon stars due to the carbon-to-oxygen ratio $>1$ in the third dredge-up of nuclear-processed material, but in a low-metallicity environment a lower limit will be considered $1.1 \, M_\odot$ (\citealp{leisenring2008}). In addition, for stars with $M/M_\odot < 1.1$, the third dredge-up has not occurred sufficiently, and for stars with $M/M_\odot > 4$ the nuclear burning of the carbon at the bottom of the convection zone prevents the carbon from enriching the surface; hence these stars are oxygen-rich stars. Therefore, it is understood that the reddening of the carbon stars differs from the oxygen stars. Having corrected the $K_s$-band magnitude and separated oxygenous and carbonaceous stars, the birth mass will be calculated assuming the maximum $K_s$-band brightness achieved by the stellar models.

\subsection{How to  calculate the SFR} \label{sec:sec6-2}

The approach we use to derive the SFH from the LPVs catalogue is outlined as follows:

\begin{itemize}

	\item[$\bullet$] LPVs have achieved peak brightness in near-infrared wavelengths, allowing us to estimate their mass using stellar evolution models. Additionally, being in the advanced stages of their evolution, their mass can be translated into age (stellar lifetime). As a result, the identification of LPVs through comprehensive long-term monitoring surveys is imperative, forming the foundation for subsequent mass and age estimations. Once LPVs are identified, it is crucial to account for two corrections in the analysis. Firstly, due to the potential incompleteness inherent in monitoring surveys, a simulated fraction of potentially missed LPVs needs to be reintegrated into the list. This correction, detailed in Section 8 following \cite{Rejkuba2003a} simulations, takes into consideration various parameters such as magnitude range, period, and amplitude. Consequently, a formula has been developed to quantify this incompleteness. Secondly, the influence of circumstellar dust must be addressed. The bending of isochrones after their peak indicates the presence of circumstellar dust, leading to the dimming and reddening of LPVs. Our methodology for determining star mass and age assumes that stars are at the pinnacle of their isochrones. To account for this, de-reddening equations derived from isochrones slopes are employed to correct magnitudes and restore them to their optimal states. Correcting for circumstellar reddening requires observations in at least two bands, enabling estimation of the star's color.  Hence the correction equation is:
\begin{equation} 
K_0 = K_s + a(1.5 - (J_s - K_s))
\label{eq:eq3}
\end{equation}
where $K_0$ is the corrected magnitude, $a$ is the slope of the isochrone, $K_s$ and $J_s-K_s$ are the observed magnitude and color of each LPV star, respectively. This correction is applied to variable stars with $ J_s - K_s > 1.5$  mag.
	
	\item[$\bullet$] In the subsequent phase, we employ Padova stellar evolutionary models to determine the peak of the isochrones (in the $K_s$ --band) across a range of different ages.  Following this, we establish a correlation between the mass of stars associated with these peaks and their luminosity (table  \ref{tab:taba1}). As mentioned earlier, the LPVs are positioned at the zenith of the isochrones (following correction for circumstellar dust). Therefore, this relationship will provide the mass estimation for each LPV star. Then the age of each LPV star is estimated using the mass--age relationship, taking into account that LPVs represent the endpoint of stellar evolution (table  \ref{tab:taba2}). Utilizing the age--metallicity relationship specific to each galaxy, we establish these correlations for diverse metallicities. Consequently, for each star, both mass and age can be estimated across all available choices of metallicity. To better understand this method, consider a star in the CMD (Fig.\,\ref{fig:fig2}) with $K_s \sim$ 20 mag and $J_s-K_s\sim$ 1 mag. While
the isochrones suggest that this star aligns well with the 100 Myr isochrone, our method assumes
that as an LPV, this star must be at the peak of the isochrone. 
Therefore, we cannot allocate this star to the 100 Myr isochrone because its brightness 
is much fainter than what the 100 Myr isochrone suggests at its peak. Instead, its magnitude suggests that this star aligns well 
with the peak of the isochrone at log t = 9.32. As we know, photometric errors
may cause stars to appear redder or bluer than their actual colors, making
it challenging to align the star with each isochrone. However, since our 
method is not CMD--based and only uses one filter to estimate mass and age,
the effect of photometric errors is negligible in the final results (see section \ref{sec:sec8-1}).\\
We must mention an important uncertainty in the Padova models regarding the evolution of super--AGB stars, which have birth masses in the range of approximately 5–10 $M_\odot$ (\citealp{siess2007}). These models do not compute the entire thermal pulsing phase of super--AGB stars, causing their evolution to appear to terminate prematurely. This uncertainty is evident as an excursion towards fainter $K_s$--band magnitudes in Fig.\,\ref{fig:fig13} for the range of $0.7 < log(M/M_\odot) < 1-1.1$. To address this, we interpolate the Mass--Luminosity relation over this mass range, providing a continuous connection between the final luminosities of AGB stars and those of massive red supergiants.
	
	\item[$\bullet$] An aspect of significant importance concerning the presentation of a SFH is the way it is binned in terms of age. The younger, more massive variable stars are often considerably fewer than the older, low-mass variable stars, and inadequate binning can either lead to spurious peaks in the SFR or mask any such real bursts. From a statistical point of view, an advantage lies in ensuring that each bin contains the same number of stars, thus providing uniform uncertainties to SFR values. To accomplish this, we initiated the process by arranging stars by mass and began counting until a predetermined number was reached. At that point, we commenced counting stars for the subsequent bin. Through this approach, it becomes evident that each bin is associated with a specific mass range, supposing stars with masses between m(t) and m(t+dt) are within one of these bins.\\
 Our methodology assumes that all stars within this mass range should now be identified as LPVs. However, due to statistical limitations (ensuring an adequate number of stars in each bin), the age bin (dt) associated with that mass bin is larger than the LPV phase duration ($\delta t$). In this case, if a star with mass m(t+dt) -- currently identified as an LPV-- formed later in that bin, it will not be recognized as an LPV since it hasn't yet reached the LPV phase when we currently observe it. A similar situation also applies to the lowest mass limit in that bin: if stars with mass m(t), currently identified as LPVs, had formed slightly earlier (beyond the LPV phase duration of those stars) within that same bin, they would not be identified as LPVs because they would have already completed this phase. These were just examples, but the same logic applies to all stars formed within that age bin. 
 \\
 To clarify, let's consider that we want to calculate the SFH between log t=8.98--9.08. During this period, our ability to identify stars is limited to those with masses between 2 and 2.2 $M_\odot$ because they have reached their final evolutionary stage and can be identified as LPVs. For example, consider a star with a mass of 2 $M_\odot$ which  formed $t\sim 1.20$ Gyr ago (log t=9.08) and has presently entered the LPV phase. This star  remains  at this evolutionary stage for $\delta t \sim 2.4$ Myr  (log $\delta t$=6.38), making it recognizable as an LPV. Therefore, since this star will remain in this phase for 2.4 Myr, if it forms 2.4 Myr later, it can still be recognized as an LPV.  However, if its formation occurred outside this timeframe, exceeding the $\delta t$ from the beginning of this age bin, it hasn't yet reached the final evolutionary stage and won't be identified as an LPV.  For the mentioned example, the age bin is almost 250 Myr, several times longer than the LPV phase duration. Hence, we can't identify all stars formed within this specific mass range because some haven't entered the LPV phase yet, or have just finished the phase. This fraction is influenced by the duration of the age bin ($dt$) and the LPV phase duration ($\delta t$) of each star. Therefore, the proportion of recognized LPVs in that bin compared to the overall count of stars formed within the range of m(t+dt) and m(t) in the same bin is equivalent to $\frac{\delta t}{dt}$. 
\\
To address this consideration in our analysis, we developed a relationship between LPV phase duration and mass (table \ref{tab:taba3}). This value serves as a weight for each star, as depicted in Eq.\,\ref{eq:eq1}. Essentially, the inverse of the LPV phase duration for each star is estimated and then aggregated over the specific bin. Finally, we apply the Initial Mass Function (IMF) correction based on the minimum and maximum mass of each bin. Substituting these parameters into Eq.\,\ref{eq:eq1} yields the star formation rate within that specific bin. The luminosity function of LPVs after applying the circumstellar dust correction for stars with J-$K_s>1.5$ mag is illustrated in Fig. \ref{fig:fig_Kmaghis}. Our method involves considering the number of LPVs across different magnitude ranges. However, it's crucial to note that two corrections are applied to the LPVs count. The first correction accounts for LPV phase duration, while the second one involves applying the IMF. Consequently, the $K_s$-band histogram does not  track the SFH (Fig. \ref{fig:fig9}), as expected.

\begin{figure}
	\centerline{
			\epsfig{figure=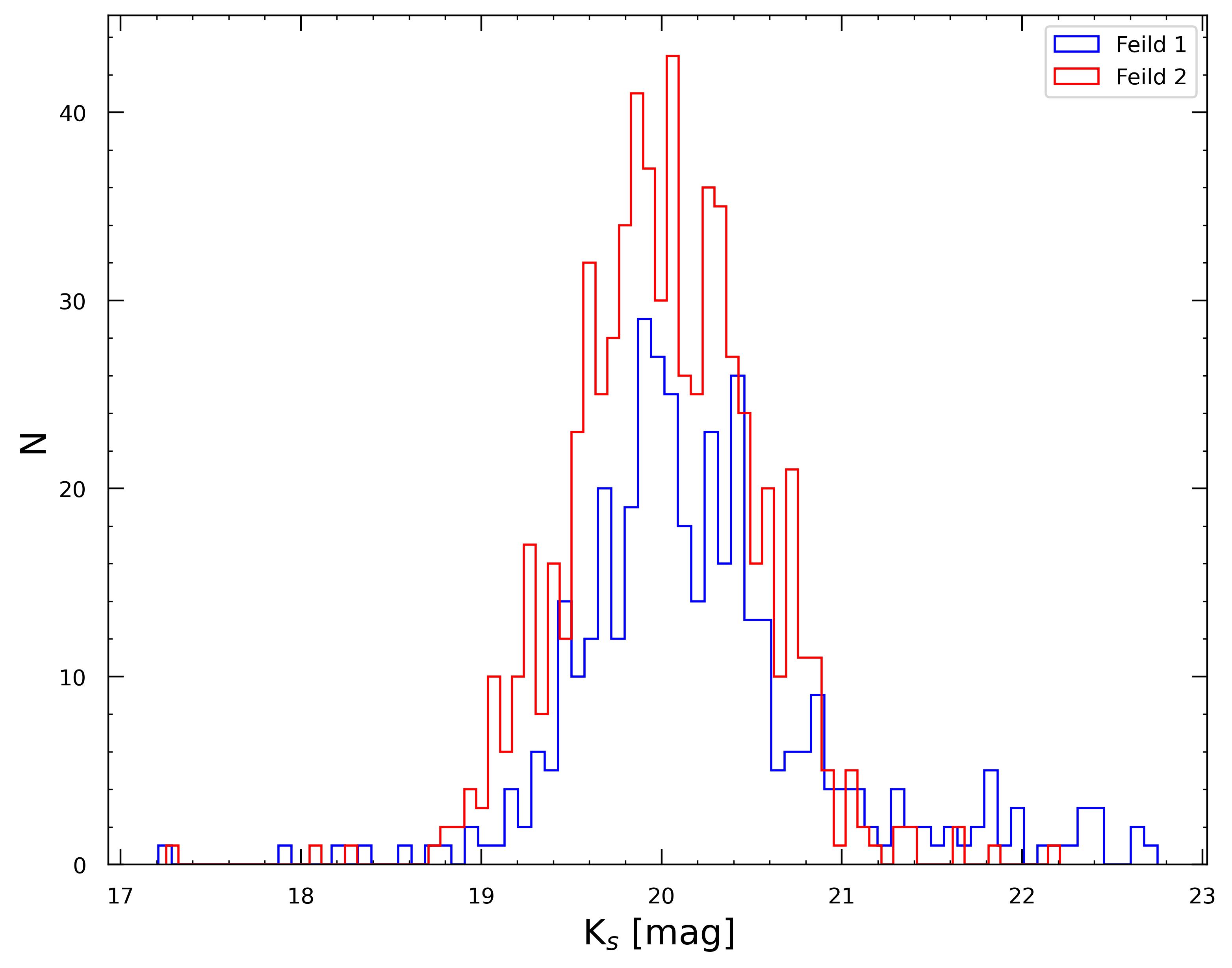,width=90mm}}
	\caption[]{$K_s$-band luminosity function for LPVs. The circumstellar dust correction is applied to stars with J-$K_s$ > 1.5 mag.}
	\label{fig:fig_Kmaghis}
\end{figure}

\item[$\bullet$] Finally, we present the SFH  in two formats. First, as shown in  Fig.\,\ref{fig:fig4}, we display various SFHs, each corresponding to a constant metallicity. It's evident that due to the galaxy's chemical enrichment, the highest metallicity suitable for recent times isn't suitable for older epochs. However, the advantage of this representation lies in the fact that, guided by age--metallicity relationships, the appropriate SFR value for each bin can be selected. These SFHs can be further consolidated into a single graph based on the assumed age--metallicity relationships, as depicted in  Fig.\,\ref{fig:fig9}. It must be noted that while these SFHs can be useful for estimating the SFR in specific age bins based on the appropriate metallicity of those age bins, complications arise when age bins encompass two or more metallicity ranges, making it challenging to estimate the exact SFR. For example, in the Woodley model (see section \ref{sec:sec8-6}), Z=0.003 is appropriate for the age bin log t$\sim$9.90--10.08. However, in the presented SFHs, the SFR for this metallicity is calculated for the age bin log t$\sim$9.70--10.20, which spans age bins suitable for different metallicities. To address this, instead of directly utilizing these SFHs, we derived all the necessary relationships required in Eq.\,\ref{eq:eq1} to construct the SFH, taking into consideration the age-metallicity relationships. For instance, to establish the $K_s$--band--age relationship, we utilized the appropriate metallicity for each age bin as outlined in Table \ref{tab:tab6} and Table \ref{tab:tab7}. Isochrones at the corresponding metallicity for each age duration were utilized to estimate the peak of the isochrone in the $K_s$--band. Finally, lines were fitted to the data points to derive metallicity--dependent age--luminosity relationships, as illustrated in Fig.\,\ref{fig:fig_rel_metallicity}.

\end{itemize}


\section{Results} \label{sec:sec7}

Having considered our sample (section \ref{sec:sec2} and \ref{sec:sec3}) and having applied the method (section \ref{sec:sec6}), we calculated the SFRs in each epoch for different metallicities based on a function of age in the galaxy. The number of stars selected in each bin is a crucial parameter. On the one hand, large numbers of stars in a bin may cause the age interval to be wider and not reveal variation within the bin. In contrast, small numbers of stars per bin can cause confusing and meaningless fluctuations.

As we expect the results to be sensitive to the metallicity, the difference between the actual and calculated SFR in each epoch depends on the difference between the actual and assumed metallicity for the epoch. To trace the change of metallicity over time, we use different metallicities corresponding to stellar populations of different ages. The effect of metallicity on SFR arises when we calculate the mass, age, and LPV phase duration of each star which are explained in Appendix \ref{app:app}. Fig.\,\ref{fig:fig4} shows there is a significant difference in the derived SFRs using different metallicities.  The oldest bin is only partially displayed as it stretches to unrealistically large ages. It is included in the plot solely to illustrate that, as anticipated, the star formation rate we calculate for ages exceeding the Hubble time is negligible for the metallicity range suitable for these epochs (Fig.\,\ref{fig:fig4} left panel). However, as depicted in the right panel of Fig.\,\ref{fig:fig4}, for higher metallicities, this value isn't negligible. Nevertheless, based on the age--metallicity relationships for this galaxy (see section \ref{sec:sec8-6}), these metallicities aren't suitable for the oldest bins; hence, the values derived based on them are not reliable.

The calculated SFRs of Field\,1 and Field\,2 for several constant metallicities are presented in Fig.\,\ref{fig:fig4}. As can be seen, applying different (but constant in time) metallicities results in very similar SFR patterns (SFR variations in time) for both fields. While one field is in the northeastern part (Field\,1) and the other in the south (Field\,2), $\sim28$ kpc apart, they show very similar SFHs. On the other hand, Field\,2 experienced higher SFRs at all ages likely due to its location closer to the centre of the galaxy.

For low to intermediate metallicities in the range $0.0003 \leq Z \leq 0.006$ in both Field\,1 and Field\,2, we do not recover stellar populations older than $\log t({\rm yr}) \ \gsim \ 9.8$ but do find a consistent pattern of two peaks in SFR, one at $\log t({\rm yr}) \sim 9.3$--$9.6$ and another at $\log t({\rm yr}) \ \sim 8.6$--$9.0$ (both older for higher metallicity) -- the latter coincides with a major merger $t\sim 800$ Myr ago (\citealp{israel1998}). For higher metallicities in the range $0.008 \leq Z \leq 0.039$, on the other hand, a significant population of older stars is recovered, with a peak in star formation around $\log t({\rm yr})\sim 9.8$--$10$ ($t\sim 8$--$10$ Gyr) -- such old stars were also reported by \cite{Rejkuba2005}, \cite{kaviraj2005}, and \cite{woodley2010b}, and we suggest here that they be relatively metal-rich. Also in this metallicity range a hike in SFR is seen to occur $t\sim 800$ Myr ago.

\cite{woodley2010b} studied 72 globular clusters (GC) in the halo of NGC\,5128 and noted that more than $85\%$ of these are old and metal-poor. Some of them, though, with an age of $10$ Gyr have metallicities $Z > 0.008$. From Fig.\,\ref{fig:fig5}, it is clear that most of the stars in the halo of NGC\,5128 are older than 1 Gyr, which is also in agreement with their results.

\cite{Rejkuba2004a} investigated recent star formation of NGC\,5128 by simulating the U--V color--magnitude diagram of a field that encompassed Field\,1. For two metallicities $Z = 0.004$ and $Z = 0.008$, changing the slope of Salpeter's initial mass function and starting and ending epochs of star formation, they found that SFRs $100$ Myr ago are in the range of $\sim 9 \times 10^{-5}$ to $\sim 10^{-3}$ M$_\odot$ yr$^{-1}$ kpc$^{-2}$. Assuming metallicity $Z = 0.003$ and $Z = 0.008$ for Field\,1 ($5.7$ kpc$^{2}$), we derive SFR $= 2.2 \times 10^{-3}$ M$_\odot$ yr$^{-1}$ kpc$^{-2}$ and SFR = $9.7 \times 10^{-4}$ M$_\odot$ yr$^{-1}$ kpc$^{-2}$, respectively, for $\log t({\rm yr})\leq 8.7$, which seems reasonable since they mentioned that their results are lower than what they had expected for a giant elliptical galaxy.

\begin{figure*}[ht!]
	\centering{\hbox{
			\epsfig{figure=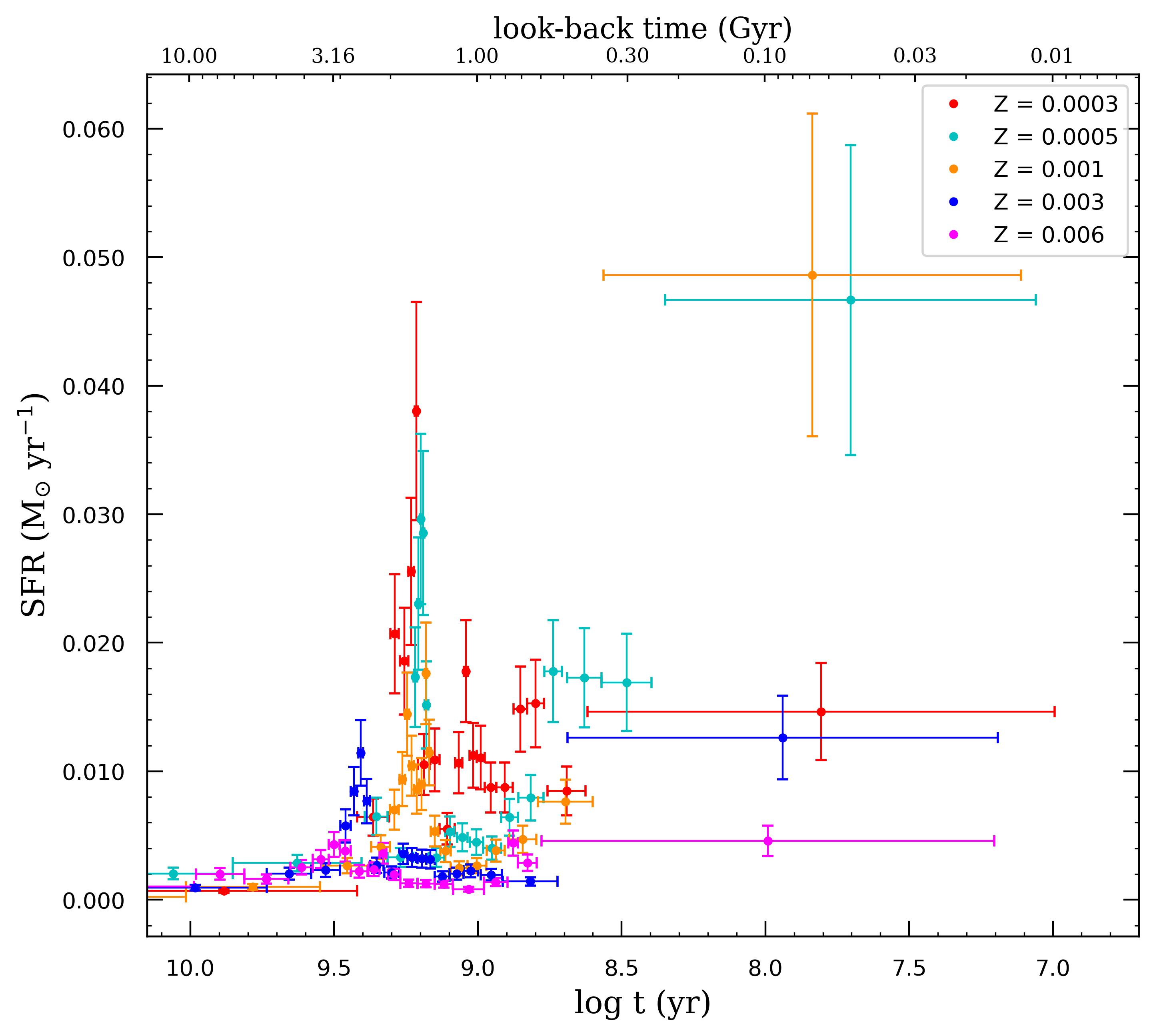,width=90mm}
			\epsfig{figure=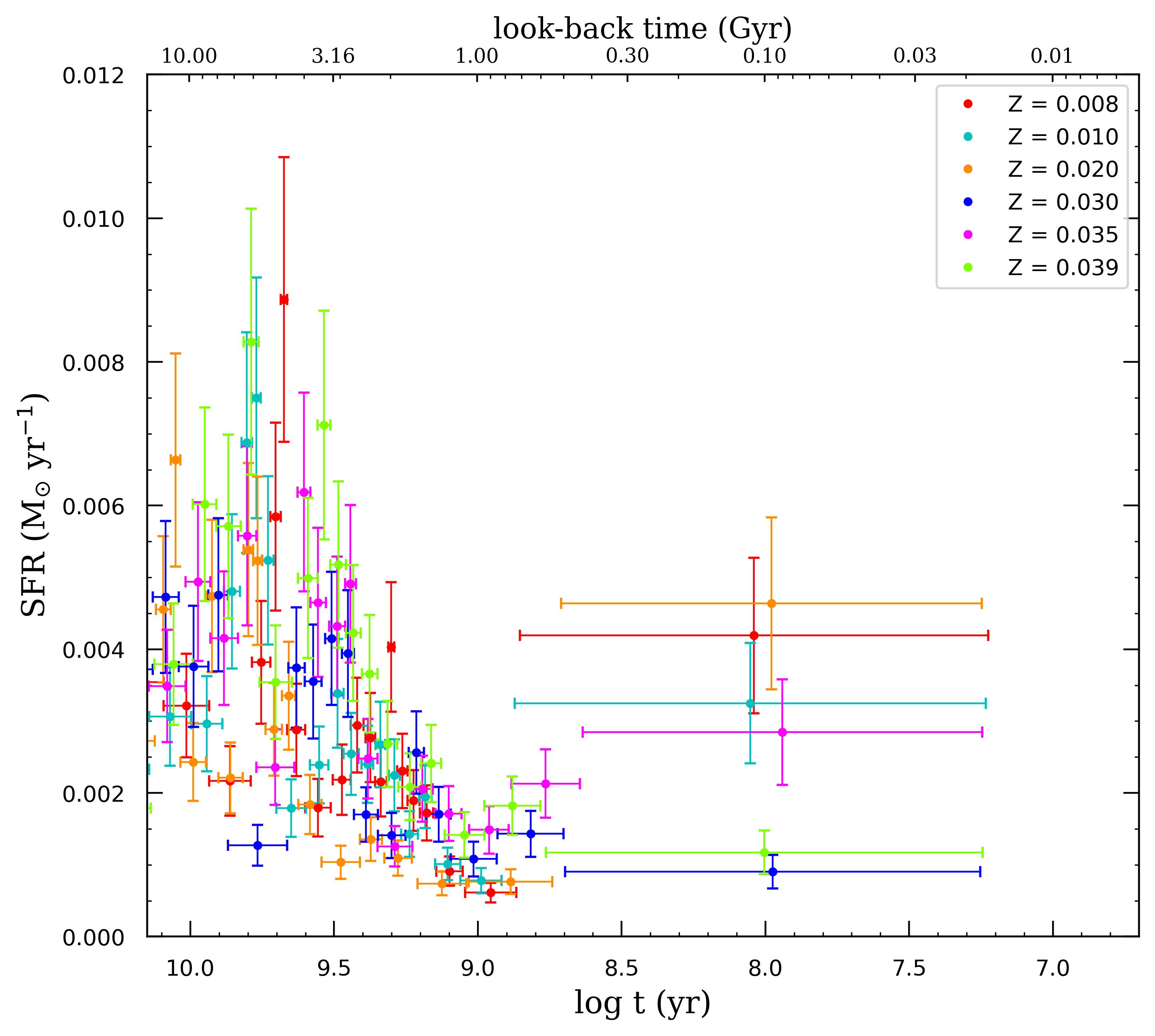,width=90mm}}}
	\centering{\hbox{
			\epsfig{figure=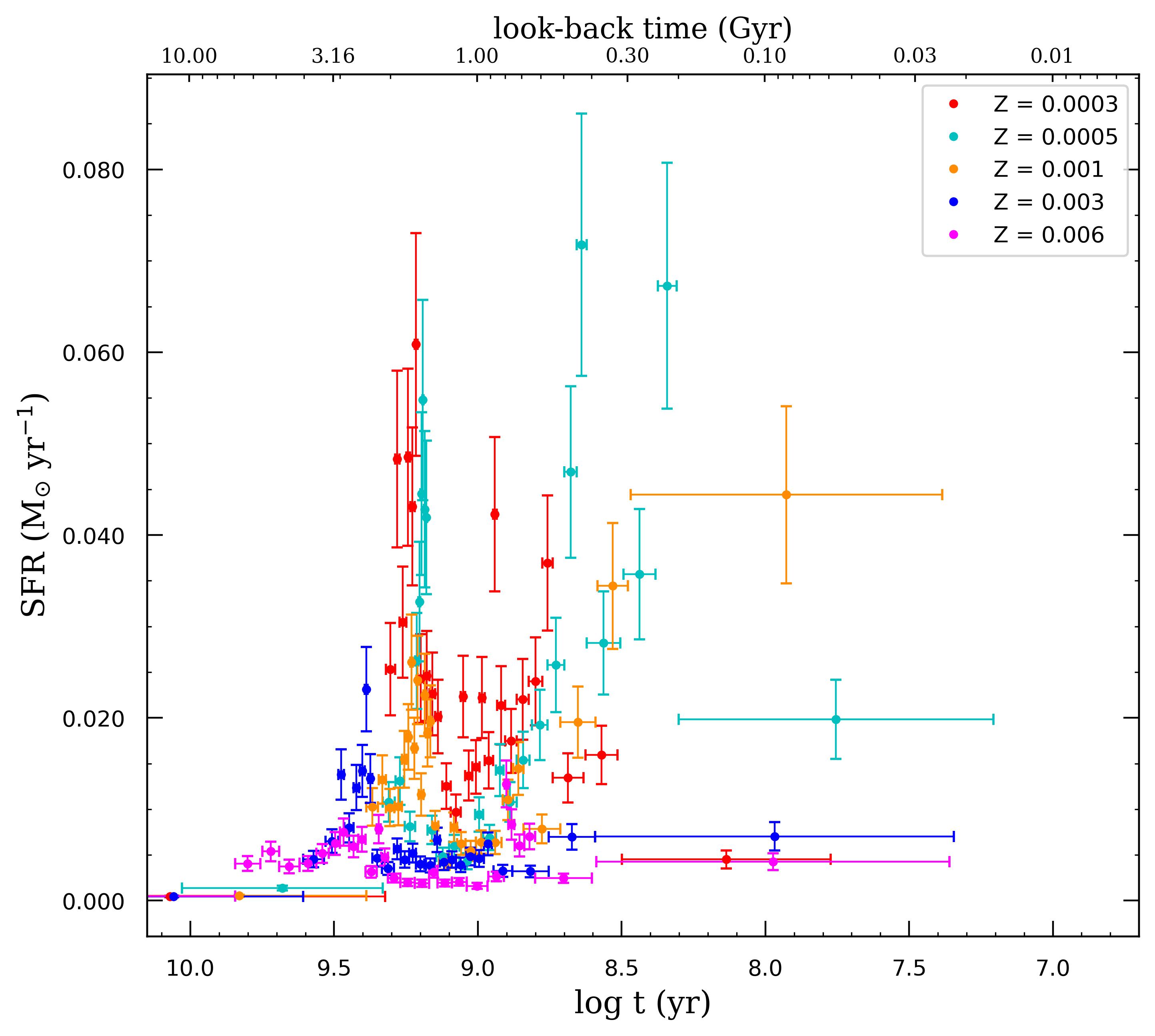,width=90mm}
			\epsfig{figure=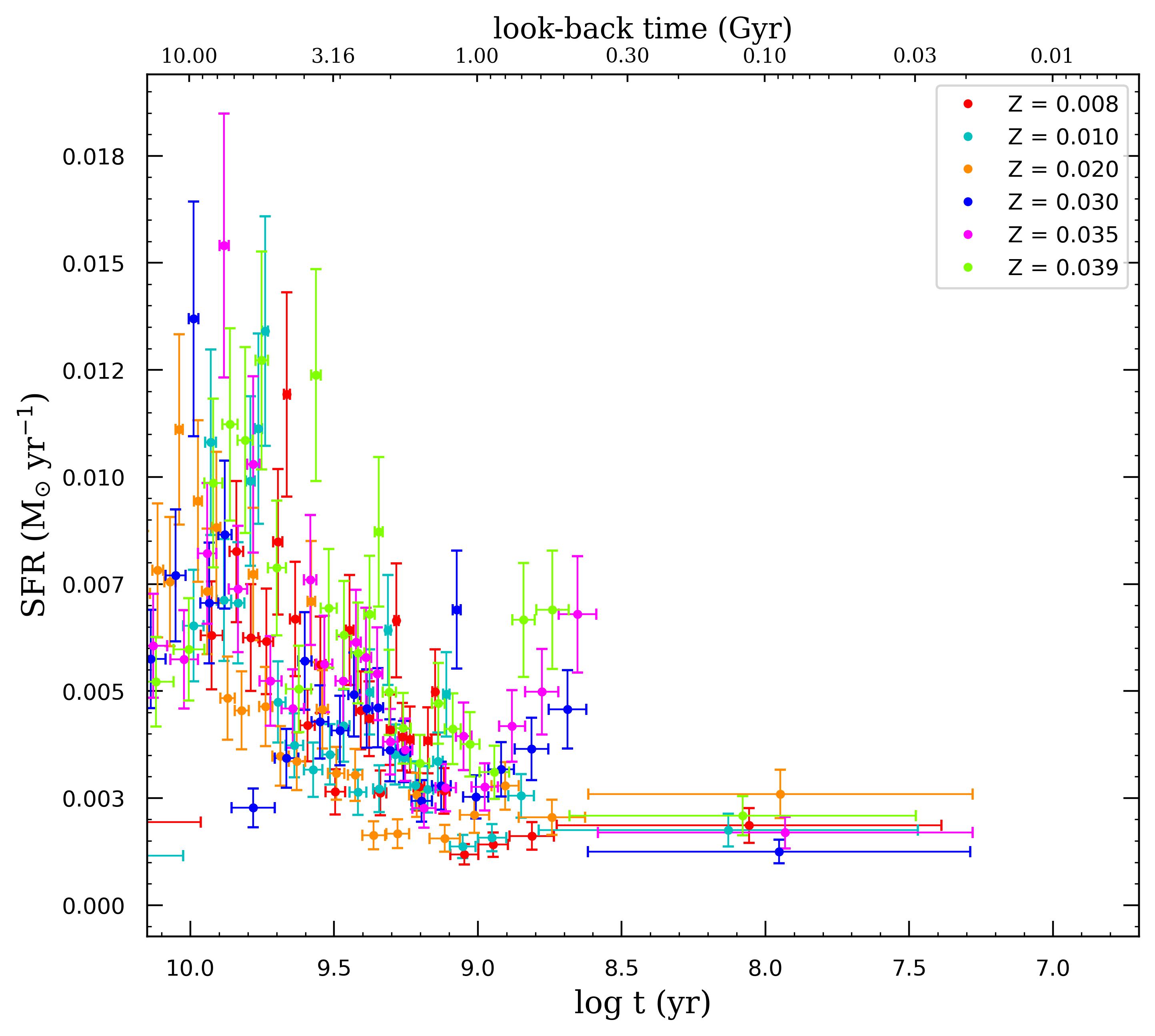,width=90mm}}}		
	\caption{Calculated SFHs for different metallicities for Field\,1 (upper panel) and Field\,2 (lower panel). The sample includes $395$ and $671$ LPVs with periods longer than $70$ days for Fields\,1 and 2, respectively.}
	\label{fig:fig4}
\end{figure*} 

Figure \ref{fig:fig5} presents the LPVs age histograms for two metallicities $Z = 0.01$ and $Z = 0.02$ for both fields. It highlights the presence of a distinct star formation epoch around $\log t({\rm yr})\sim9.3$ at sub-solar metallicity, which however disappears if the populations have solar metallicity.

\begin{figure*}
	\centerline{\hbox{
			\epsfig{figure=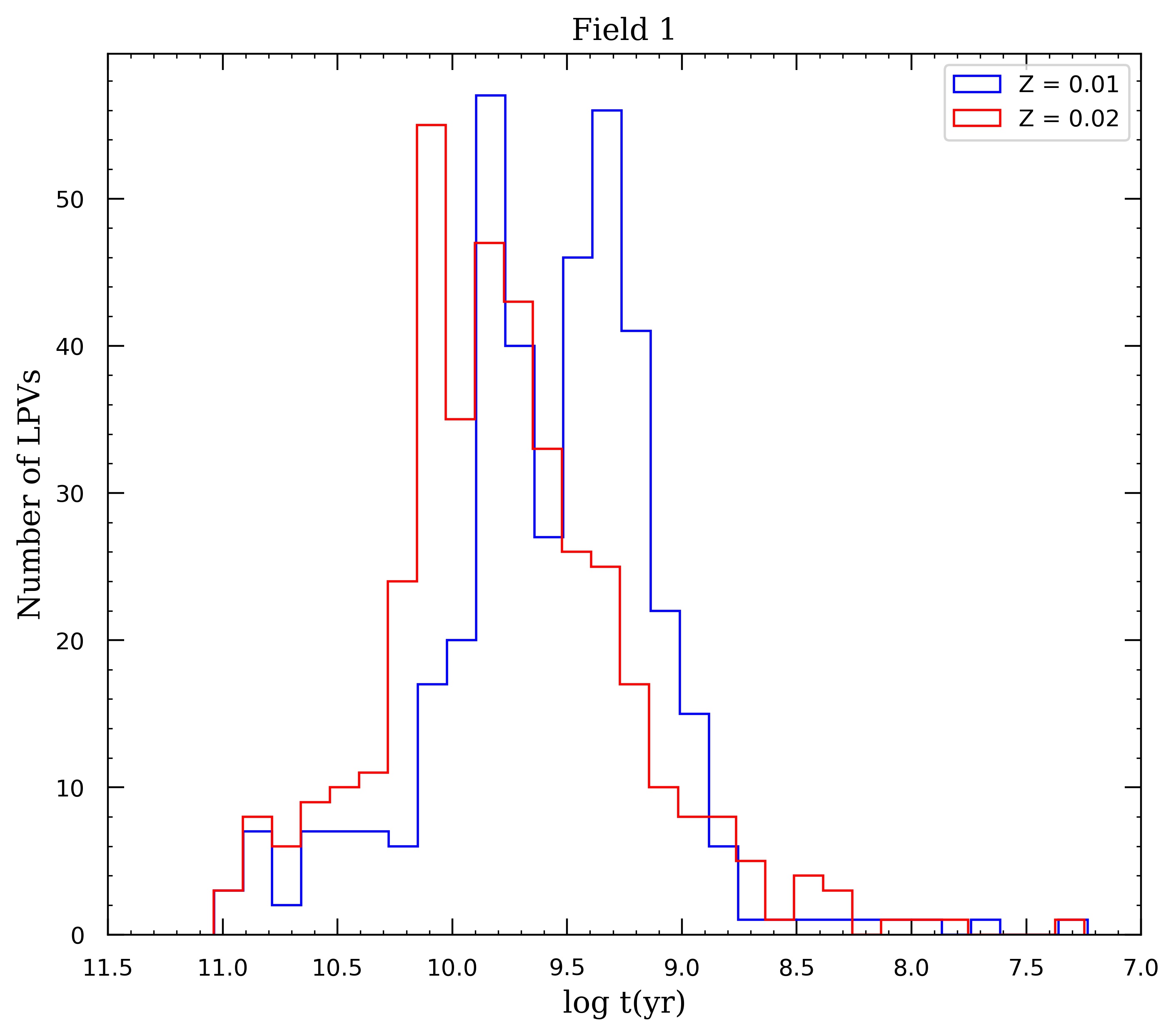,width=90mm}
			\epsfig{figure=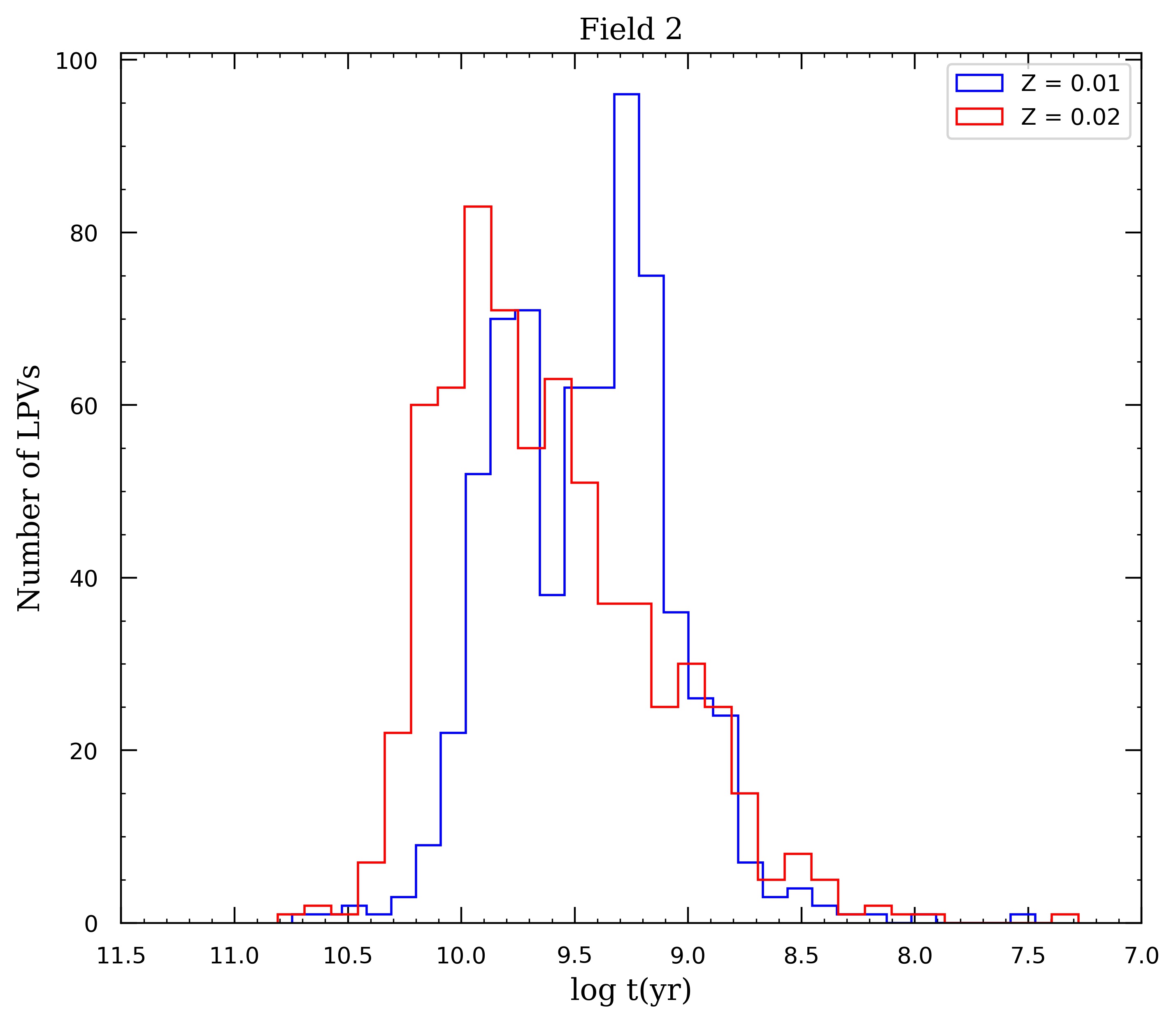,width=90mm}}}
	\caption[]{Age histogram of LPVs for two metallicities ($Z = 0.01$ and $Z = 0.02$) in Field\,1 and Field\,2 where $t$ is look--back time.}
	\label{fig:fig5}
\end{figure*}


\section{Accounting for Observational Bias} \label{sec:sec8}

\subsection{Photometric Uncertainties} \label{sec:sec8-1}

In this section, we investigate the impact of accounting for photometric uncertainties 
on the derived SFH. Our method is not based on CMD studies;
instead, it focuses on the  brightness of stars in the $K_s$--band. When we consider the error
in magnitude, the star's brightness undergoes slight variations, resulting in a
corresponding adjustment in its assigned mass (and consequently, its age).

\begin{figure*}
	\centerline{\hbox{
			\epsfig{figure=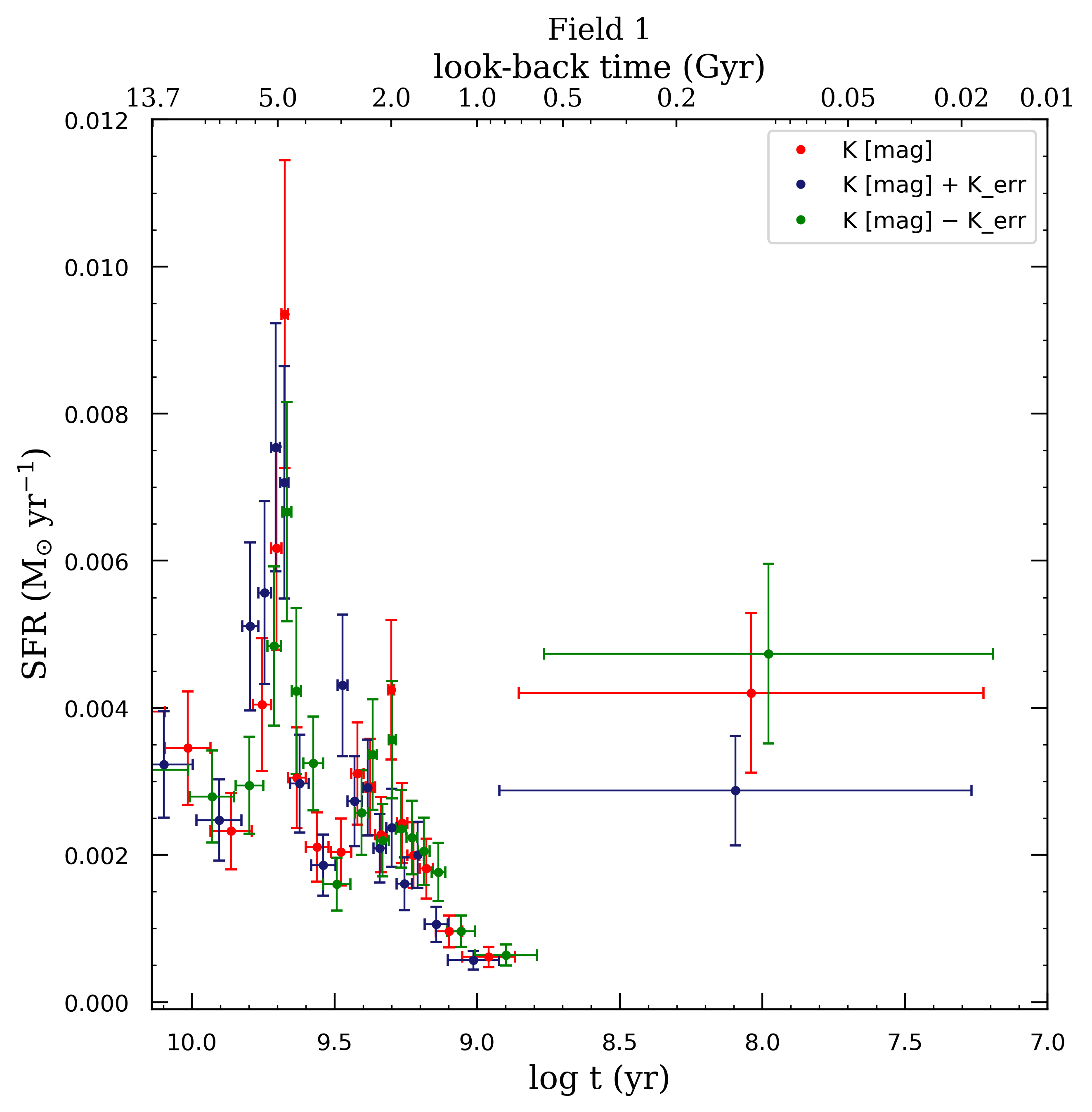,width=90mm}
			\epsfig{figure=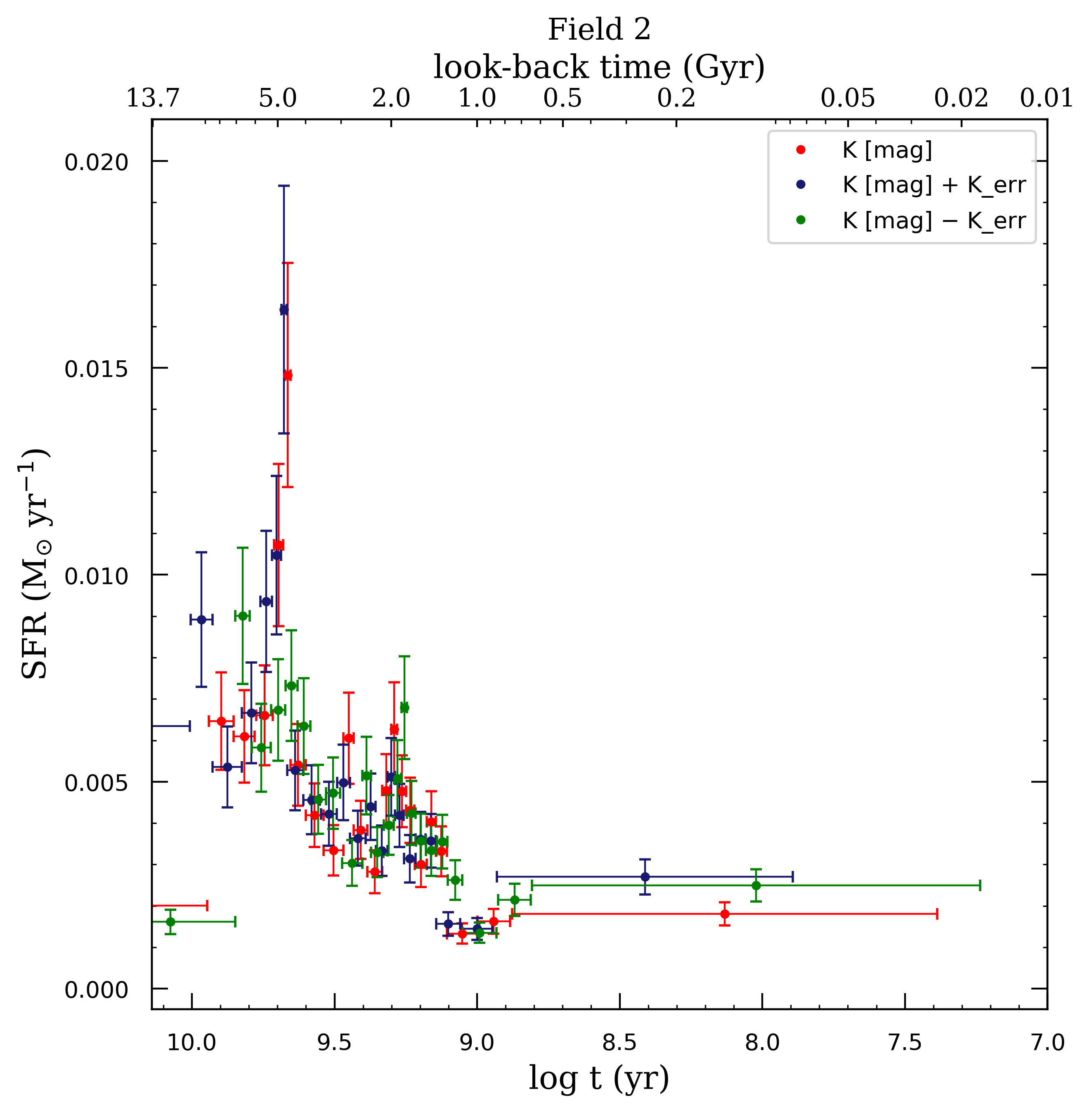,width=90mm}}}
	\caption[]{SFHs  derived by considering the photometric errors for Field\,1 ({\it left panel}) and Field\,2 ({\it right panel}) at Z=0.008 }
	\label{fig:fig_er}
\end{figure*}

Fig.\,\ref{fig:fig_er} illustrates the SFH  for Z=0.008 when we incorporate
photometric errors in the $K_s$--band from the mean magnitude. Overall, we observe a
similar behavior in the SFH with comparable epochs of star formation as seen when
using the mean magnitude (table \ref{tab:tab4} and table \ref{tab:tab5}). However, it's not 
surprising to find that the SFH shifts towards earlier or later epochs when
we add or subtract photometric errors. Nevertheless, the overall effect
of these errors on the SFH is negligible.

Given our focus on LPVs, whose brightness typically exceeds that of the RGB--tip
in surveys deep enough to cover red giant branch stars, photometric errors for LPVs
are generally lower (mostly less than 0.1 mag). As mentioned, our method doesn't rely 
on CMD--based studies and solely utilizes LPVs'brightness to estimate their
mass and age. Consequently, photometric errors cannot lead to stars being 
associated with the wrong isochrone, a common challenge encountered
in CMD--based studies. This highlights a strength of our approach. For example,
consider an LPV star with $K_s$=21 mag. The models suggest that this star should
have a mass of   0.95 $M_\odot$ and an age of 10.8 Gyr (at Z=0.008). Accounting
for a photometric error of 0.1 mag in the $K_s$--band, the mass varies 
between 0.91--0.99 $M_\odot$ and the age varies between 9.5--12.4 Gyr (log t=9.97-10.09).
Notably, the width of the age bins used to derive the star formation during these
times is log dt$\sim 0.2$ (see Fig.\,\ref{fig:fig_er}), significantly larger than the age shift caused by
accounting for photometric errors.

\subsection{Comparison with BaSTI models}\label{sec:sec8-2}

It's important to emphasize that our method relies on model predictions, and among the available models, the Padova models are the only ones providing essential information about the pulsation behavior, a key component of our approach. While there are other models that simulate the thermal--pulsing AGB evolution (such as the BaSTI models by \citealp{pietrinferni2004}, \citeyear{pietrinferni2021}) these models do not cover super--AGB stars like the Padova models. 

\begin{figure*}[ht!]
	\centering{\hbox{
			\epsfig{figure=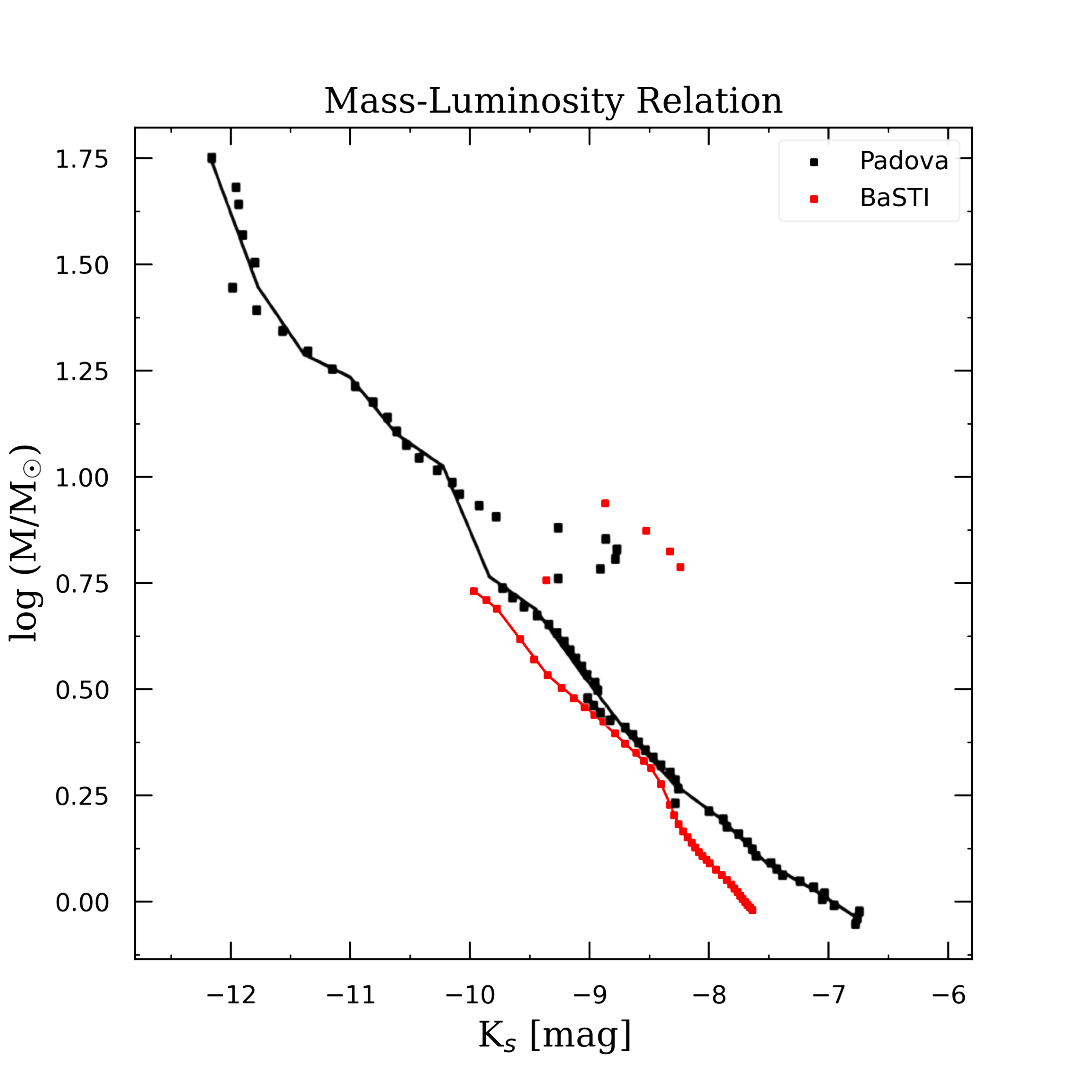,width=90mm}
			\epsfig{figure=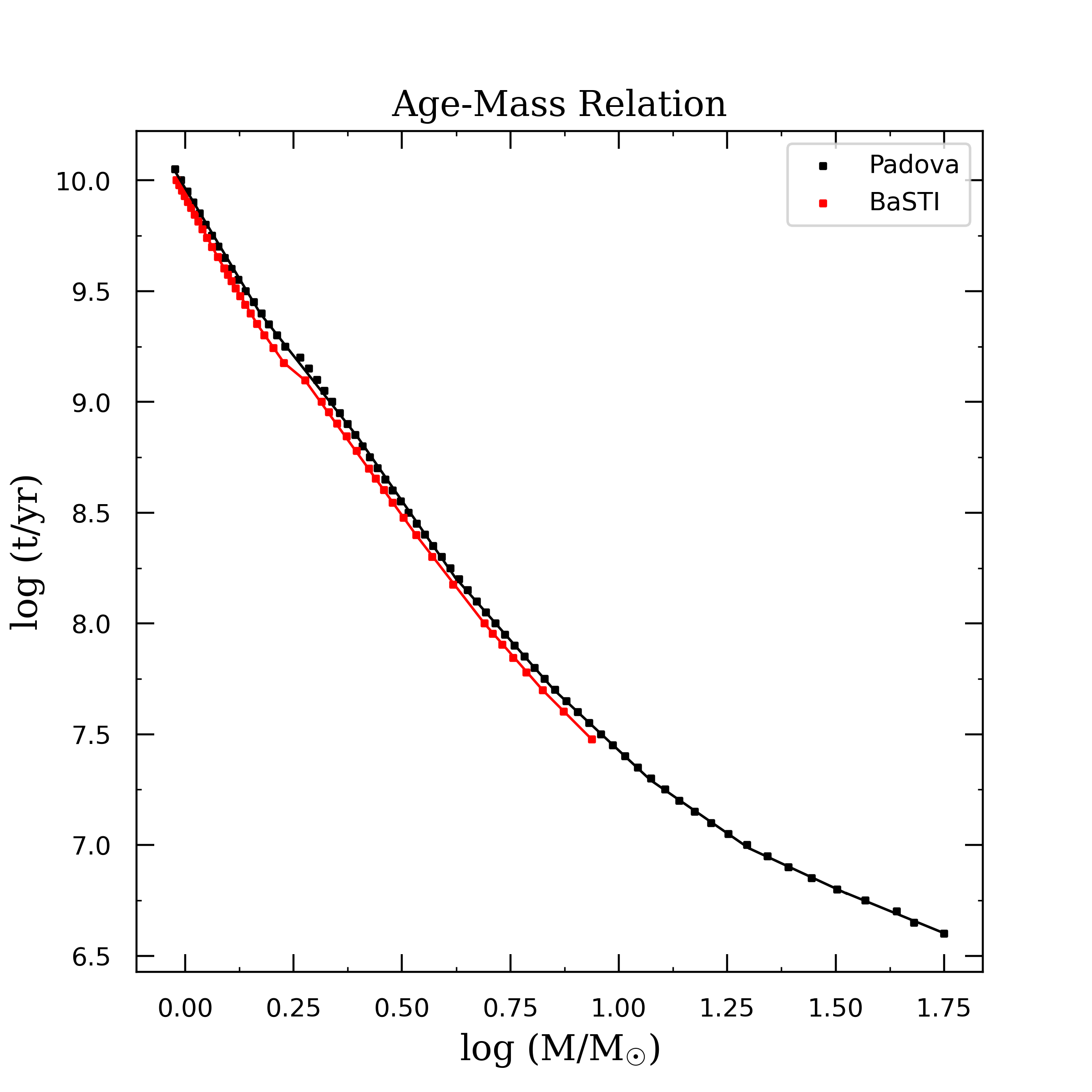,width=90mm}}}
	\caption{The mass--luminosity ({\it left panel}) and mass--age ({\it right panel}) relations are depicted, considering Padova models ({\it black line}) and BaSTI models ({\it red line}) at Z=0.008.}
	\label{fig:fig_BaSTI}
\end{figure*}

Here we compare the Mass--Luminosity and Mass--Age relations between the BaSTI and Padova models (Fig.\,\ref{fig:fig_BaSTI}). The main difference is that the faintest stars, around $\sim -7.63$ mag, are less massive and older when using the BaSTI models ($\sim 9.9$ Gyr compared to 3.8 Gyr when using the Padova models), but the difference is negligible for intermediate--age stars (about a Gyr old). Additionally, as can be seen in the right panel of Fig.\,\ref{fig:fig_BaSTI}, the BaSTI models don't cover the mass range of RSGs, which is important when deriving the recent star formation history from our method.

As indicated in Eq.\,\ref{eq:eq1}, one of the parameters necessary for deriving the SFH is the LPV phase duration of LPVs ($\delta t$). This parameter is essential for converting the count of identified LPVs to the SFR. However, BaSTI models do not provide any estimation of $\delta t$, hence we cannot derive the SFH using the BaSTI models.

\subsection{Simulated Data and the Probability Function} \label{sec:sec8-3}

NGC\,5128 is a distant galaxy, hence it is unlikely to detect all of its LPVs. The likelihood for an LPV to be identified depends on its mean brightness, period, and amplitude. Finding this probability is a key point in the method since the number of detected LPVs directly affects the SFH results. 

To determine the completeness limit, \cite{Rejkuba2003a} simulated LPVs and compared them with the catalog for $K_s < 20.5$ mag and $K_s > 20.5$ mag separately.
Their simulated variable stars span a magnitude range of $19.5 < K_s < 21.5$, with $K_s$-band amplitudes between $0.1$ and $1.4$ mag for two constant periods of $100$ and $450$ days, and for an amplitude of $\Delta K_s = 0.7$ mag for periods from $50$ to $1100$ days.

To find the multi-dimensional probability function, we used the simulation results from \cite{Rejkuba2003a} (detailed information is presented in Appendix \ref{app:appB}). The first part is a period-probability diagram for $\Delta K_s = 0.7$ mag, while the next parts are amplitude-probability diagrams for $P = 100$ and $P = 450$ days, respectively. We now need to determine a function that maintains the dependence on period and amplitude since they will affect one another. The cross-terms are expected to satisfy the condition since this is what makes statistical sense. By using functions for $f(A)$ and $f(P)$, we obtain $f(A,P) = f(A) \times f(P)$.

The shape of the period-probability diagram (\citealp{Rejkuba2003a} Figs.\ 12 and 14) suggests the need for a high-order polynomial function, while the amplitude function might be represented by an error function -- it has both the statistical significance and the functional shape we need. The error function (also called the Gaussian error function), $erf$, is a complex function of a complex variable:

\begin{equation}
erf\ z = \frac{2}{\sqrt{\pi}}\ {\int_{0}^{z}e^{-t^2}\ dt}
\label{eq:eq3}
\end{equation}

It is not possible to fit the integral form of $erf$ to the data. Instead, we use an approximation (\citealp{Abromowitz1972}):

\begin{equation}
\begin{gathered}
erf\ A = 1 - [\alpha_1(\frac{1}{1 + \beta A}) + \alpha_2(\frac{1}{1 + \beta A})^2 \\
+ \alpha_3(\frac{1}{1 + \beta A})^3] e^{-A^2} 
\label{eq:eq4}
\end{gathered}
\end{equation}

\begin{figure*}[ht!]
	\centering{\hbox{
			\epsfig{figure=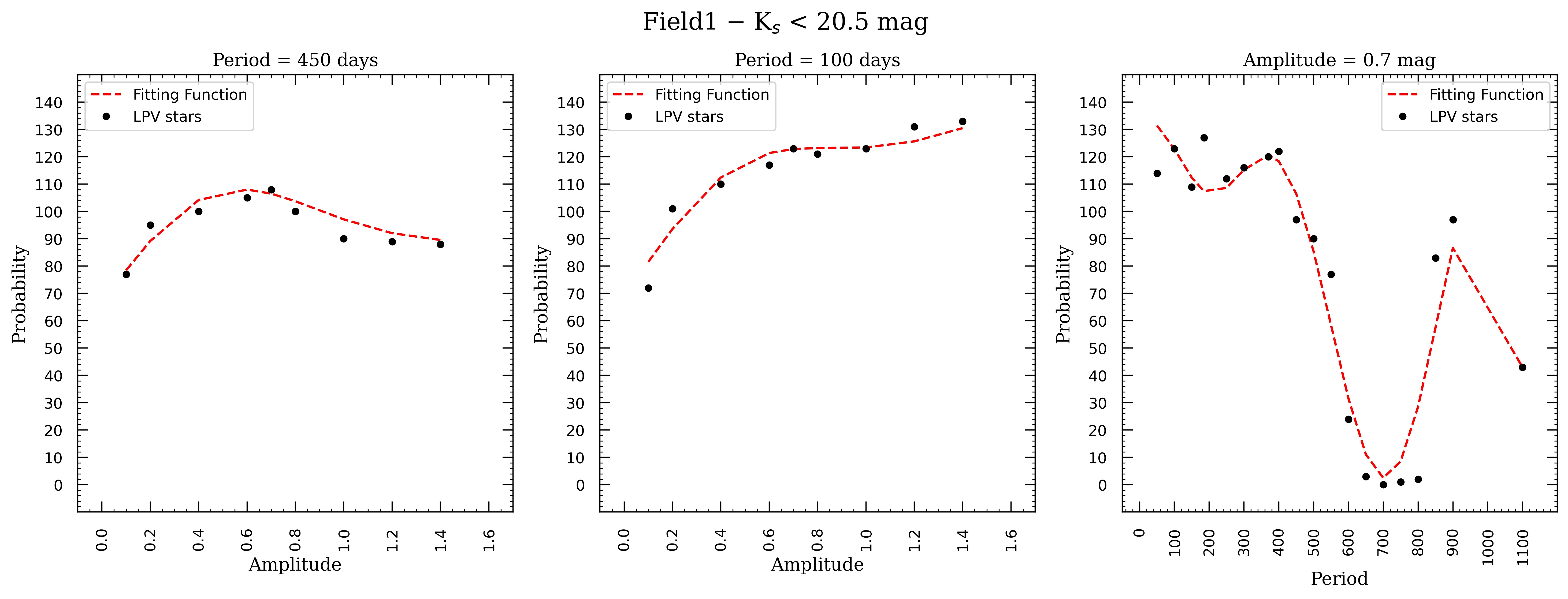,width=180mm}}}
	\caption{The result of fitting a probability function to the data with $K_s < 20.5$ mag in Field\,1. More details are presented in Appendix \ref{app:appB}.}
	\label{fig:fig6}
\end{figure*}

We thus find two functions, Eq.\,\ref{eq:eq5}, valid for $K_s > 20.5$ mag, and Eq.\,\ref{eq:eq6}, valid for $K_s < 20.5$ mag, whose coefficients are presented in table \ref{tab:tab2} and \ref{tab:tab3}. Fig.\,\ref{fig:fig6} illustrates the accuracy of the fitting function on the data belonging to Field\,1 for $K_s < 20.5$ mag -- see Appendix \ref{app:appB} for the other fits. We must note that the probability of bright stars exceeds $100\%$. There are two critical reasons assigned by \cite{Rejkuba2003a}. The first one is about the moving of some faint stars into brighter magnitude bins (Malmquist bias). The second one is related to false detection or migration which happens because of blending with original stars in the images.

\begin{equation}
\begin{gathered}
f(A,P) = (1 - [\alpha_1(\frac{1}{1 + \beta A}) + \alpha_2(\frac{1}{1 + \beta A})^2 \\
+ \alpha_3(\frac{1}{1 + \beta A})^3] e^{-\eta A^2}) \times (\gamma_1 P + \gamma_2 P^2 + \gamma_3 P^3
+ \gamma_4 P^4 \\+ \gamma_5 P^5 + \gamma_6 P^6 + \gamma_7 P^7) + \delta_1 \frac{A^2}{P^2} + \delta_2 \frac{A^3}{P^3} \\
+ \delta_3 \frac{A^4}{P^4} + \delta_4 \frac{A^2}{P}
\label{eq:eq5}
\end{gathered}
\end{equation}

\begin{equation}
\begin{gathered}
f(A,P) = (1 - [\alpha_1(\frac{1}{1 + \beta A}) + \alpha_2(\frac{1}{1 + \beta A})^2 \\
+ \alpha_3(\frac{1}{1 + \beta A})^3] e^{-\eta A^2}) \times (\gamma_1 P + \gamma_2 P^2 + \gamma_3 P^3
+ \gamma_4 P^4 \\+ \gamma_5 P^5 + \gamma_6 P^6 + \gamma_7 P^7) + \delta_1 \frac{A^2}{P^2} + \delta_2 \frac{A^3}{P^3} + \delta_3 \frac{A^4}{P^4}
\label{eq:eq6}
\end{gathered}
\end{equation}

\begin{table}[]
	\centering
	\caption{Coefficients of the probability function of Field\,1 and Field\,2 for simulated variable stars of $K_s > 20.5$ mag related to Eq.\,\ref{eq:eq5}.}
	\label{tab:tab2}
	\begin{tabular}{cccc}
		\hline
		N  & Symbol     & Coefficient -- Field\,1      & Coefficient -- Field\,2      \\ \hline
		1  & $\alpha_1$ & $-$1.305 $\times$ 10$^2$     & $-$3.782 $\times$ 10$^{3}$   \\
		2  & $\alpha_2$ & $-$96.808                    & $-$2.639 $\times$ 10$^{3}$   \\
		3  & $\alpha_3$ & 2.385 $\times$ 10$^2$        & 6.424 $\times$ 10$^{3}$      \\
		4  & $\beta$    & 0.322                        & 6.483 $\times$ 10$^{-4}$     \\
		5  & $\eta$     & 2.008                        & 1.935                        \\
		6  & $\gamma_1$ & 8.166 $\times$ 10$^{-2}$     & 7.553 $\times$ 10$^{-1}$     \\
		7  & $\gamma_2$ & $-$7.452 $\times$ 10$^{-4}$  & $-$7.021 $\times$ 10$^{-3}$  \\
		8  & $\gamma_3$ & 3.344 $\times$ 10$^{-6}$     & 3.006 $\times$ 10$^{-5}$     \\
		9  & $\gamma_4$ & $-$7.911 $\times$ 10$^{-9}$  & $-$6.722 $\times$ 10$^{-8}$  \\
		10 & $\gamma_5$ & 9.873 $\times$ 10$^{-12}$    & 8.036 $\times$ 10$^{-11}$    \\
		11 & $\gamma_6$ & $-$6.133 $\times$ 10$^{-15}$ & $-$4.868 $\times$ 10$^{-14}$ \\
		12 & $\gamma_7$ & 1.493 $\times$ 10$^{-18}$    & 1.174 $\times$ 10$^{-17}$    \\
		13 & $\delta_1$ & 3.588 $\times$ 10$^5$        & $-$1.592 $\times$ 10$^5$     \\
		14 & $\delta_2$ & $-$2.256 $\times$ 10$^7$     & 1.034 $\times$ 10$^7$        \\
		15 & $\delta_3$ & 1.000                        & 1.000                        \\
		16 & $\delta_4$ & 2.038 $\times$ 10$^3$        & 1.106 $\times$ 10$^3$        \\ \hline
	\end{tabular}
\end{table}

\begin{table}[]
	\centering
	\caption{Coefficients of the probability function of Field\,1 and Field\,2 for simulated variable stars of $K_s < 20.5$ mag related to Eq.\,\ref{eq:eq6}.}
	\label{tab:tab3}
	\begin{tabular}{cccc}
		\hline
		N  & Symbol     & Coefficient -- Field\,1         & Coefficient -- Field\,2         \\ \hline
		1  & $\alpha_1$ & $-$37.589                       & 6.004 $\times$ 10$^{-1}$        \\
		2  & $\alpha_2$ & $-$29.358                       & $-$1.414                        \\
		3  & $\alpha_3$ & 67.154                          & 2.561                           \\
		4  & $\beta$    & 1.276 $\times$ 10$^{-2}$        & 3.817                           \\
		5  & $\eta$     & 2.122                           & 5.385 $\times$ 10$^{-2}$        \\
		6  & $\gamma_1$ & 2.432                           & 4.144                           \\
		7  & $\gamma_2$ & $-$2.497 $\times$ 10$^{-2}$     & $-$4.522 $\times$ 10$^{-2}$     \\
		8  & $\gamma_3$ & 1.197 $\times$ 10$^{-4}$        & 2.204 $\times$ 10$^{-4}$        \\
		9  & $\gamma_4$ & $-$2.927 $\times$ 10$^{-7}$     & $-$5.411 $\times$ 10$^{-7}$     \\
		10 & $\gamma_5$ & 3.721 $\times$ 10$^{-10}$       & 6.936 $\times$ 10$^{-10}$       \\
		11 & $\gamma_6$ & $-$2.338 $\times$ 10$^{-13}$    & $-$4.432 $\times$ 10$^{-13}$    \\
		12 & $\gamma_7$ & 5.740 $\times$ 10$^{-17}$       & 1.114 $\times$ 10$^{-16}$       \\
		13 & $\delta_1$ & 3.359 $\times$ 10$^5$           & 2.456 $\times$ 10$^4$           \\
		14 & $\delta_2$ & $-$9.117 $\times$ 10$^6$        & 3.267 $\times$ 10$^6$           \\
		15 & $\delta_3$ & 1.000                           & 1.000                           \\ \hline
	\end{tabular}
\end{table}

\subsection{Effect of Probability Function on SFR} \label{sec:sec8-4}

\begin{figure*}[ht!]
	\centering{\hbox{
			\epsfig{figure=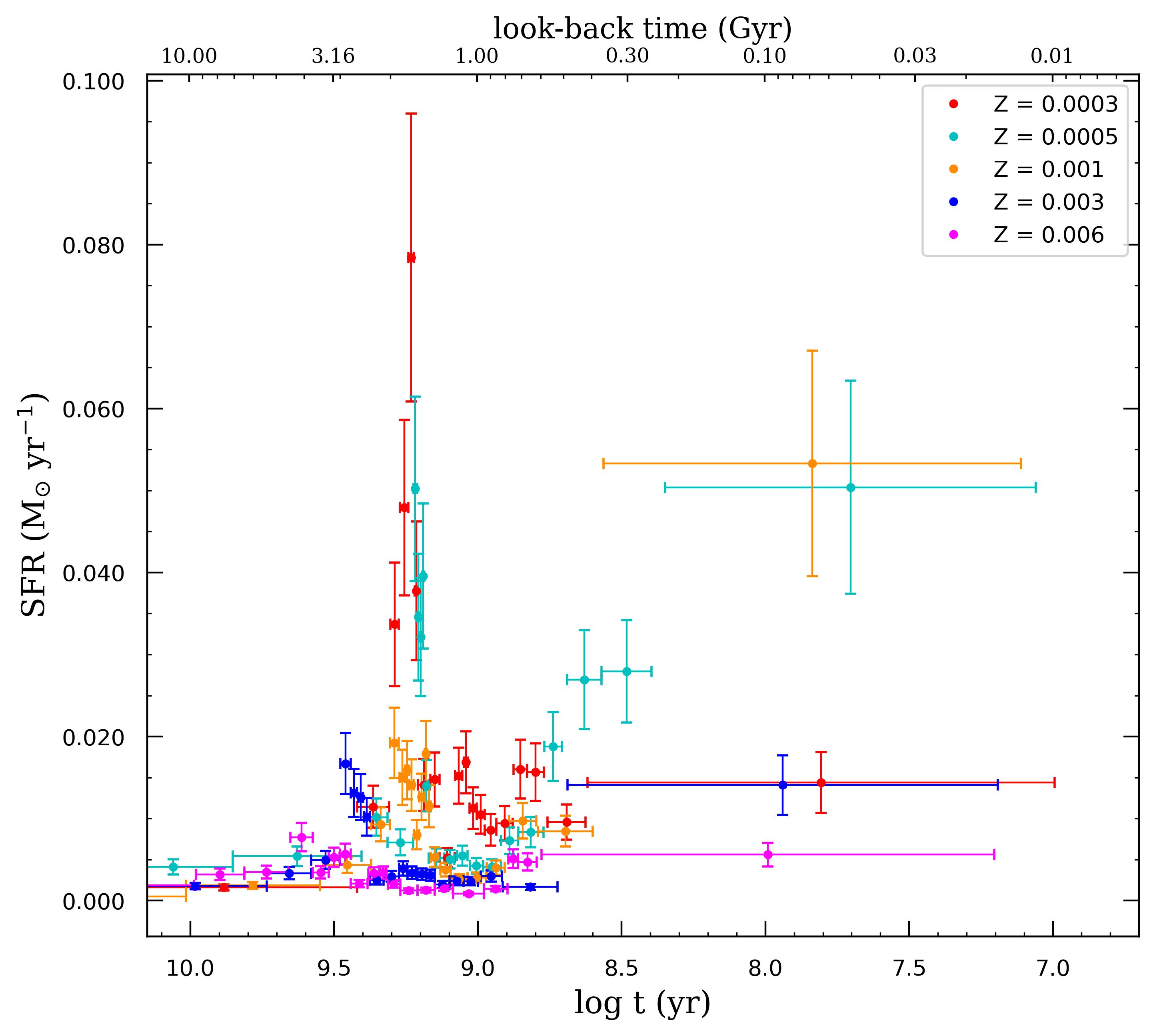,width=90mm}
			\epsfig{figure=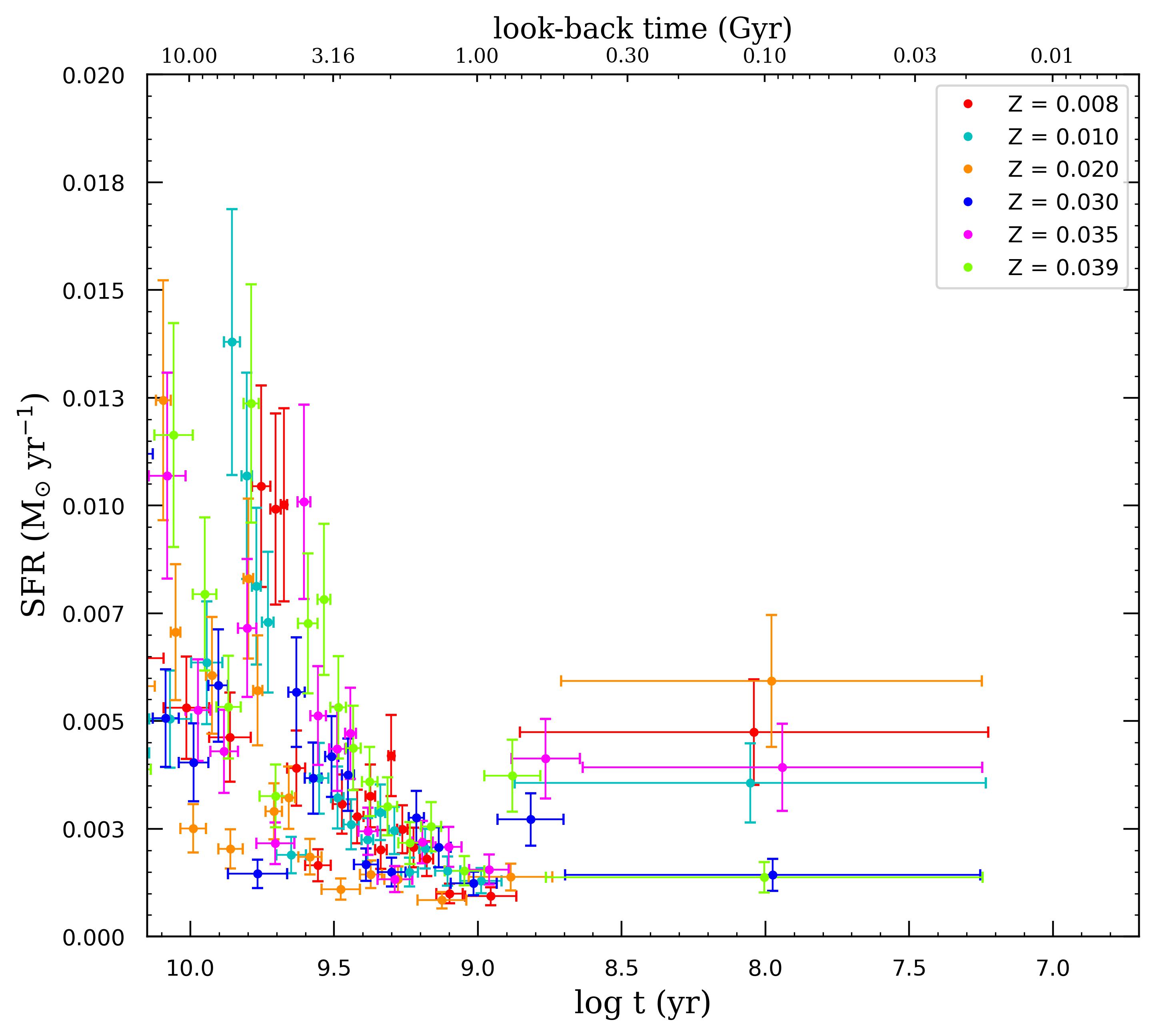,width=90mm}}}
	\centering{\hbox{
			\epsfig{figure=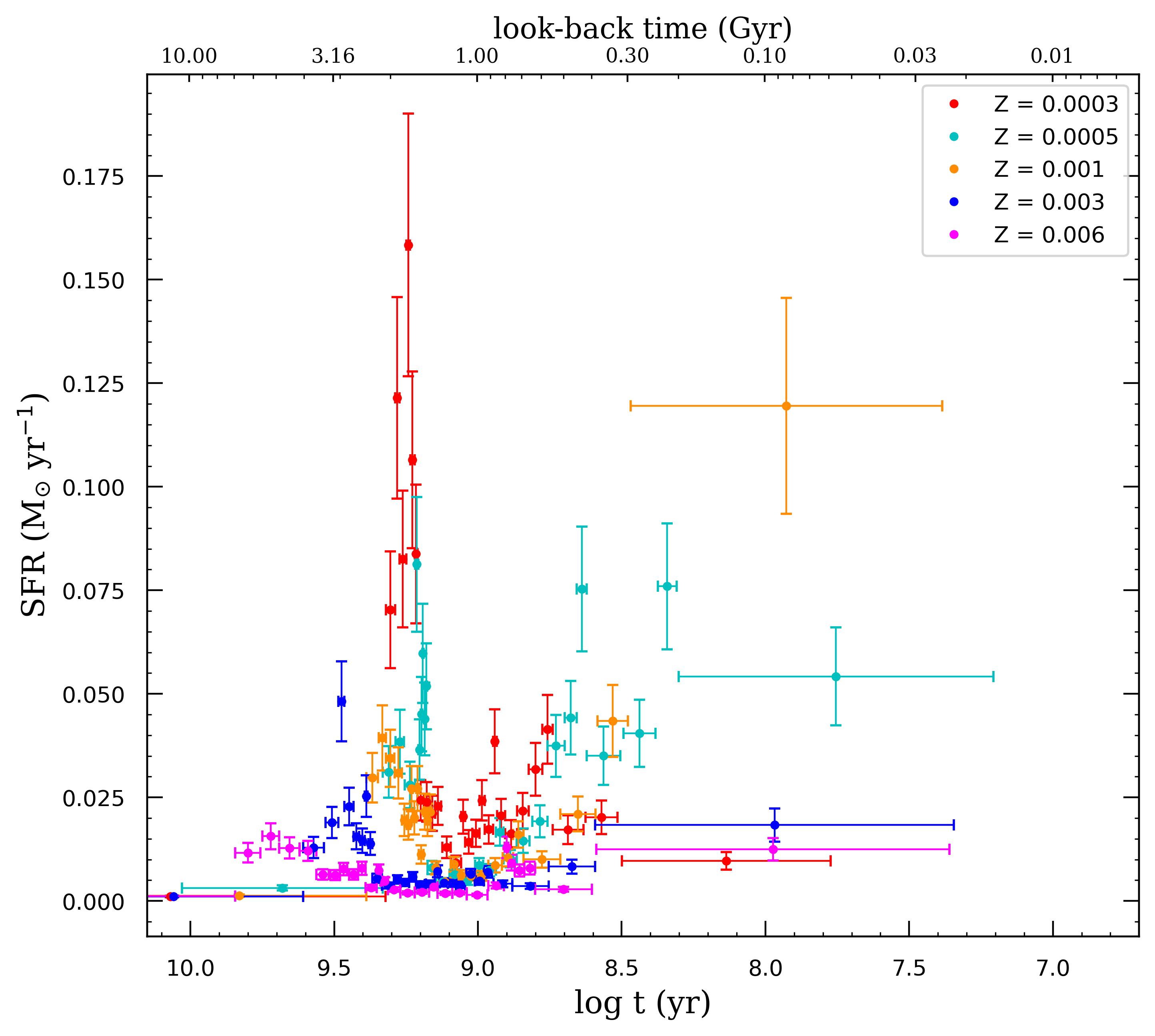,width=90mm}
			\epsfig{figure=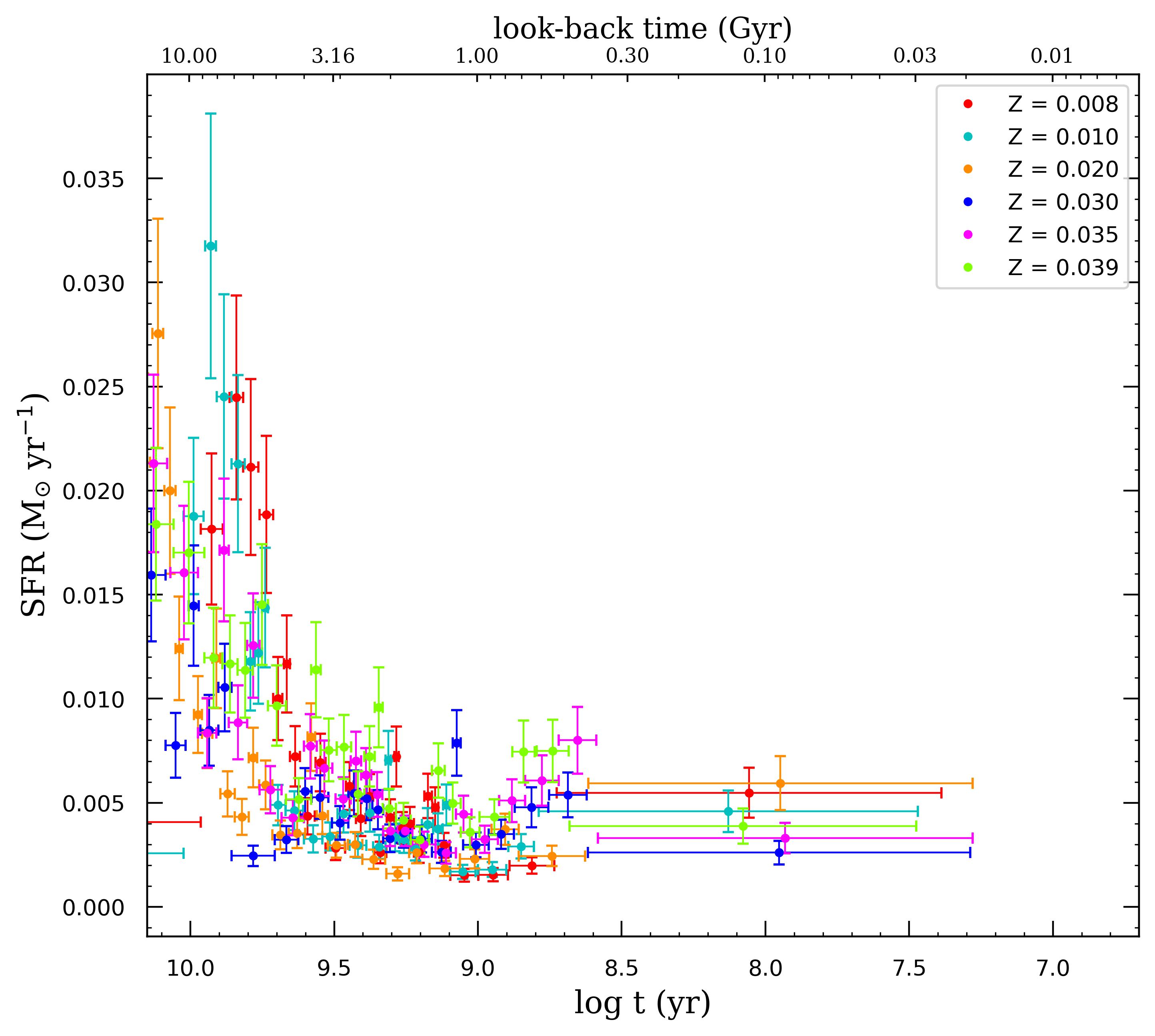,width=90mm}}}		
	\caption{Obtained SFHs for Field\,1 ({\it upper panels}) and Field\,2 ({\it lower panels}) where each color represents a metallicity. The probability function was applied to the data, thus, it is seen that the SFR is raised. Each metallicity is scaled to $1$ and the yellow regions are the desired epochs of SFR.}
	\label{fig:fig7}
\end{figure*}

The effect of applying the probability functions on the LPVs is illustrated in Fig.\,\ref{fig:fig7} for Field\,1 and Field\,2.  Furthermore, Table \ref{tab:tab4} and Table \ref{tab:tab5} represent the star formation epochs and the total mass formed for Field\,1 and Field\,2, respectively. The SFHs are very similar to those obtained without applying the corrections, but the SFRs are increased. 

Having applied the probability function on each star, we derived a recent SFR of $2.56 \times 10^{-3}$ M$_\odot$ yr$^{-1}$ kpc$^{-2}$ and $10^{-3}$ M$_\odot$ yr$^{-1}$ kpc$^{-2}$ for metallicities $Z = 0.003$ (closest match for which there are models) and $Z = 0.008$, respectively. These results are in the range of what \cite{Rejkuba2004a} derived for various slopes of a Salpeter IMF, viz.\ from $\sim 9 \times 10^{-5}$ to $\sim 10^{-3}$ M$_{\odot}$ yr$^{-1}$ kpc$^{-2}$ (in an area of $45$ kpc$^2$).

The SFHs we reconstruct depend on the adopted metallicity (Fig.\,\ref{fig:fig20} and \ref{fig:fig21} in Appendix \ref{app:appC}). At high metallicity, $Z = 0.010$--$0.039$, the SFH generally exhibits a first epoch of star formation around $\log\,t({\rm yr})\sim10$, followed by peaks in star formation around $\log\,t({\rm yr})\sim 9.5$ and $\log\,t({\rm yr})\sim8.9$. At lower metallicities, the first peak vanishes, but the other two epochs of star formation persist -- if at somewhat shifted ages.

\begin{table}
\caption[]{Description of the catalogue containing all the parameters needed to derive the SFHs (Fig.\,\ref{fig:fig7}): star
ID, x and y positions with respect to the reference epoch (\citealp{Rejkuba2003a}), single epoch $J_s$ and H--band magnitudes,  mean
$K_s$-band magnitude from the sine--curve fit, probability of detection value, mass of the stars derived from the mass-luminosity relation, age of the stars derived from the age-mass relation, and LPV phase duration derived from the LPV phase duration ($\delta t$)--mass relation. (The table is available electronically.)} 
\label{tab:tab_SFHs}
\begin{tabular}{cl}
\hline\hline
Column No. & Descriptor \\
\hline
 1         & ID \\
 2         & x position  \\
 3         & y position \\
 4         & $J_s$--band magnitude\\
 5         &  $H$--band magnitude \\
 6         &$K_s$--band  mean magnitude \\
 7         &  Probability\\
8        &  log Mass ($M_\odot$)\\
9         &  log t (yr)\\
10         & log $\delta t$ (yr)\\
\hline
\end{tabular}
\end{table}

\subsection{Mass of the Formed Stars} \label{sec:sec8-5}

\begin{figure*}[ht!]
	\centering{\hbox{
			\epsfig{figure=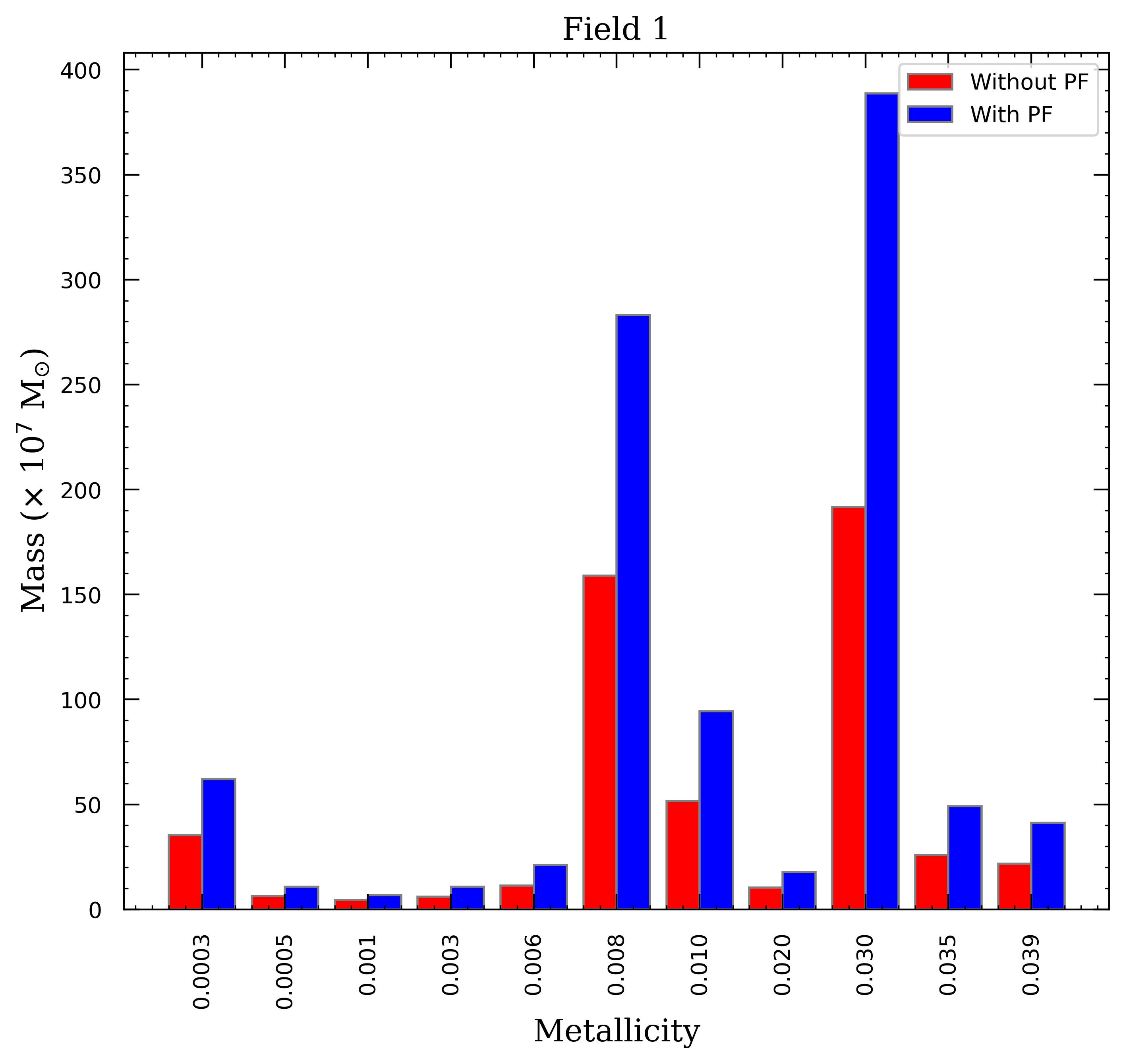,width=90mm}
			\epsfig{figure=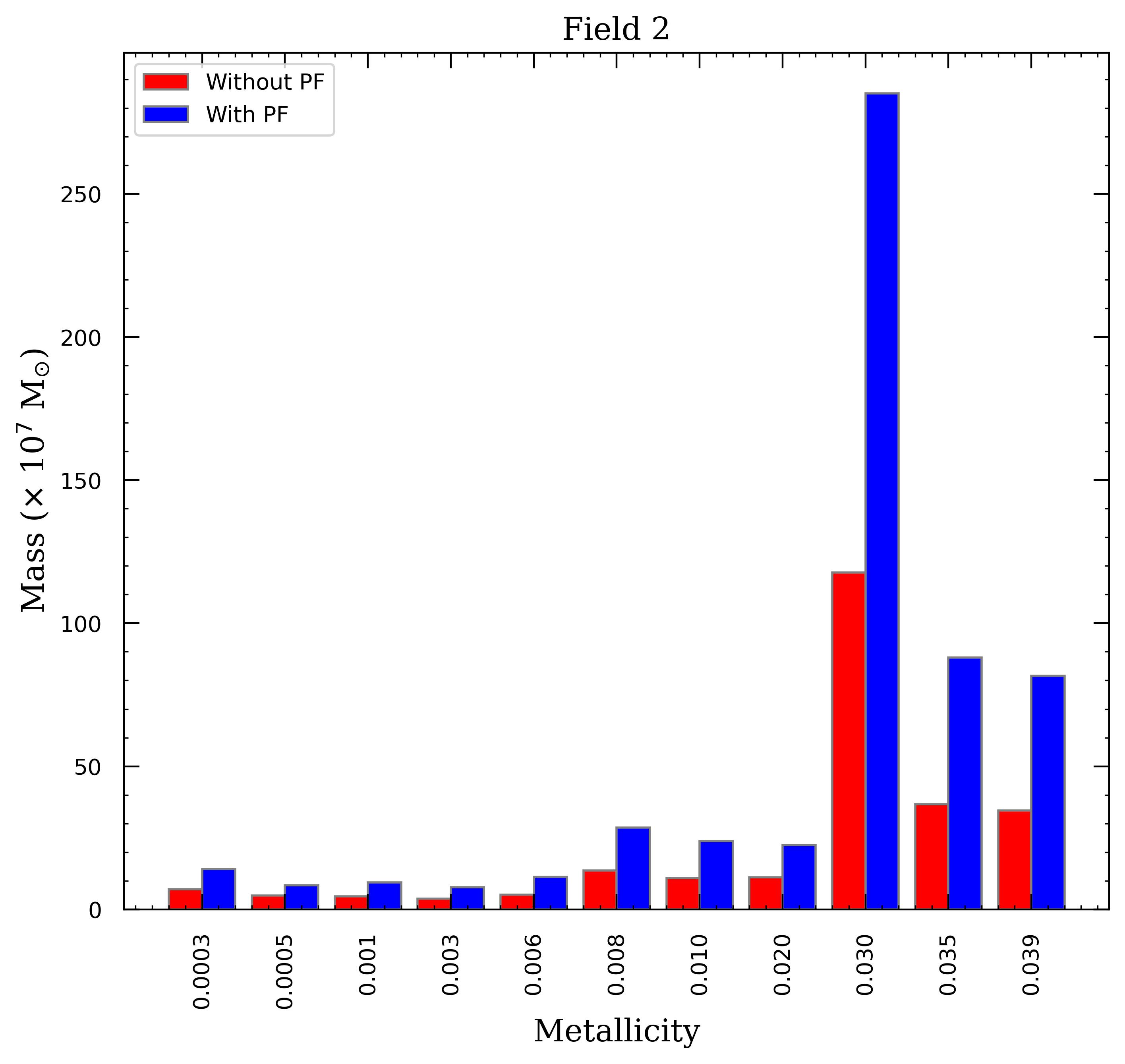,width=90mm}}}
	\caption{Total stellar mass derived from the SFH of Field\,1 and Field\,2 without and with applying probability function (PF).}
	\label{fig:fig8}
\end{figure*}

From the SFH shown in Fig.\,\ref{fig:fig7}, we can determine the total stellar mass produced (Fig.\,\ref{fig:fig8}). Most stars in the two fields we study were formed up to $400$ Myr ago when $93\%$ and $96\%$ of stars were formed, respectively for $Z = 0.003$ and $Z = 0.008$ in Field\,1 ($80\%$ and $91\%$ in Field\,2). This is very similar to what \cite{Rejkuba2011} derived for a region close to Field\,2, viz.\ $\sim80\%$ for these metallicities. We note that the old stellar populations have metallicities ranging from $Z = 0.0003$--$0.04$, while the younger stars have metallicities ranging from $Z \sim 0.002$--$0.004$ (\citealp{Rejkuba2011}).

We derived $(2.83 \pm 0.61) \times 10^9$ M$_\odot$ and $(2.86 \pm 0.61) \times 10^9$ M$_\odot$ for $Z = 0.008$ for Field\,1 and Field\,2, respectively. Our derived values for Field\,2 are higher than the $4 \times 10^7$ M$_\odot$ reported by \cite{Rejkuba2011} for a region close to Field\,2. They used the observed metallicity distribution function (MDF) for a single age population, and then compared the observed luminosity function (LF) with the theoretical one produced by the {\sc BASTI} stellar evolutionary tracks. Their results are based on a double-burst-fitting model. We, instead, integrate the entire SFH.

\subsection{Age--Metallicity Relation} \label{sec:sec8-6}

As a galaxy ages, the metallicity of the ISM -- and consequently that of new generations of stars -- undergoes changes due to nucleosynthesis and feedback from dying stars. Therefore, we anticipate that older stars were formed in environments with lower metallicity compared to the environments of younger stars. Until now, we have computed the  SFH assuming a constant metallicity and have not accounted for the chemical evolution of the galaxy. In this section, we will incorporate the age-metallicity relations estimated for NGC\,5128 to apply this temporal variation in metallicity to the presented SFH of the galaxy.

\begin{table}[]
	\centering
	\caption{The detected epochs of star formation for each metallicity in Field\,1}
	\label{tab:tab4}
	\begin{tabular}{ccccc} \hline
		Z      & epoch\,1 & epoch\,2 & epoch\,3 & Mass\,($\times 10^8$ M$_\odot$) \\ \hline
		0.0003 & 9.23    & 9.04    & 8.85    & 6.20                           \\
		0.0005 & 9.22    & 8.48    &         & 1.08                          \\
		0.001  & 9.29    & 9.18    & 8.84    & 0.68                         \\
		0.003  & 9.46    & 9.23    & 8.95    & 1.07                          \\
		0.006  & 9.61    & 9.46    & 8.88    & 2.12                          \\
		0.008  & 9.75    & 9.30    &         & 28.32                      \\
		0.010  & 9.85    & 9.55    &         & 9.44                         \\
		0.020  & 10.09   & 9.80    & 9.37    & 1.78                        \\
		0.030  & 10.20   & 9.63    & 8.82    & 38.86                        \\
		0.035  & 10.08   & 9.60    & 8.76    & 4.93                            \\
		0.039  & 10.06   & 9.79    & 8.88    & 4.12                        \\ \hline
	\end{tabular}
\end{table}

\begin{table}[]
	\centering
	\caption{The detected epochs of star formation for each metallicity in Field\,2}
	\label{tab:tab5}
	\begin{tabular}{ccccc}
		\hline
		Z      & epoch\,1 & epoch\,2 & epoch\,3 & Mass\,($\times 10^8$ M$_\odot$) \\ \hline
		0.0003 & 9.24    & 8.94    & 8.76    & 1.41                           \\
		0.0005 & 9.21    & 8.64    & 8.34    & 0.84                             \\
		0.001  & 9.33    & 8.86    & 8.53    & 0.94                             \\
		0.003  & 9.47    & 9.14    & 8.67    & 0.78                             \\
		0.006  & 9.72    & 9.47    & 8.90    & 1.14                            \\
		0.008  & 9.84    & 9.28    & 9.17    & 28.60                            \\
		0.010  & 9.93    & 9.31    & 9.11    & 2.39                            \\
		0.020  & 10.11   & 9.58    & 8.91    & 2.24                            \\
		0.030  & 10.25   & 9.60    & 9.07    & 28.50                           \\
		0.035  & 10.13   & 9.58    & 8.65    & 8.80                            \\
		0.039  & 10.12   & 9.75    & 8.74    & 8.16                            \\ \hline
	\end{tabular}
\end{table}

\begin{figure*}[ht!]
	\centering{\hbox{
			\epsfig{figure=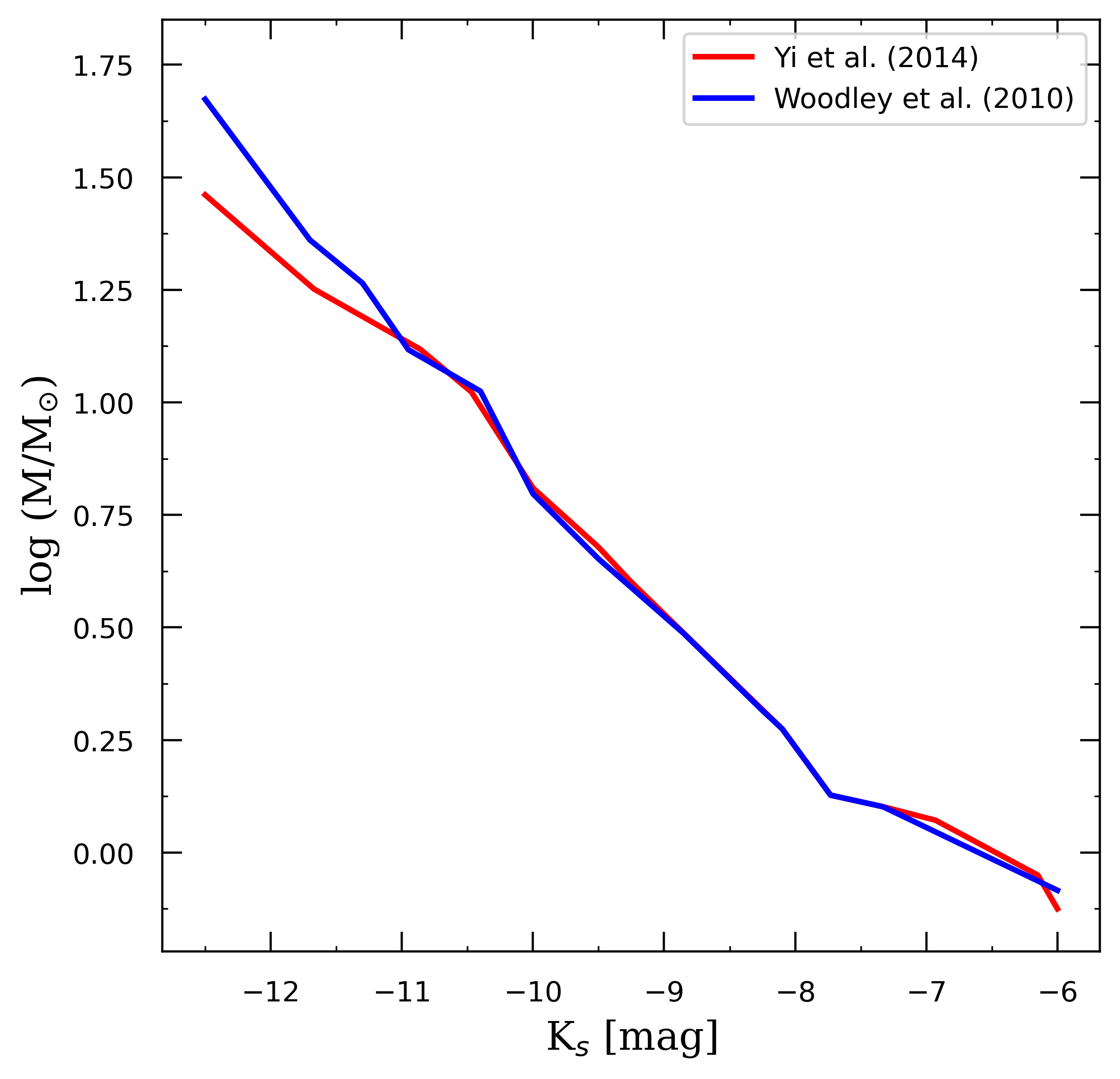,width=90mm}
			\epsfig{figure=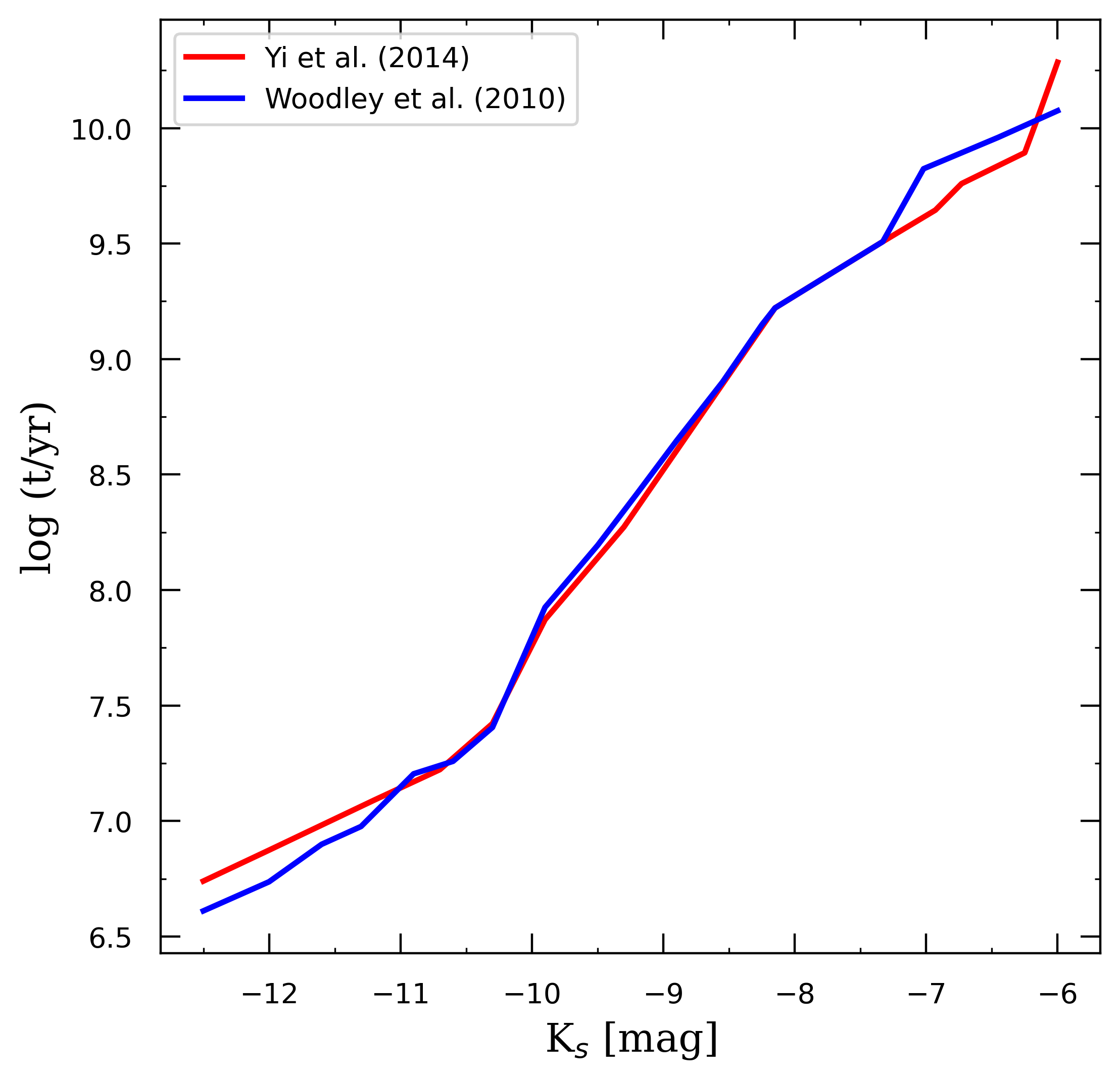,width=90mm}}}
	\caption{The mass--luminosity ({\it left panel}) and age--luminosity ({\it right panel}) relations by considering AMRs of \cite{yi2004} ({\it red line}) and \cite{woodley2010b} ({\it blue line}) for Field\,1 ({\it left panel}) and Field\,2 ({\it right panel}).}
	\label{fig:fig_rel_metallicity}
\end{figure*}

\begin{figure*}[ht!]
	\centering{\hbox{
			\epsfig{figure=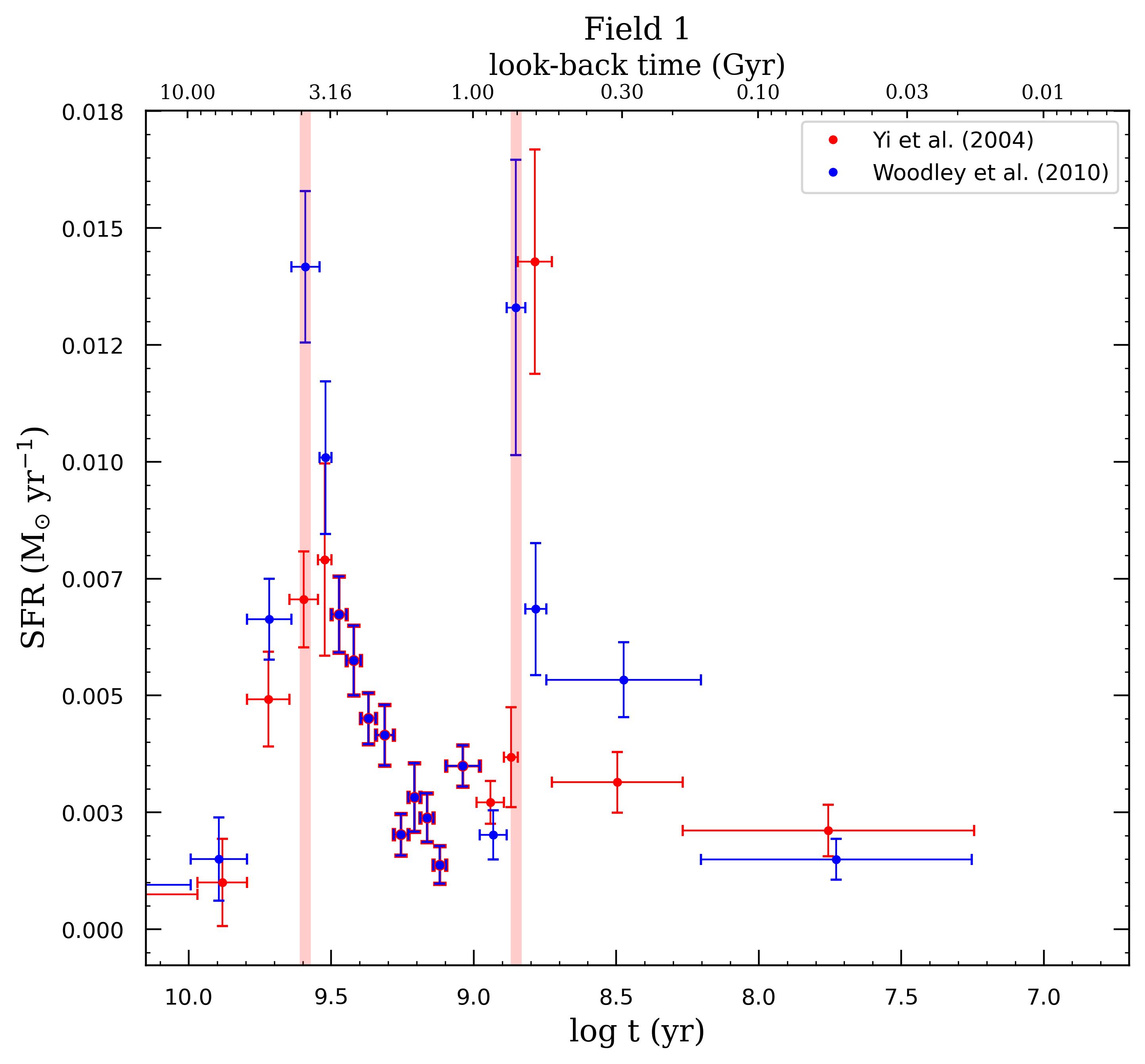,width=90mm}
			\epsfig{figure=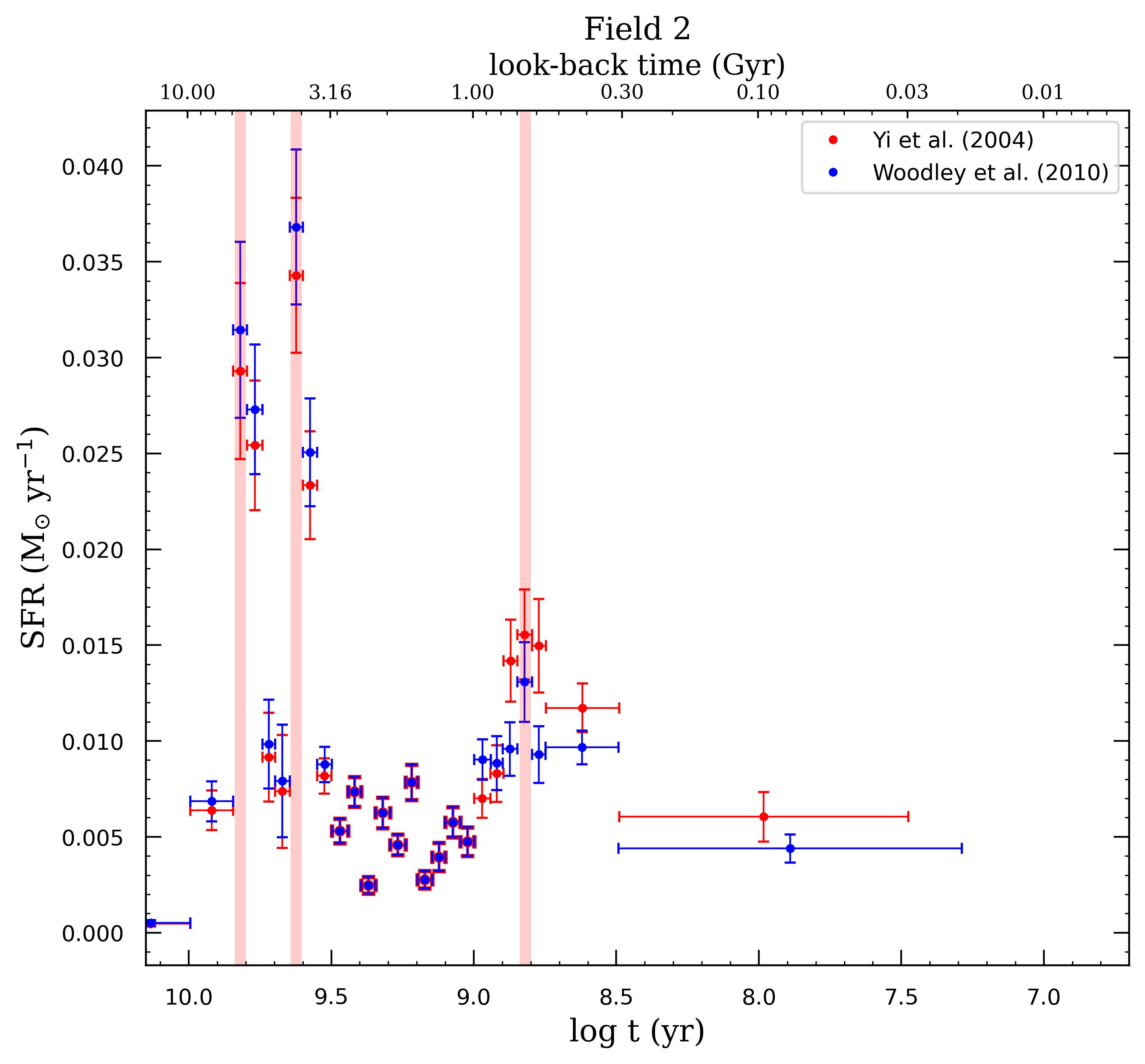,width=90mm}}}
	\caption{Star formation history derived by considering AMRs of \cite{yi2004} ({\it red line}) and \cite{woodley2010b} ({\it blue line}) for Field\,1 ({\it left panel}) and Field\,2 ({\it right panel}). The highlighted regions represent the peaks of star formation during the major epochs.}
	\label{fig:fig9}
\end{figure*}

The age--metallicity relations (AMRs) has been derived using globular clusters (GCs) in the NGC\,5128 halo, as demonstrated in \cite{woodley2010b} and \cite{yi2004}. The appropriate metallicities for various age bins, obtained from \cite{woodley2010b}, are listed in Table \ref{tab:tab6}. In addition, \cite{yi2004} presented a metallicity range of $-2.0 \leq$ [Fe/H] $\leq +0.3$, which aligns with our considerations for this galaxy. Referring to Table 2 in \cite{yi2004}, we adopt the AMR as summarized in Table \ref{tab:tab7}. Utilizing these metallicity relationships, based on the suitable metallicity for different age bins, the appropriate Padova models are selected, and the mass--luminosity and age--luminosity relations are derived, as can be seen in Fig.\,\ref{fig:fig_rel_metallicity}. Instead of employing mass--luminosity and age--luminosity relations for a constant metallicity, these new relationships, which have considered the metallicity changes of the galaxy over time, have been applied to derive the SFH, as depicted in  Fig.\,\ref{fig:fig9}. In Field\,1, SFRs exhibited significant increases around $t\sim 3.8$ Gyr  and $t\sim 800$ Myr. Meanwhile, in Field 2, we observe three epochs of star formation. The first two peaked around $t\sim 6.3$ Gyr ago, and $t\sim 3.8$ Gyr ago, respectively. The third epoch began around  $t\sim 800$ Myr  ago and peaked around  $t\sim 700$ Myr ago.
   The total mass produced in Field\,1  using the models of \cite{yi2004} and  \cite{woodley2010b}  is $(1.03 \pm 0.12) \times 10^8$ M$_\odot$   and $(1.04 \pm 0.12) \times 10^8$ M$_\odot$, respectively. For Field\,2, these values are $(1.60 \pm 0.23) \times 10^8$ M$_\odot$  and $(1.50 \pm 0.23) \times 10^8$  M$_\odot$. These estimates  are only a few times higher than the value of $4\times 10^7$ M$_\odot$  obtained by \cite{Rejkuba2011} for a region close to Field\,2.

\begin{table}[]
	\centering
	\caption{The age--metallicity relation  investigated by \cite{woodley2010b} assuming Z$_\odot$ = 0.0198 (\citealp{Rejkuba2011}).}
	\label{tab:tab6}
	\begin{tabular}{cc}\hline
		age range (Gyr)              & Z     \\ \hline
		age $\geq$ 12                & 0.001 \\
		8 $\leq$ age \textless 12    & 0.003 \\
		6.5 $\leq$ age \textless 8   & 0.006 \\
		5.5 $\leq$ age \textless 6.5 & 0.008 \\
		3 $\leq$ age \textless 5.5   & 0.010 \\
		2 $\leq$ age \textless 3     & 0.020 \\
		age \textless 2              & 0.030 \\ \hline
	\end{tabular}
\end{table}

\begin{table}[]
	\centering
	\caption{The age--metallicity relation  investigated by \cite{yi2004} assuming Z$_\odot$ = 0.0198 (\citealp{Rejkuba2011}).}
	\label{tab:tab7}
	\begin{tabular}{cc}\hline
		age range (Gyr)           & Z      \\ \hline
		age $\geq$ 10             & 0.0003 \\
		6 $\leq$ age \textless 10 & 0.001  \\
		4 $\leq$ age \textless 6  & 0.003  \\
		3 $\leq$ age \textless 4  & 0.010  \\
		2 $\leq$ age \textless 3  & 0.020  \\
		age \textless 2           & 0.039  \\ \hline
	\end{tabular}
\end{table}

\begin{figure*}[ht!]
	\centering{\hbox{
		\epsfig{figure=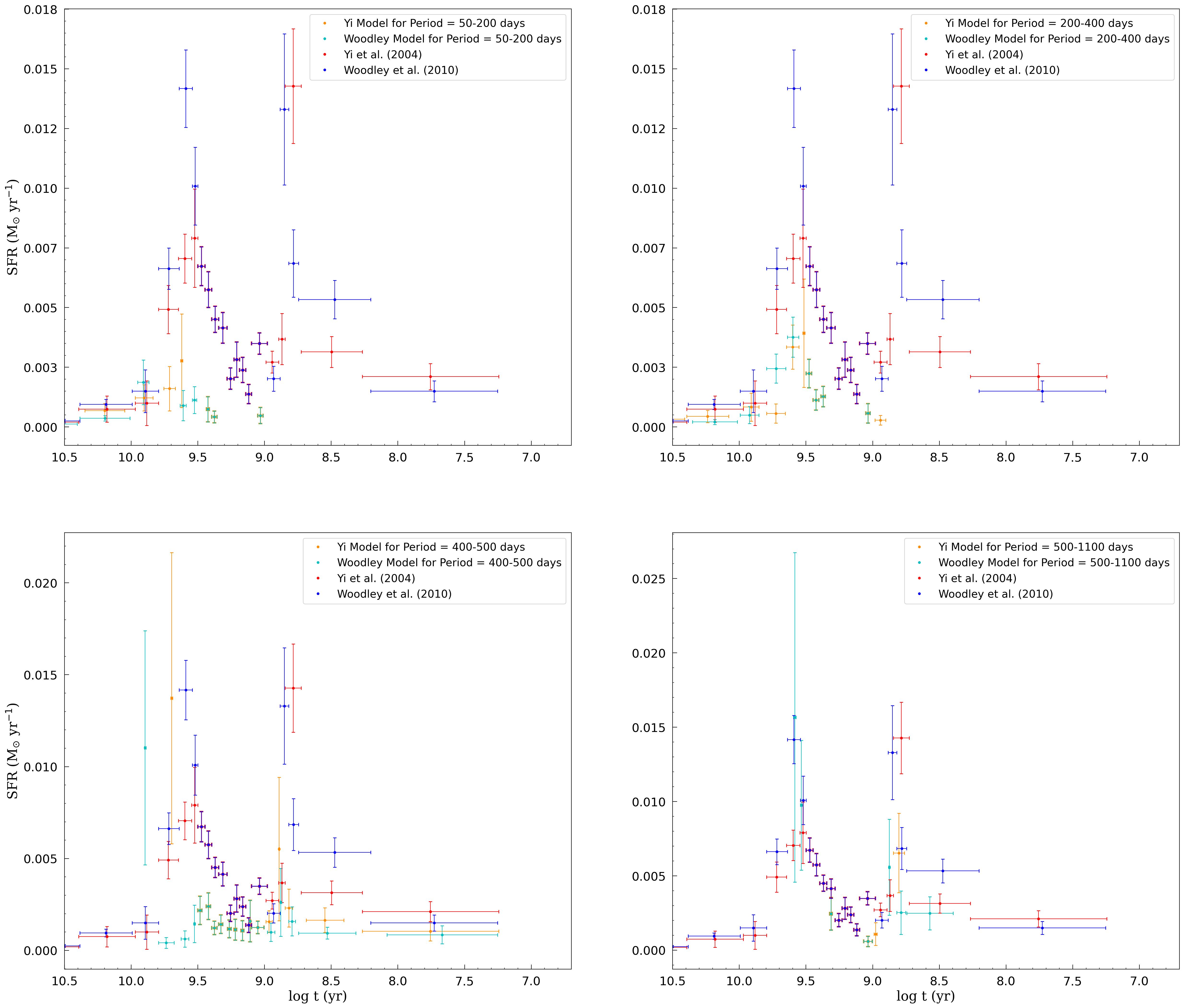,width=180mm}}}
	\caption{SFHs in Field\,1 for four different period bins using AMRs from \cite{yi2004} and \cite{woodley2010b}: P=50-200 days ({\it Top left}), P=200-400 days ({\it Top right}), P=400-500 days ({\it Bottom left}) and P=500-1100 days ({\it Bottom right}).}
	\label{fig:fig10}
\end{figure*}

\begin{figure*}[ht!]
	\centering{\hbox{
		\epsfig{figure=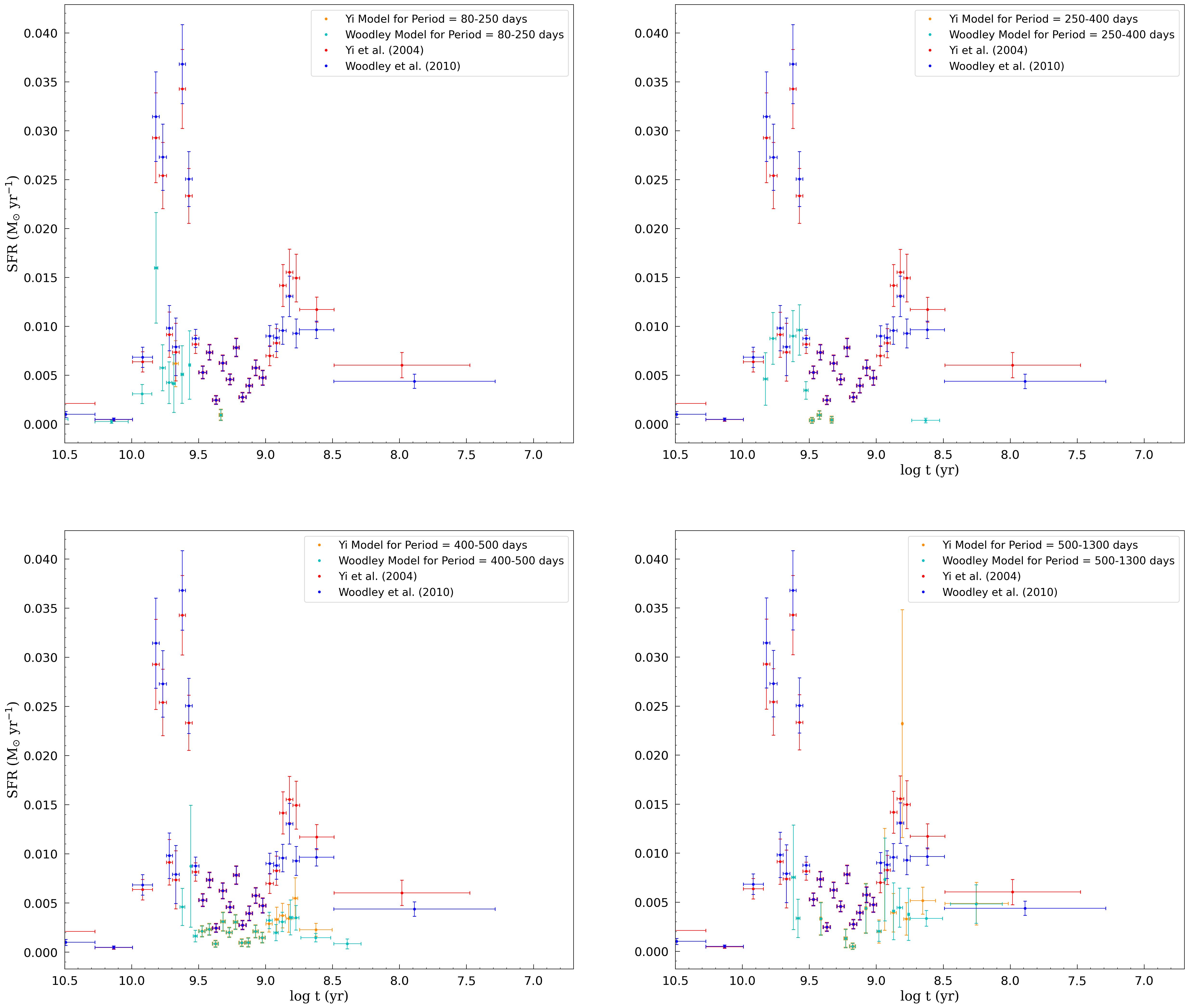,width=180mm}}}
	\caption{SFHs in Field\,2 for four different period bins using AMRs from \cite{yi2004} and \cite{woodley2010b}: P=80-250 days ({\it Top left}), P=250-400 days ({\it Top right}), P=400-500 days ({\it Bottom left}) and P=500-1300 days ({\it Bottom right}).}
	\label{fig:fig11}
\end{figure*}

\begin{figure*}[ht!]
	\centering{\hbox{
			\epsfig{figure=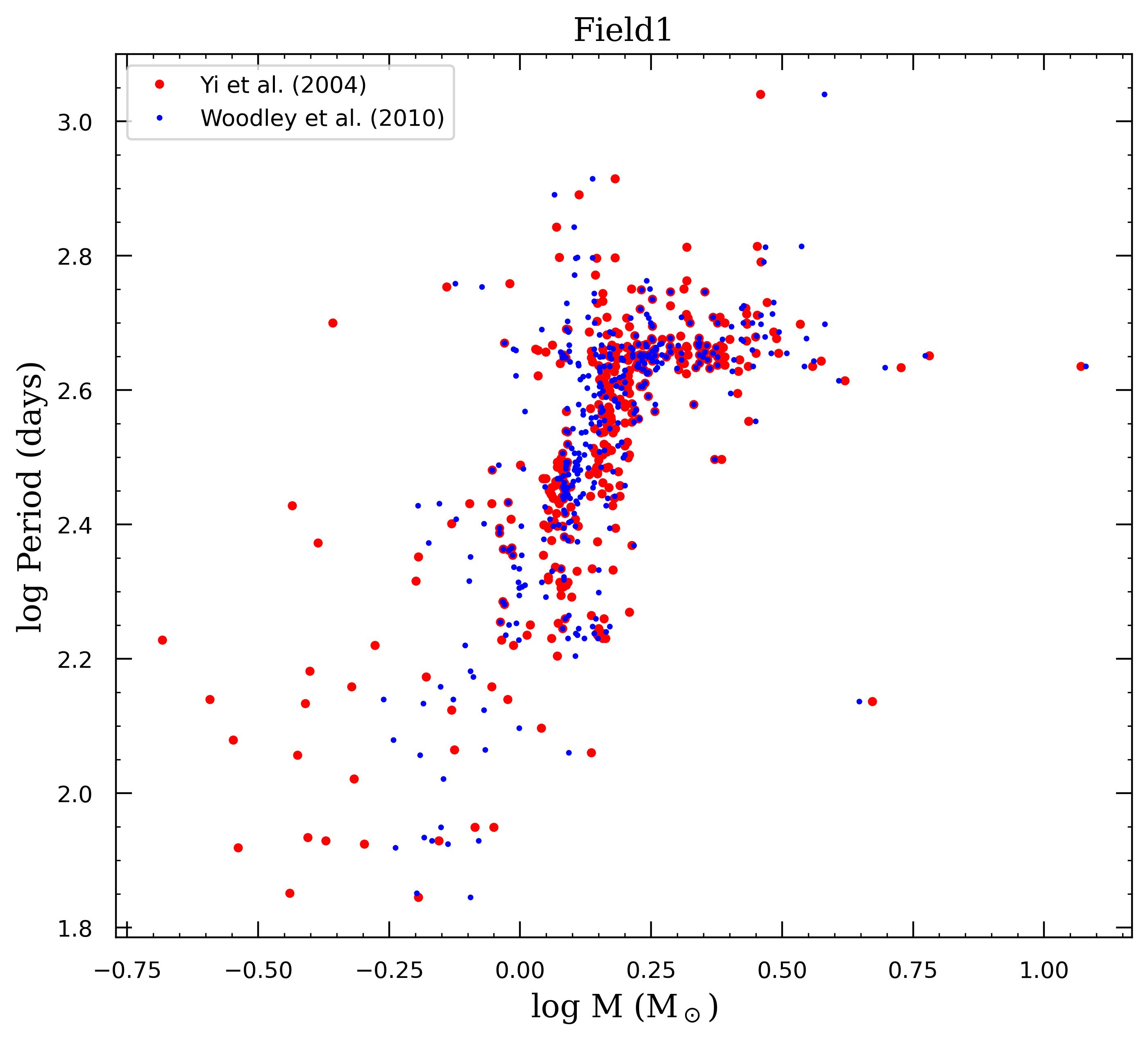,width=90mm}
			\epsfig{figure=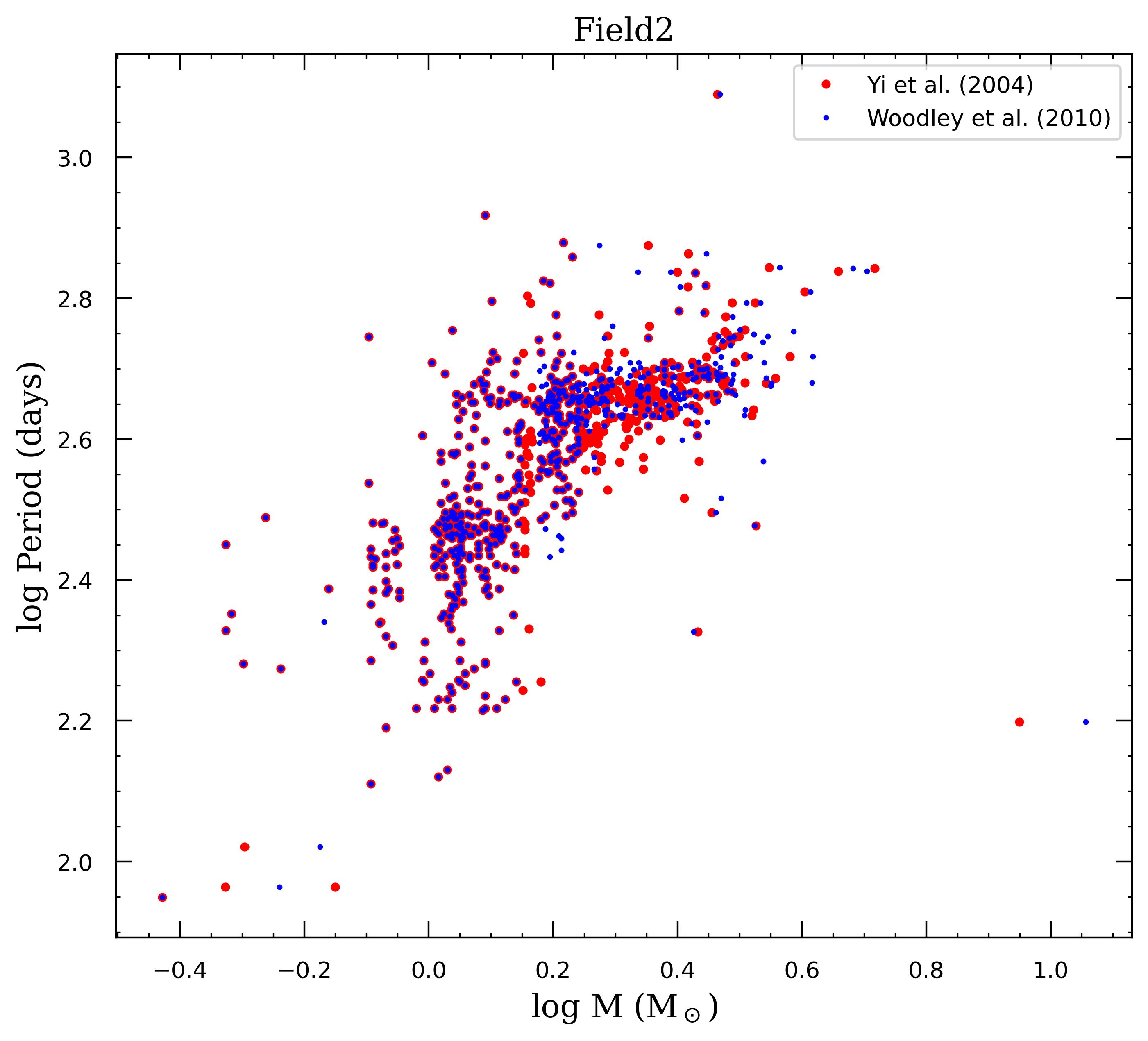,width=90mm}}}
	\caption{Pulsation period  of LPVs vs.  mass for Field\,1 ({\it left panel}) and Field\,2 ({\it right panel}). The blue and red dots refer to AMR models of \cite{woodley2010b}) and \cite{yi2004}, respectively.}
	\label{fig:fig12}
\end{figure*}

To further understand the correlation between different populations and distinct epochs of SFHs, we have partitioned the LPVs into four distinct period bins: 50-200, 200-400, 400-500, and 500-1100 days for Field\,1; and 80-250, 250-400, 400-500, and 500-1300 days for Field\,2. Utilizing the AMR models, we have constructed SFH diagrams for each of these period bins, which are depicted in  Fig.\,\ref{fig:fig10} and Fig.\,\ref{fig:fig11} for Field\,1 and Field\,2, respectively.\\
It is evident that the oldest epoch of star formation, occurring at $\log t({\rm yr}) > 9.5$ (t is look--back time), predominantly involves stars with periods ranging from 50 to 400 days in both Field\,1 and Field\,2. By closely examining the period--mass diagram presented in Fig.\,\ref{fig:fig12}, we can deduce that the majority of stars within the $50 < \text{Period (days)}  < 400$ range are not significantly massive, possessing masses spanning from 0.2 to 1.8 M$_\odot$. Notably, LPVs with periods exceeding 400 days primarily trace intermediate and recent star formation activity.\\
It's important to mention that certain studies propose deviations from the Period-Luminosity (PL) relation for longer periods (P > 400 days). In particular, low-mass stars may experience reduced mass due to sustained heavy mass loss (\cite{wood2000}). Conversely, the most massive AGB stars could exhibit heightened luminosity owing to the effects of Hot Bottom Burning (HBB) (\cite{whitelock2003}). Furthermore, even more massive Red Supergiants (RSGs) might not adhere to a similar PL relation as AGB stars; \cite{yang2012} suggest that RSGs could pulsate in the first overtone. As a result, it's important to recognize that not all stars may conform exactly to the expected behavior based solely on their periods. 
Hence, identifying the populations responsible for the epoch of star formation that occurred around $\sim 800$ Myr ago is not straightforward. This is evident from Fig. \ref{fig:fig10} and Fig. \ref{fig:fig11}, which illustrates that both of the last period bins somewhat trace this epoch.

\section{Discussion} \label{sec:sec9}
\subsection{Merger Tracers} \label{sec:sec9-1}

In addition to gas accretion from the cosmic web (e.g., \citealp{dekel2006}; \citealp{dekel2009}; \citealp{sanchez2014}), mergers play critical roles in the growth of galaxies (e.g., \citealp{guo2011}). Based on their progenitors' mass ratio ($\frac{M1}{M2}$, $ M1\geq M2$), galaxy mergers are classified into two main categories. Major merger (usually $\frac{M1}{M2} < 4$) is violent and the remnant can be very different from its progenitors. Minor merger ($\frac{M1}{M2} > 4$) is not powerful enough to disrupt the host galaxy in each interaction. However, since giant galaxies have a significant number of satellites, minor mergers are expected to be more common and highly affect the formation and evolution of the host galaxy on cosmic timescales (\citealp{bournaud2011}). Gas content of progenitors is also another key factor in these processes. Wet mergers (in contrast to dry mergers) -- mergers among gas-rich progenitors -- are expected to be responsible for enhanced star formation and/or AGN activity (\citealp{mo2010}). According to numerical simulations, the remnant can be determined using parameters such as mass ratio, morphology, orbital properties, and gas content of the progenitors (\citealp{barnes1992}; \citealp{hernquist1993}; \citealp{barnes1996}; \citealp{dubinski1996}; \citealp{springel1999}; \citealp{naab2003}; \citealp{boylan2005}; \citealp{cox2006}).

The peculiar appearance of NGC\,5128, especially in the form of the central disk of dust and gas (\citealp{graham1979}), optical shells (\citealp{malin1983}; see also Fig.\,2 of \citealp{peng2002}), neutral hydrogen shells (\citealp{schim1994}), and recent star formation (\citealp{blanco1975}; \citealp{graham1981}; \citealp{fassett2000}; \citealp{mould2000}; \citealp{Rejkuba2001}) suggest that the stream of young stars was formed based as a consequence of the merging of the giant elliptical galaxy with a small gas-rich galaxy (see \citealp{israel1998} and references therein).

Based on our results, a significant increase in star formation in Field\,1 and Field\,2 around $800$ Myr ago (Fig.\,\ref{fig:fig7}) supports the idea of a recent merger. There are numerous works that have suggested the likely time of merger agrees with our findings of an amplified star formation. \cite{malin1983} interpreted the optical shell system as the aftermath of a merger with a small late-type spiral galaxy (like M\,33) about one Gyr ago. \cite{sparke1996} found that three-quarters of a Gyr had elapsed since the formation of the disk from the captured galaxy. Furthermore, \cite{tubbs1980} and \cite{malin1983} suggested a timescale of $2$--$8 \times 10^8$ yr (\citealp{israel1998}) that is supported also by \cite{sparke1996}. \cite{israel1998} also proposed that the appearance of the galaxy and outcome of various dynamical models are strong indicators of at least one major merger between $10^8$--$10^9$ years ago. All of these estimates are also in full agreement with the study of globular clusters (GCs) by \cite{kaviraj2005} where they suggested that a starburst within $< 1$ Gyr is responsible for $\sim 5$ per cent of the stellar mass of the GCs in NGC\,5128. The stellar mass of $\sim 3\%$ we found to have formed in the last Gyr fits perfectly with their results.

Within the Centaurus group are a large number of dwarf irregular galaxies that are gas-rich and host young stellar populations. As a result of dynamical friction, they transfer kinetic energy and momentum onto the environment. This causes them to slow down and fall towards the host galaxy on a spiral path. As the density of the environment (halo of the host galaxy) increases, the dynamical friction grows, and infalling galaxies encounter stronger resistance from the environment. Tidal forces disrupt the infalling dIrr resulting in a stream of stars -- including young stars -- in the halo of the host galaxy (\citealp{peng2002}; \citealp{israel1998}; \citealp{tubbs1980}).

In addition, it appears that recent star formation in both fields, and specifically in Field\,1, may be due to a minor merger $\sim 400$ Myr ago according to \cite{peng2002}. The lower rate of star formation and lack of morphological and kinematic distortions imply this was a lesser merger than the one that had occurred $\sim 800$ Myr ago (\citealp{wang2020}).


We see another peak in star formation, between $\log\,t({\rm yr})\sim9.4$--$9.6$, in both fields, that might also have resulted from mergers or galaxy--galaxy interactions (Fig.\,\ref{fig:fig9}). There is another peak in Field\,2 which is between $\log\,t({\rm yr})\sim9.7$--$9.8$. Many works confirm the existence of distinct generations of stars with similar ages, $t\sim3$ and $8$ Gyr (\citealp{Rejkuba2005}; \citealp{Rejkuba2011}; \citealp{kaviraj2005}; \citealp{woodley2010b}). Higher rates of star formation in Field\,2 can be interpreted as a result of more dynamical friction effects in this field closer to the center of NGC\,5128.

By comparing simulated CMDs of a remote halo field located $38$ kpc south from the center of NGC\,5128 (\citealp{Rejkuba2005}) and the CMDs obtained from {\it Hubble} Space Telescope (HST) observations, \cite{Rejkuba2011} found that the closest matches to the data are models with two bursts of star formation and a metallicity distribution function. They also suggest that $80$\% of formed stars are old ($\sim12$ Gyr), while $20$\% of them are younger ($\sim2$--$3$ Gyr) with lower metallicity. \cite{Rejkuba2011} also considered models with a combination of three bursts and found that in this case, using the data studied by \cite{woodley2010b}, fractions of old ($11$--$13$ Gyr), intermediate-age ($6$--$9$ Gyr), and young ($3$--$5$ Gyr) components in the halo range between $60$\%--$80$\%, $5$\%--$15$\%, and $10$\%--$20$\%, respectively. We obtained for Field\,2 that $75$\% of stars were formed $8$--$13$ Gyr ago (for $Z=0.008$), $18$\% $5$--$8$ Gyr ago, and $4$\% $2$--$5$ Gyr ago. This is consistent with their two-burst scenario but also their three-burst scenario. It shows that most of the stars in the halo of NGC\,5128 were formed before $400$ Myr ago.

\subsection{AGN Activity} \label{sec:sec9-2}

Besides major and minor mergers, NGC\,5128 has experienced AGN activity, given its central supermassive black hole (SMBH) and ample amounts of cold gas to fuel the AGN (\citealp{mo2010}). In most cases, gas within galaxies is either too hot or has too much angular momentum to accrete onto the SMBH.
Mergers and interactions among gas-rich galaxies (minor and major) offer ways to channel gas into the direction of the SMBH (\citealp{mo2010}).

The radio jet morphology of NGC\,5128 is one of the most spectacular and complicated astrophysical jets known. A pair of inner lobes -- a northern inner lobe (NIL) at a distance of $\sim 5$ kpc, and a southern inner lobe (SIL) $\sim 11$ kpc from the center (\citealp{neff2015}) -- is symmetric around the AGN and oriented at an angle of $55^\circ$ (anti-clockwise) from the north--south axis (\citealp{schreier1981}; \citealp{clarke1992}). Further out, the NIL is followed by the northern middle lobe (NML; \citealp{morganti1999}). In Fig.\,\ref{fig:fig1}, the radio emission contours from the inner lobes, as well as the NML are shown (see also Fig.\,1 in \cite{morganti2010}). The NML lobe is located $\sim 30$ kpc from the center at a position angle of $45^\circ$ and does not show a southern counterpart in total intensity maps (\citealp{feain2011}). Finally, at the largest scale, northern and southern outer lobes are seen with a total extent of $\sim 500$ kpc end-to-end. The position angles of the outer lobes are lower than those of the middle and inner ones and also decrease gradually when their distances increase from the center (\citealp{mckinley2013}).

Here we suggest that some of the observed epochs of enhanced star formation may have been triggered by the jets. As a positive AGN feedback, jet-driven shocks can cool the gas efficiently and trigger the formation of stars (\citealp{best2012}; \citealp{ivison2012}; \citealp{salome2016b}; \citealp{santoro2016}; \citealp{wang2020}; \citealp{joseph2022}). Such positive feedback has been confirmed by observations (e.g., \citealp{emonts2014}) and simulations (e.g., \citealp{fragile2004}; \citealp{gaibler2012}). Accepting this scenario gives us the approximate times of AGN activity or the age of the jets.

\cite{saxton2001} estimated an age of the NML of $\log\,t({\rm yr})\sim 8.2$ ($\sim 150$ Myr), whilst \cite{hardcastle2009} determined the age of the NIL ans SIL to be $\log\,t({\rm yr})\sim 7.5$ ($\sim 30$ Myr). Also, they found a jet through the NIL to intrude the NML, indicating that activity has continued and enriched the NML. Therefore, recent AGN activity should be considered as a potential factor in recent star formation in the halo of NGC\,5128 (\citealp{hardcastle2009}), in agreement with the SFRs from $\log\,t({\rm yr})\sim 7.2$ to $8.2$ for the NIL/SIL and NML lobes. We thus propose that the enhanced star formation around $800$ Myr ago might be associated with the outer lobes, that resulted from enhanced AGN activity due to the wet minor merger around that time.

\subsection{NGC\,5128 Evolution} \label{sec:sec9-3}

Deriving the SFH in the host galaxy resulting from a merger is complicated. As we explained in Section \ref{sec:sec6}, the SFRs we derive are based on identified LPV stars. These may have formed in NGC\,5128 (in-situ) or outside of it (ex-situ, in a merging galaxy). Therefore, both in-situ and ex-situ stars can amplify the apparent SFR in the halo of NGC\,5128. The time of amplification by in-situ or ex-situ stars, that are formed {\em during the merger process} (for a review of star formation during mergers and interactions, see \citealp{bournaud2011}) is equal to the time interval over which the host galaxy captures the infalling galaxy. Amplification by ex-situ stars can, however, occur also in another, independent way, namely {\em within the infalling galaxy before it started merging}. In this case, it occurs much earlier than the accretion event, and reflects the evolution of the infalling galaxy, not NGC\,5128 or the merger.

The stars of a small galaxy falling into a more massive halo will form shells (\citealp{quinn1984}; \citealp{dupraz1987}). The dynamics of the gas from such merging galaxy is however not well understood. Based on some simple simulations, \cite{weil1993} concluded that dissipation would result in the rapid concentration in the center. On the other hand, the detection of $4 \times 10^8$ M$_\odot$ of H\,{\sc i} gas associated with the stellar shells (with a small displacement) by \cite{schim1994} contradicts their claim. \cite{combes1999} and \cite{charmandaris2000} suggested that if the ISM of the merging galaxy is clumpy then the collision and dissipation rates are low. There is clear evidence of star formation going on in the halo, but it only happens along the jet (e.g., \citealp{Rejkuba2001}; \citealp{joseph2022}) and it is relatively inefficient (\citealp{salome2016b}, \citeyear{salome2017}). Therefore, it seems that any elevated levels of star formation that occur during the merger process will likely happen within the secondary galaxy in addition to the halo of the host galaxy.

As illustrated in \cite{schim1994}, there are $3$ main H\,{\sc i} clouds in NGC\,5128, two of which are in a similar position to Field\,1 and Field\,2. They found that star formation continues at the location of Field\,1. \cite{oosterloo2004} and \cite{mould2000} reported a recent SFR in the H\,{\sc i} cloud near Field\,1 of a few times $10^{-3}$ M$_\odot$ yr$^{-1}$, while \citet{salome2016b} obtained $4 \times 10^{-3}$ M$_\odot$ yr$^{-1}$ (at low metallicity); this is similar to what we derived for metallicities in the range $0.008 < Z < 0.039$. Field\,2 has not been studied as much as Field\,1, but given our results agree with for Field\,1 this lends some credibility for our results in Field\,2. \citealp{santoro2015a}, \citet{santoro2015b} and \citet{santoro2016} suggested that ionization by young stars ($<4$ Myr) and AGN shocks traces the interaction between the AGN jet and gas from an accreted galaxy.


We identified two epochs of increased star formation for Field\,1 around $\log\,t({\rm yr})\sim 8.9$, $9.6$, and for Field\,2 at $\log\,t({\rm yr})\sim 8.9$, $9.6$, and $9.8$ (Fig.\,\ref{fig:fig9}). The latest epoch of enhanced star formation at $\log\,t({\rm yr)}\sim8.9$ can be linked with a minor wet merger, where star formation took place primarily within the merging galaxy rather than in the halo of NGC\,5128 -- timescales of $<1$ Gyr are in agreement with such scenario \citep{israel1998}. There is no such evidence for previous mergers that could be responsible for the more ancient epochs of elevated SFRs around $\log\,t({\rm yr})\sim9.6$ and $9.8$. While we cannot rule out past mergers, it is also conceivable that those older stars actually had formed in the galaxy before it merged with NGC\,5128 around $\log\,t({\rm yr})\sim8.9$ (\citealp{israel1998}). This would make it a relatively massive merger.

\cite{peng2002} referred to the large number of dIrr galaxies in the Centaurus Group, rendering it plausible that a gas-rich merger and a supply of already formed relatively youthful stellar population enriched the halo of NGC\,5128. \cite{Rejkuba2022}, too, argued that the extended, $2$--$3$-Gyr old stellar population points at the accretion of a small gas-rich spiral galaxy having provided the fuel for ongoing star formation at the center of NGC\,5128. The recent merger, $< 2 \times 10^8$ yr ago, requires that the intermediate-age stars came from either that accreted galaxy or a previous merger (\citealp{Rejkuba2022}).

Our analysis is based on fields located far from the center. Gas and stars have different dynamics during mergers or interactions within the halo; gas cannot form stars efficiently and eventually ends up in the center (\citealp{Rejkuba2002}; \citealp{israel1998}). A complementary study in the central part of the galaxy could provide more certainty about the history of the halo. Unfortunately, to date, there is no LPV catalog in the inner parts of the galaxy. Conducting a survey for LPVs in the outer part of the central disk (where it is less crowded and obscured) could be a promising approach.

Finally, we note that thus far the application of our LPV method was limited to galaxies within the LG. Now, for the first time, we have used it to find the SFH of a galaxy outside the LG. Our results strongly confirm the reliability of the method even for such distant galaxies and using ground-based observations.


\section{Summary of Conclusion} \label{sec:sec10}

NGC\,5128 is the nearest giant elliptical galaxy at a distance of $3.8$ Mpc. Based on its resolved stellar populations and prominent features such as AGN activity, this galaxy has been extensively studied. It is expected to become a popular target for future large telescopes. In this paper, we have used the novel method to find SFHs of two fields in the halo of the NGC\,5128 galaxy -- Field\,1 is in the northern-eastern part which is $\sim17^\prime$ ($\sim 18.8$ kpc) away from the center of the galaxy. Field\,2 is located in the southern part at a distance of $\sim 9^\prime$ ($\sim 9.9$ kpc) from the center. Our method is based on the identification of $395$ LPVs in Field\,1 and $671$ LPVs in Field\,2. Our main results are:

\begin{itemize}
	
	\item[$\bullet$] Even though the two fields are located $28$ kpc away from each other on different sides of the galaxy, they show similar SFHs. In Field\,1, star formation rates (SFRs) increased significantly around $t\sim 800$ Myr and $t\sim 3.8$ Gyr; and in Field\,2, SFRs increased considerably  around $t\sim 800$ Myr, $t\sim 3.8$ Gyr, and $t\sim 6.3$ Gyr, where $t$ is look--back time.\\

	\item[$\bullet$] To account for incompleteness, we constructed a probability function with dependency on amplitude, period, and magnitude. Applying this correction by giving a weight to each identified LPV results in a more accurate SFR.\\
	
	\item[$\bullet$] We postulate that the enhanced SFR around $t\sim 800$ Myr ago may have been the result of triggered star formation by a merger that happened $t\sim 1$ Gyr ago.\\
	
\end{itemize}


\section*{Acknowledgments}

We  thank the referee for her/his useful report which prompted us to improve the
manuscript.

\clearpage

\appendix \section{Supplementary Material} \label{app:app}

\subsection{Padova Evolutionary Model} \label{app:appA}

In order to find the SFH based on the LPVs counts, we need to determine mass, age, and LPV phase duration of the LPVs (\citealp{javadi2011b}, \citeyear{javadi2011c}, \citeyear{javadi2017}; \citealp{Rezaei2014}; \citealp{golshan2017}; \citealp{hashemi2019}; \citealp{navabi2021}; \citealp{saremi2021}). First we assume metallicity, then we link observational fluxes (for $K_s$-band magnitude) of each star to its mass, and finally, we estimate the age and LPV phase duration from the derived mass. In this paper, we have assumed $11$ different metallicities (covering the whole range of metallicity in the fields of study) to extract associated Padova evolutionary models (see section \ref{sec:sec4}) and obtain the equations for describing fitted curves and lines. Preferably, we use the final version of the Padova model (\citealp{marigo2017}), but given that the final version does not consider the variability of massive stars, thus we use the previous release (\citealp{marigo2008}) only for LPV phase duration of massive stars ($\log(M/M_{\odot})>0.8$). 

\begin{figure*}
	\plotone{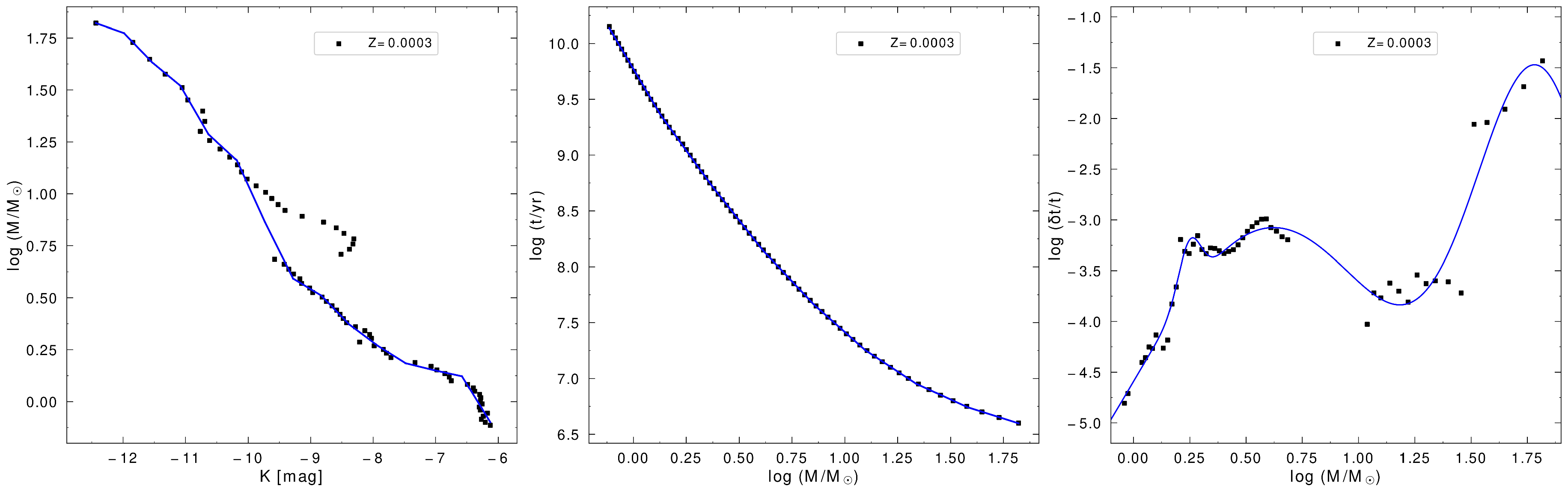}
	\caption{{\it Left panel}: The mass-luminosity relation in the $K_s$ band for a metallicity of $Z = 0.0003$. The solid lines are the linear spline fits, for the case in which the function is interpolated across the super--AGB phase to massive red supergiants, i.e. for  0.7 < $\log (M/M_\odot)$ < 1-1.1. {\it Middle panel}: The mass-age relation for a metallicity of $Z = 0.0003$ along with linear spline fits. {\it Right panel}: The mass-LPV phase duration relation with the same metallicity, where the points show the ratio of LPV phase duration to age vs. mass; the solid lines are multiple-Gaussian fits.
	\label{fig:fig13}}
\end{figure*}

The constant values of mass, age, and LPV phase duration relations are investigated by using the IRAF software. The related coefficients can be found in tables \ref{tab:taba1}, \ref{tab:taba2}, and \ref{tab:taba3} for various ranges, and the set of Fig.\,\ref{fig:fig13} depict linear and multiple-Gaussian fits on the obtained data which are the peak of magnitude in each age of theoretical isochrones of Padova evolutionary models (\citealp{marigo2017}).

\figsetstart
\figsetnum{9}
\figsettitle{The mass-luminosity, age-mass, and mass-LPV phase duration relations for metallicities in range of $0.0003 < Z < 0.04$.}

\figsetgrpstart
\figsetgrpnum{1.1}
\figsetgrptitle{Z = 0.0003}
\figsetplot{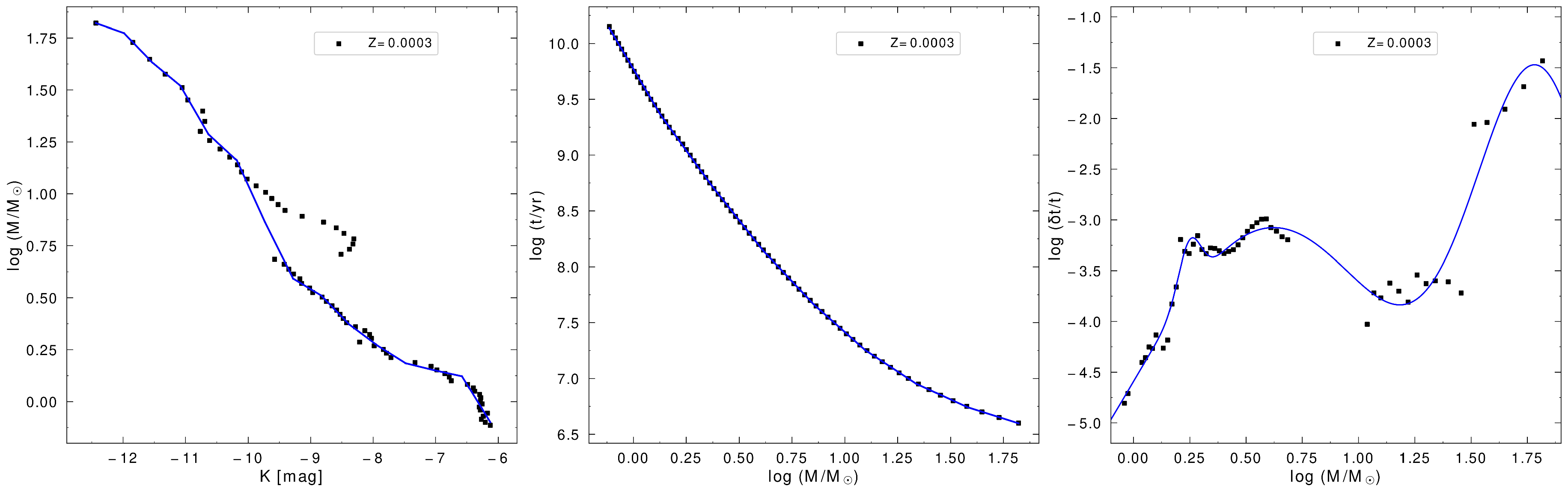}
\figsetgrpnote{{\it Left panel}: The mass-luminosity relation in the $K_s$ band for a metallicity of $Z = 0.0003$. The solid lines are the linear spline fits, for the case in which the function is interpolated across the super--AGB phase to massive red supergiants, i.e. for  0.7 < $\log (M/M_\odot)$ < 1-1.1. {\it Middle panel}: The mass-age relation for a metallicity of $Z = 0.0003$ along with linear spline fits. {\it Right panel}: The mass-LPV phase duration relation with the same metallicity, where the points show the ratio of LPV phase duration to age vs. mass; the solid lines are multiple-Gaussian fits.}

\figsetgrpend

\figsetgrpstart
\figsetgrpnum{1.2}
\figsetgrptitle{Z = 0.0005}
\figsetplot{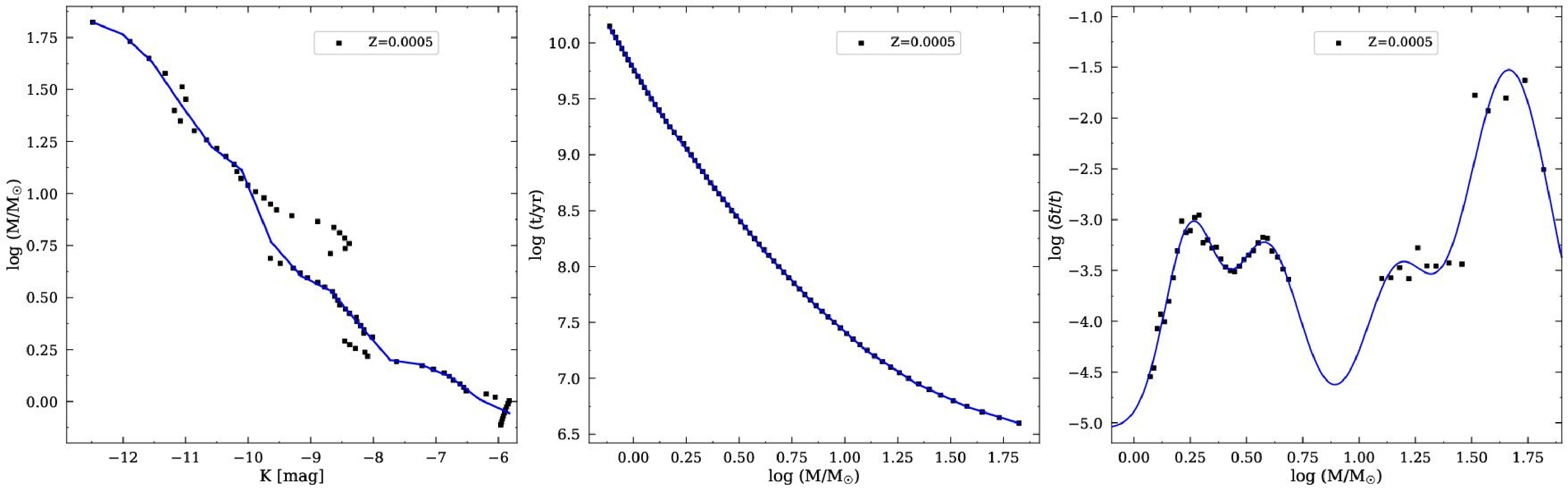}
\figsetgrpnote{Same as metallicity of $Z = 0.0003$}
\figsetgrpend

\figsetgrpstart
\figsetgrpnum{1.3}
\figsetgrptitle{Z = 0.001}
\figsetplot{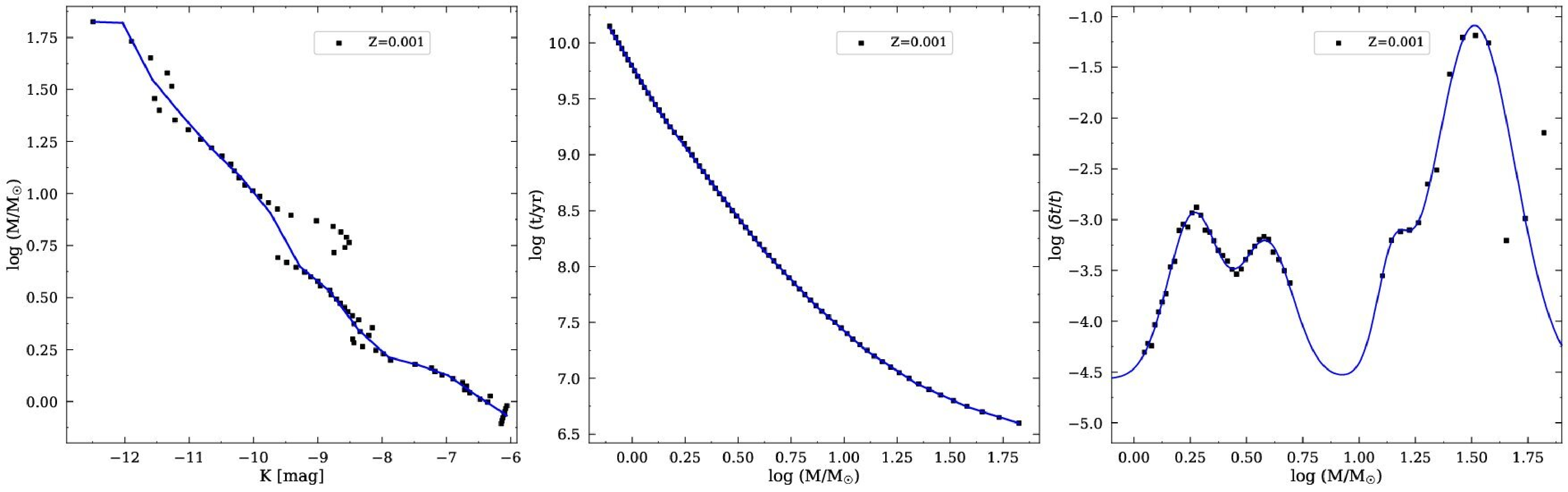}
\figsetgrpnote{Same as metallicity of $Z = 0.0003$}
\figsetgrpend

\figsetgrpstart
\figsetgrpnum{1.4}
\figsetgrptitle{Z = 0.003}
\figsetplot{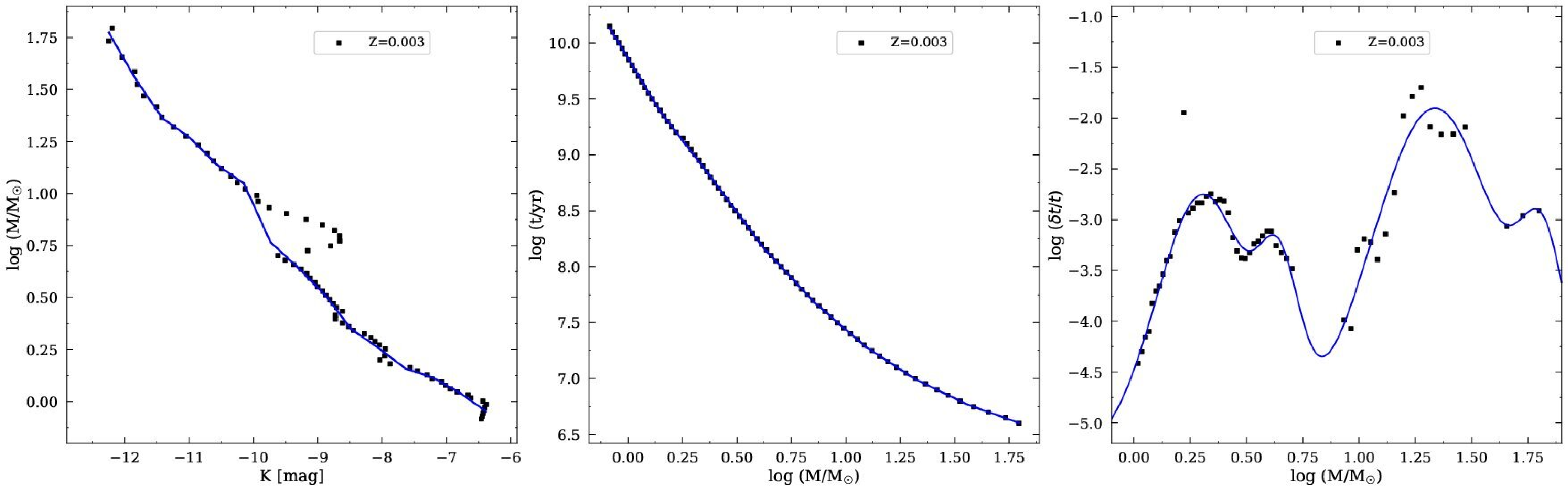}
\figsetgrpnote{Same as metallicity of $Z = 0.0003$}
\figsetgrpend

\figsetgrpstart
\figsetgrpnum{1.5}
\figsetgrptitle{Z = 0.006}
\figsetplot{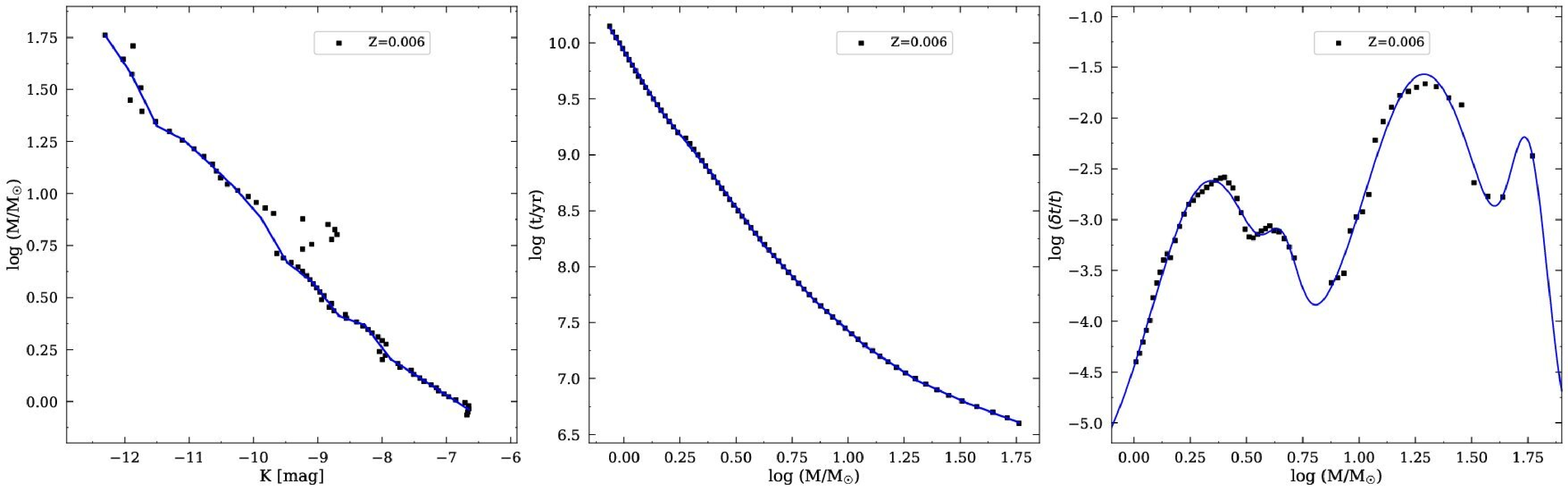}
\figsetgrpnote{Same as metallicity of $Z = 0.0003$}
\figsetgrpend

\figsetgrpstart
\figsetgrpnum{1.6}
\figsetgrptitle{Z = 0.008}
\figsetplot{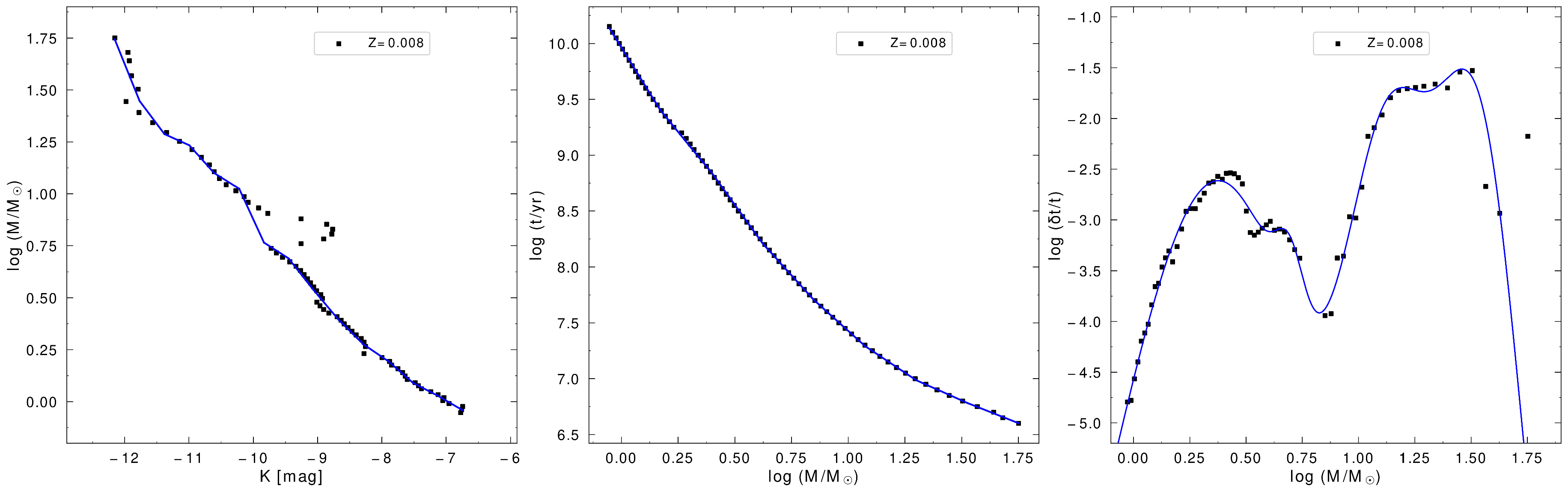}
\figsetgrpnote{Same as metallicity of $Z = 0.0003$}
\figsetgrpend

\figsetgrpstart
\figsetgrpnum{1.7}
\figsetgrptitle{Z = 0.010}
\figsetplot{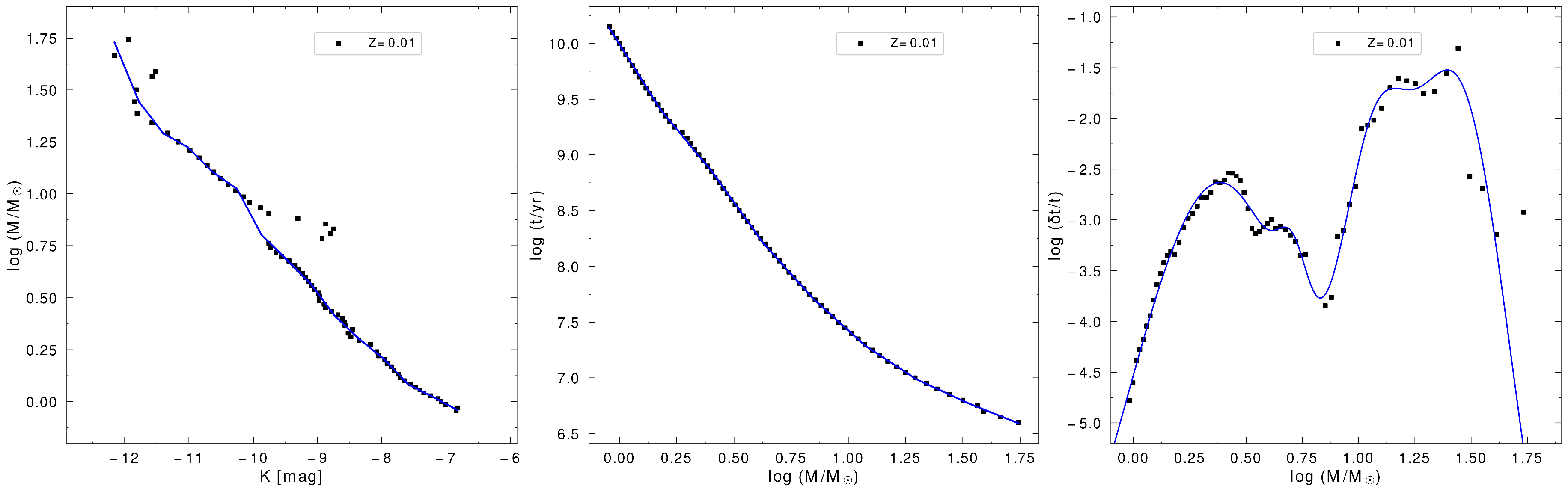}
\figsetgrpnote{Same as metallicity of $Z = 0.0003$}
\figsetgrpend

\figsetgrpstart
\figsetgrpnum{1.8}
\figsetgrptitle{Z = 0.020}
\figsetplot{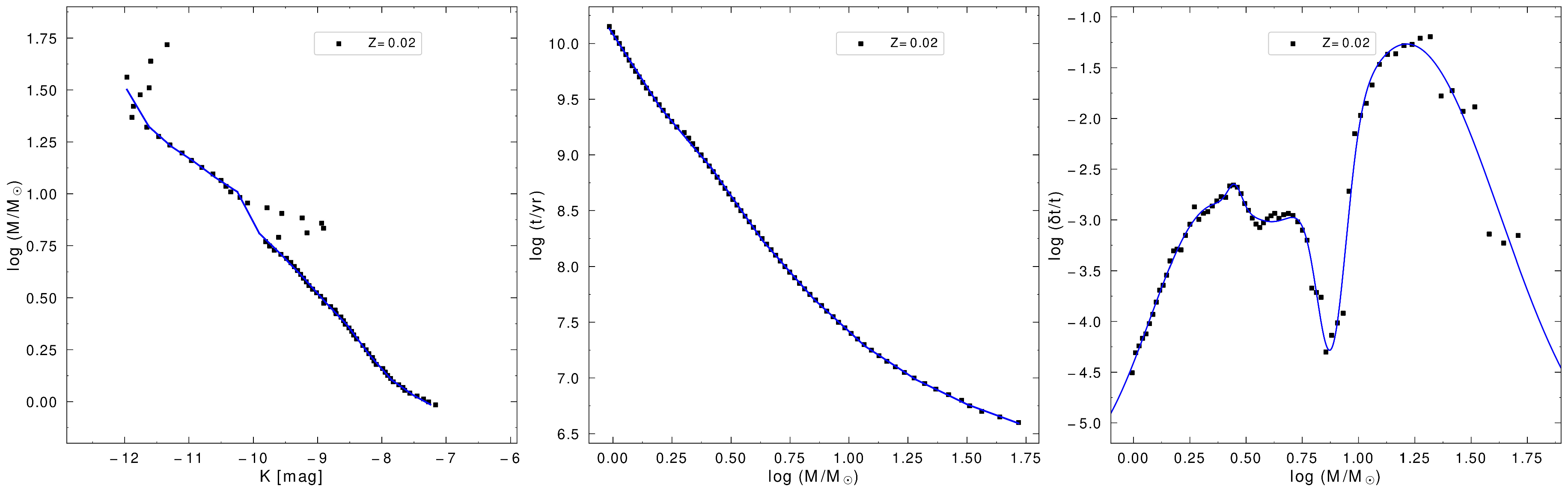}
\figsetgrpnote{Same as metallicity of $Z = 0.0003$}
\figsetgrpend

\figsetgrpstart
\figsetgrpnum{1.9}
\figsetgrptitle{Z = 0.030}
\figsetplot{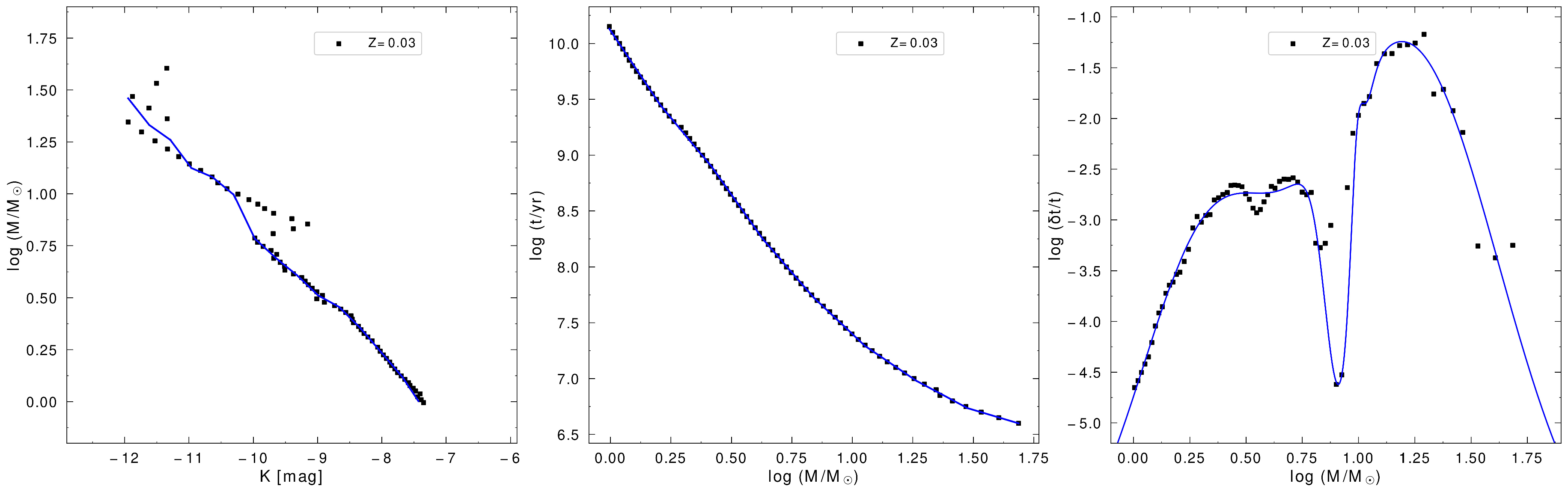}
\figsetgrpnote{Same as metallicity of $Z = 0.0003$}
\figsetgrpend

\figsetgrpstart
\figsetgrpnum{1.10}
\figsetgrptitle{Z = 0.035}
\figsetplot{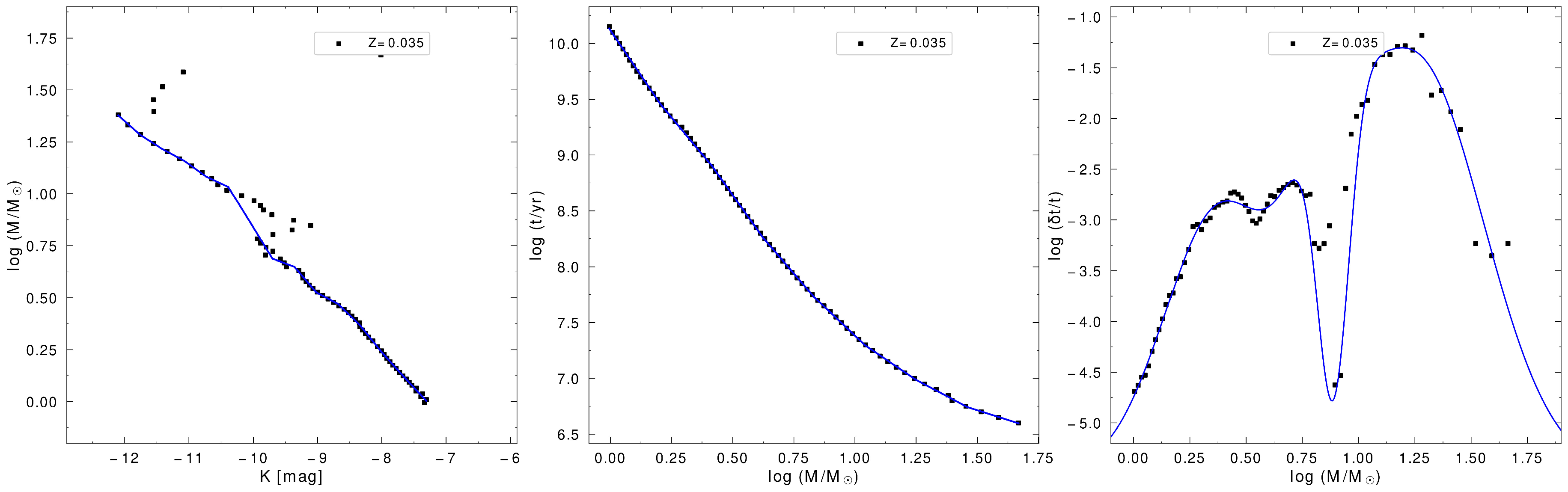}
\figsetgrpnote{Same as metallicity of $Z = 0.0003$}
\figsetgrpend

\figsetgrpstart
\figsetgrpnum{1.11}
\figsetgrptitle{Z = 0.039}
\figsetplot{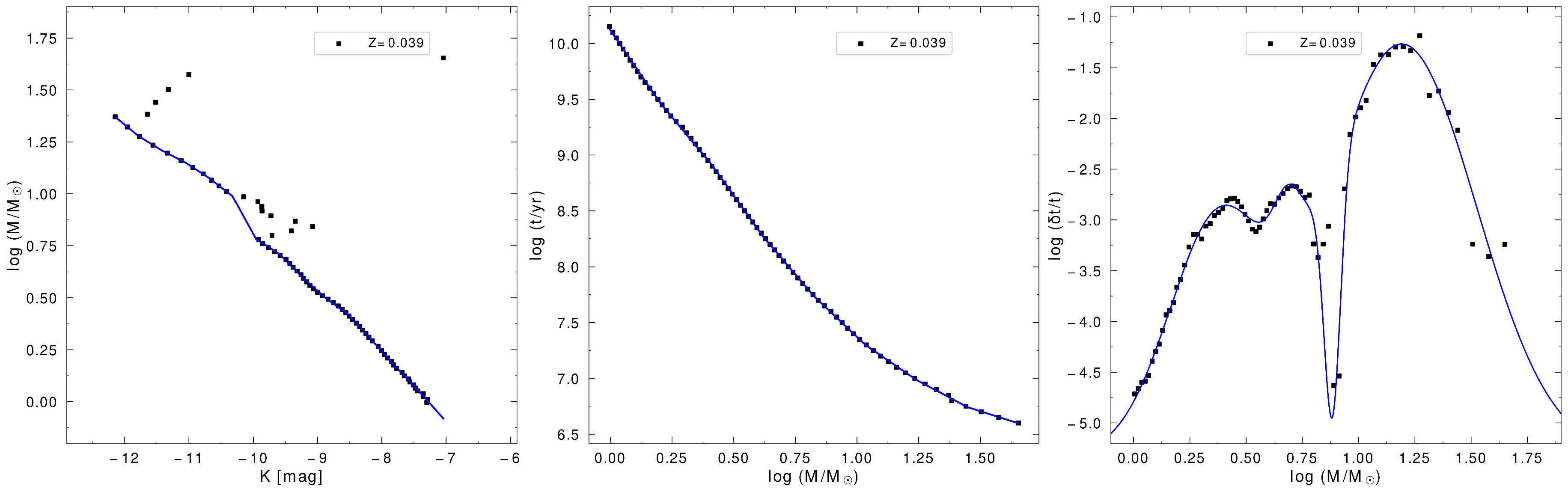}
\figsetgrpnote{Same as metallicity of $Z = 0.0003$}
\figsetgrpend

\figsetend

\begin{table}[]
	\centering
	\caption{Fitting equations of the relation between birth mass and $K_s$-band magnitude, $\log \, M/M_\odot = a \, K + b$, for a distance modulus of $\mu = 27.87$ mag \cite{Rejkuba2004a}.}
	\label{tab:taba1}
	\begin{tabular}{ccc}
		\doubleRule
		a                  & b                   & validity range               \\ \hline
		\multicolumn{3}{c}{Z = 0.0003}                                          \\ \hline
		-0.107 $\pm$ 0.181 & 0.497 $\pm$ 2.124  & $K_s \leq$ -11.984           \\
		-0.312 $\pm$ 0.182 & -1.965 $\pm$ 2.062 & -11.984 $< K_s \leq$ -11.534 \\
		-0.249 $\pm$ 0.128 & -131.238 $\pm$ 1.390 & -11.534 $< K_s \leq$ -11.083 \\
		-0.520 $\pm$ 0.098 & -4.239 $\pm$ 1.028 & -11.083 $< K_s \leq$ -10.633 \\
		-0.278 $\pm$ 0.087 & -1.669 $\pm$ 0.863 & -10.633 $< K_s \leq$ -10.182 \\
		-0.660 $\pm$ 0.118 & -5.564 $\pm$ 1.112 & -10.182 $< K_s \leq$ -9.732  \\
		-0.602 $\pm$ 0.113 & -4.993 $\pm$ 1.031 & -9.732 $< K_s \leq$ -9.281   \\
		-0.181 $\pm$ 0.083 & -1.088 $\pm$ 0.717 & -9.281 $< K_s \leq$ -8.830   \\
		-0.310 $\pm$ 0.084 & -2.226 $\pm$ 0.687 & -8.830 $< K_s \leq$ -8.380   \\
		-0.225 $\pm$ 0.083 & -1.514 $\pm$ 0.640 & -8.380 $< K_s \leq$ -7.929   \\
		-0.189 $\pm$ 0.129 & -1.227 $\pm$ 0.922 & -7.929 $< K_s \leq$ -7.479   \\
		-0.074 $\pm$ 0.137 & -0.367 $\pm$ 0.943 & -7.479 $< K_s \leq$ -7.028   \\
		-0.066 $\pm$ 0.107 & -0.310 $\pm$ 0.676 & -7.028 $< K_s \leq$ -6.578   \\
		-0.491 $\pm$ 0.091 & -3.106 $\pm$ 0.544 & $K_s >$ -6.578               \\ \doubleRule
		\multicolumn{3}{c}{Z = 0.0005}                                          \\ \hline
		-0.121 $\pm$ 0.227 & 0.309 $\pm$ 2.663  & $K_s \leq$ -12.011           \\
		-0.281 $\pm$ 0.226 & -1.614 $\pm$ 2.558 & -12.011 $< K_s \leq$ -11.535 \\
		-0.443 $\pm$ 0.143 & -3.475 $\pm$ 1.554 & -11.535 $< K_s \leq$ -11.060 \\
		-0.414 $\pm$ 0.123 & -3.160 $\pm$ 1.265 & -11.060 $< K_s \leq$ -10.584 \\
		-0.231 $\pm$ 0.126 & -1.217 $\pm$ 1.247 & -10.584 $< K_s \leq$ -10.108 \\
		-0.729 $\pm$ 0.118 & -6.259 $\pm$ 1.103 & -10.108 $< K_s \leq$ -9.632  \\
		-0.344 $\pm$ 0.126 & -2.547 $\pm$ 1.123 & -9.632 $< K_s \leq$ -9.156   \\
		-0.151 $\pm$ 0.111 & -0.774 $\pm$ 0.941 & -9.156 $< K_s \leq$ -8.681   \\
		-0.354 $\pm$ 0.093 & -2.538 $\pm$ 0.740 & -8.681 $< K_s \leq$ -8.205   \\
		-0.345 $\pm$ 0.152 & -2.463 $\pm$ 1.123 & -8.205 $< K_s \leq$ -7.729   \\
		-0.045 $\pm$ 0.196 & -0.144 $\pm$ 1.383 & -7.729 $< K_s \leq$ -7.253   \\
		-0.122 $\pm$ 0.144 & -0.710 $\pm$ 0.947 & -7.253 $< K_s \leq$ -6.778   \\
		-0.231 $\pm$ 0.140 & -1.443 $\pm$ 0.840 & -6.778 $< K_s \leq$ -6.302   \\
		-0.139 $\pm$ 0.125 & -0.866 $\pm$ 0.706 & $K_s >$ -6.302               \\ \doubleRule
		\multicolumn{3}{c}{Z = 0.001}                                           \\ \hline
		-0.011 $\pm$ 0.303 & 1.684 $\pm$ 3.566  & $K_s \leq$ -12.033           \\
		-0.590 $\pm$ 0.262 & -5.274 $\pm$ 2.996 & -12.033 $< K_s \leq$ -11.573 \\
		-0.375 $\pm$ 0.176 & -2.795 $\pm$ 1.915 & -11.573 $< K_s \leq$ -11.114 \\
		-0.347 $\pm$ 0.187 & -2.484 $\pm$ 1.948 & -11.114 $< K_s \leq$ -10.654 \\
		-0.300 $\pm$ 0.165 & -1.979 $\pm$ 1.646 & -10.654 $< K_s \leq$ -10.195 \\
		-0.376 $\pm$ 0.150 & -2.752 $\pm$ 1.428 & -10.195 $< K_s \leq$ -9.735  \\
		-0.561 $\pm$ 0.155 & -4.552 $\pm$ 1.399 & -9.735 $< K_s \leq$ -9.276   \\
		-0.255 $\pm$ 0.140 & -1.714 $\pm$ 1.204 & -9.276 $< K_s \leq$ -8.816   \\
		-0.414 $\pm$ 0.115 & -3.117 $\pm$ 0.941 & -8.816 $< K_s \leq$ -8.357   \\
		-0.280 $\pm$ 0.139 & -1.996 $\pm$ 1.056 & -8.357 $< K_s \leq$ -7.897   \\
		-0.082 $\pm$ 0.203 & -0.432 $\pm$ 1.453 & -7.897 $< K_s \leq$ -7.438   \\
		-0.108 $\pm$ 0.208 & -0.626 $\pm$ 1.409 & -7.438 $< K_s \leq$ -6.978   \\
		-0.216 $\pm$ 0.170 & -1.381 $\pm$ 1.070 & -6.978 $< K_s \leq$ -6.518   \\
		-0.204 $\pm$ 0.130 & -1.305 $\pm$ 0.764 & $K_s >$ -6.518               \\ \doubleRule
	\end{tabular}
\end{table}

\begin{table}[]
	\centering
	\vspace{1.4cm}
	\begin{tabular}{ccc}
		\doubleRule
		a                  & b                   & validity range                \\ \hline
		\multicolumn{3}{c}{Z = 0.003}                                           \\ \hline
		-0.538 $\pm$ 0.085 & -4.810 $\pm$ 0.983 & $K_s \leq$ -11.827           \\
		-0.449 $\pm$ 0.086 & -3.763 $\pm$ 0.967 & -11.827 $< K_s \leq$ -11.408 \\
		-0.219 $\pm$ 0.101 & -1.134 $\pm$ 1.090 & -11.408 $< K_s \leq$ -10.990 \\
		-0.311 $\pm$ 0.101 & -2.151 $\pm$ 1.047 & -10.990 $< K_s \leq$ -10.571 \\
		-0.211 $\pm$ 0.093 & -1.093 $\pm$ 0.926 & -10.571 $< K_s \leq$ -10.152 \\
		-0.676 $\pm$ 0.115 & -5.818 $\pm$ 1.088 & -10.152 $< K_s \leq$ -9.733  \\
		-0.287 $\pm$ 0.110 & -2.030 $\pm$ 1.006 & -9.733 $< K_s \leq$ -9.315   \\
		-0.311 $\pm$ 0.073 & -2.251 $\pm$ 0.639 & -9.315 $< K_s \leq$ -8.896   \\
		-0.401 $\pm$ 0.067 & -3.049 $\pm$ 0.552 & -8.896 $< K_s \leq$ -8.477   \\
		-0.216 $\pm$ 0.067 & -1.481 $\pm$ 0.525 & -8.477 $< K_s \leq$ -8.058   \\
		-0.235 $\pm$ 0.090 & -1.636 $\pm$ 0.664 & -8.058 $< K_s \leq$ -7.639   \\
		-0.102 $\pm$ 0.102 & -0.623 $\pm$ 0.718 & -7.639 $< K_s \leq$ -7.221   \\
		-0.173 $\pm$ 0.085 & -1.130 $\pm$ 0.560 & -7.221 $< K_s \leq$ -6.802   \\
		-0.218 $\pm$ 0.069 & -1.436 $\pm$ 0.431 & $K_s >$ -6.802               \\ \doubleRule
		\multicolumn{3}{c}{Z = 0.006}                                          \\ \hline
		-0.459 $\pm$ 0.136 & -3.888 $\pm$ 1.602 & $K_s \leq$ -11.902           \\
		-0.623 $\pm$ 0.127 & -5.844 $\pm$ 1.426 & -11.902 $< K_s \leq$ -11.498 \\
		-0.159 $\pm$ 0.156 & -0.508 $\pm$ 1.698 & -11.498 $< K_s \leq$ -11.094 \\
		-0.280 $\pm$ 0.142 & -1.851 $\pm$ 1.493 & -11.094 $< K_s \leq$ -10.689 \\
		-0.298 $\pm$ 0.146 & -2.036 $\pm$ 1.474 & -10.689 $< K_s \leq$ -10.285 \\
		-0.355 $\pm$ 0.244 & -2.623 $\pm$ 2.344 & -10.285 $< K_s \leq$ -9.881  \\
		-0.529 $\pm$ 0.230 & -4.347 $\pm$ 2.148 & -9.881 $< K_s \leq$ -9.477   \\
		-0.240 $\pm$ 0.112 & -1.605 $\pm$ 0.999 & -9.477 $< K_s \leq$ -9.073   \\
		-0.397 $\pm$ 0.102 & -3.031 $\pm$ 0.862 & -9.073 $< K_s \leq$ -8.669   \\
		-0.111 $\pm$ 0.113 & -0.555 $\pm$ 0.915 & -8.669 $< K_s \leq$ -8.265   \\
		-0.405 $\pm$ 0.110 & -2.981 $\pm$ 0.841 & -8.265 $< K_s \leq$ -7.860   \\
		-0.198 $\pm$ 0.108 & -1.355 $\pm$ 0.785 & -7.860 $< K_s \leq$ -7.456   \\
		-0.200 $\pm$ 0.107 & -1.371 $\pm$ 0.734 & -7.456 $< K_s \leq$ -7.052   \\
		-0.200 $\pm$ 0.095 & -1.372 $\pm$ 0.614 & $K_s >$ -7.052               \\ \doubleRule
		\multicolumn{3}{c}{Z = 0.008}                                           \\ \hline
		-0.767 $\pm$ 0.131 & -7.575 $\pm$ 1.522 & $K_s \leq$ -11.767           \\
		-0.412 $\pm$ 0.140 & -3.402 $\pm$ 1.563 & -11.767 $< K_s \leq$ -11.381 \\
		-0.139 $\pm$ 0.165 & -0.298 $\pm$ 1.781 & -11.381 $< K_s \leq$ -10.994 \\
		-0.346 $\pm$ 0.139 & -2.565 $\pm$ 1.449 & -10.994 $< K_s \leq$ -10.608 \\
		-0.195 $\pm$ 0.126 & -0.972 $\pm$ 1.265 & -10.608 $< K_s \leq$ -10.221 \\
		-0.671 $\pm$ 0.173 & -5.838 $\pm$ 1.661 & -10.221 $< K_s \leq$ -9.835  \\
		-0.200 $\pm$ 0.167 & -1.198 $\pm$ 1.550 & -9.835 $< K_s \leq$ -9.448   \\
		-0.390 $\pm$ 0.101 & -2.999 $\pm$ 0.896 & -9.448 $< K_s \leq$ -9.061   \\
		-0.368 $\pm$ 0.089 & -2.796 $\pm$ 0.756 & -9.061 $< K_s \leq$ -8.675   \\
		-0.307 $\pm$ 0.095 & -2.269 $\pm$ 0.770 & -8.675 $< K_s \leq$ -8.288   \\
		-0.210 $\pm$ 0.102 & -1.461 $\pm$ 0.780 & -8.288 $< K_s \leq$ -7.902   \\
		-0.271 $\pm$ 0.110 & -1.948 $\pm$ 0.807 & -7.902 $< K_s \leq$ -7.515   \\
		-0.161 $\pm$ 0.107 & -1.117 $\pm$ 0.740 & -7.515 $< K_s \leq$ -7.129   \\
		-0.180 $\pm$ 0.108 & -1.258 $\pm$ 0.705 & $K_s >$ -7.129               \\ \doubleRule
	\end{tabular}
\end{table}

\begin{table}[]
	\centering
	\begin{tabular}{ccc}
		\doubleRule
		a                  & b                   & validity range                \\ \hline
		\multicolumn{3}{c}{Z = 0.010}                                           \\ \hline
		-0.750 $\pm$ 0.139 & -7.390 $\pm$ 1.617 & $K_s \leq$ -11.777           \\
		-0.402 $\pm$ 0.143 & -3.296 $\pm$ 1.599 & -11.777 $< K_s \leq$ -11.396 \\
		-0.173 $\pm$ 0.165 & -0.677 $\pm$ 1.788 & -11.396 $< K_s \leq$ -11.016 \\
		-0.310 $\pm$ 0.141 & -2.189 $\pm$ 1.479 & -11.016 $< K_s \leq$ -10.635 \\
		-0.217 $\pm$ 0.137 & -1.196 $\pm$ 1.382 & -10.635 $< K_s \leq$ -10.254 \\
		-0.583 $\pm$ 0.162 & -4.950 $\pm$ 1.562 & -10.254 $< K_s \leq$ -9.873  \\
		-0.305 $\pm$ 0.150 & -2.210 $\pm$ 1.399 & -9.873 $< K_s \leq$ -9.492   \\
		-0.309 $\pm$ 0.104 & -2.245 $\pm$ 0.929 & -9.492 $< K_s \leq$ -9.112   \\
		-0.410 $\pm$ 0.093 & -3.168 $\pm$ 0.790 & -9.112 $< K_s \leq$ -8.731   \\
		-0.292 $\pm$ 0.108 & -2.137 $\pm$ 0.883 & -8.731 $< K_s \leq$ -8.350   \\
		-0.249 $\pm$ 0.105 & -1.776 $\pm$ 0.816 & -8.350 $< K_s \leq$ -7.969   \\
		-0.324 $\pm$ 0.099 & -2.376 $\pm$ 0.735 & -7.969 $< K_s \leq$ -7.589   \\
		-0.156 $\pm$ 0.109 & -1.102 $\pm$ 0.765 & -7.589 $< K_s \leq$ -7.208   \\
		-0.166 $\pm$ 0.117 & -1.170 $\pm$ 0.776 & $K_s >$ -7.208               \\ \doubleRule
		\multicolumn{3}{c}{Z = 0.02}                                            \\ \hline
		-0.521 $\pm$ 0.104 & -4.732 $\pm$ 1.192 & $K_s \leq$ -11.622           \\
		-0.274 $\pm$ 0.117 & -1.866 $\pm$ 1.295 & -11.622 $< K_s \leq$ -11.279 \\
		-0.207 $\pm$ 0.117 & -1.101 $\pm$ 1.258 & -11.279 $< K_s \leq$ -10.936 \\
		-0.230 $\pm$ 0.110 & -1.357 $\pm$ 1.151 & -10.936 $< K_s \leq$ -10.593 \\
		-0.203 $\pm$ 0.107 & -1.068 $\pm$ 1.077 & -10.593 $< K_s \leq$ -10.250 \\
		-0.585 $\pm$ 0.124 & -4.991 $\pm$ 1.207 & -10.250 $< K_s \leq$ -9.907  \\
		-0.298 $\pm$ 0.115 & -2.147 $\pm$ 1.085 & -9.907 $< K_s \leq$ -9.564   \\
		-0.319 $\pm$ 0.077 & -2.348 $\pm$ 0.699 & -9.564 $< K_s \leq$ -9.221   \\
		-0.344 $\pm$ 0.067 & -2.577 $\pm$ 0.580 & -9.221 $< K_s \leq$ -8.878   \\
		-0.328 $\pm$ 0.065 & -2.429 $\pm$ 0.548 & -8.878 $< K_s \leq$ -8.535   \\
		-0.414 $\pm$ 0.064 & -3.166 $\pm$ 0.514 & -8.535 $< K_s \leq$ -8.192   \\
		-0.349 $\pm$ 0.067 & -2.638 $\pm$ 0.515 & -8.192 $< K_s \leq$ -7.849   \\
		-0.217 $\pm$ 0.080 & -1.601 $\pm$ 0.583 & -7.849 $< K_s \leq$ -7.506   \\
		-0.130 $\pm$ 0.099 & -0.958 $\pm$ 0.696 & $K_s >$ -7.506               \\ \doubleRule
		\multicolumn{3}{c}{Z = 0.030}                                           \\ \hline
		-0.391 $\pm$ 0.166 & -3.214 $\pm$ 1.910 & $K_s \leq$ -11.618           \\
		-0.219 $\pm$ 0.128 & -1.210 $\pm$ 1.427 & -11.618 $< K_s \leq$ -11.290 \\
		-0.413 $\pm$ 0.140 & -3.407 $\pm$ 1.514 & -11.290 $< K_s \leq$ -10.961 \\
		-0.133 $\pm$ 0.147 & -0.338 $\pm$ 1.535 & -10.961 $< K_s \leq$ -10.633 \\
		-0.257 $\pm$ 0.195 & -1.649 $\pm$ 1.965 & -10.633 $< K_s \leq$ -10.305 \\
		-0.647 $\pm$ 0.183 & -5.668 $\pm$ 1.806 & -10.305 $< K_s \leq$ -9.977  \\
		-0.284 $\pm$ 0.104 & -2.050 $\pm$ 0.985 & -9.977 $< K_s \leq$ -9.649   \\
		-0.249 $\pm$ 0.106 & -1.713 $\pm$ 0.970 & -9.649 $< K_s \leq$ -9.320   \\
		-0.296 $\pm$ 0.106 & -2.148 $\pm$ 0.937 & -9.320 $< K_s \leq$ -8.992   \\
		-0.171 $\pm$ 0.102 & -1.023 $\pm$ 0.868 & -8.992 $< K_s \leq$ -8.664   \\
		-0.307 $\pm$ 0.100 & -2.208 $\pm$ 0.822 & -8.664 $< K_s \leq$ -8.336   \\
		-0.360 $\pm$ 0.087 & -2.646 $\pm$ 0.682 & -8.336 $< K_s \leq$ -8.007   \\
		-0.360 $\pm$ 0.087 & -2.649 $\pm$ 0.656 & -8.007 $< K_s \leq$ -7.679   \\
		-0.353 $\pm$ 0.089 & -2.617 $\pm$ 0.649 & $K_s >$ -7.679               \\ \doubleRule
	\end{tabular}
\end{table}

\begin{table}[]
	\centering
	\begin{tabular}{ccc}
		\doubleRule
		a                  & b                   & validity range               \\ \hline
		\multicolumn{3}{c}{Z = 0.035}                                           \\ \hline
		-0.275 $\pm$ 0.079 & -1.949 $\pm$ 0.920 & $K_s \leq$ -11.758           \\
		-0.198 $\pm$ 0.088 & -1.040 $\pm$ 0.991 & -11.758 $< K_s \leq$ -11.415 \\
		-0.168 $\pm$ 0.094 & -0.698 $\pm$ 1.025 & -11.415 $< K_s \leq$ -11.073 \\
		-0.222 $\pm$ 0.086 & -1.296 $\pm$ 0.904 & -11.073 $< K_s \leq$ -10.731 \\
		-0.146 $\pm$ 0.081 & -0.481 $\pm$ 0.827 & -10.731 $< K_s \leq$ -10.388 \\
		-0.502 $\pm$ 0.086 & -4.179 $\pm$ 0.847 & -10.388 $< K_s \leq$ -10.046 \\
		-0.505 $\pm$ 0.075 & -4.214 $\pm$ 0.717 & -10.046 $< K_s \leq$ -9.704  \\
		-0.113 $\pm$ 0.063 & -0.408 $\pm$ 0.574 & -9.704 $< K_s \leq$ -9.361   \\
		-0.361 $\pm$ 0.059 & -2.732 $\pm$ 0.525 & -9.361 $< K_s \leq$ -9.019   \\
		-0.164 $\pm$ 0.054 & -0.951 $\pm$ 0.462 & -9.019 $< K_s \leq$ -8.676   \\
		-0.304 $\pm$ 0.048 & -2.169 $\pm$ 0.392 & -8.676 $< K_s \leq$ -8.334   \\
		-0.383 $\pm$ 0.043 & -2.830 $\pm$ 0.336 & -8.334 $< K_s \leq$ -7.992   \\
		-0.342 $\pm$ 0.045 & -2.502 $\pm$ 0.337 & -7.992 $< K_s \leq$ -7.649   \\
		-0.331 $\pm$ 0.047 & -2.418 $\pm$ 0.333  & $K_s >$ -7.649              \\ \doubleRule
		\multicolumn{3}{c}{Z = 0.039}                                           \\ \hline
		-0.254 $\pm$ 0.025 & -1.714 $\pm$ 0.286 & $K_s \leq$ -11.783           \\
		-0.190 $\pm$ 0.027 & -0.963 $\pm$ 0.301 & -11.783 $< K_s \leq$ -11.419 \\
		-0.159 $\pm$ 0.028 & -0.611 $\pm$ 0.304 & -11.419 $< K_s \leq$ -11.054 \\
		-0.202 $\pm$ 0.025 & -1.077 $\pm$ 0.264 & -11.054 $< K_s \leq$ -10.689 \\
		-0.241 $\pm$ 0.029 & -1.499 $\pm$ 0.292 & -10.689 $< K_s \leq$ -10.325 \\
		-0.564 $\pm$ 0.029 & -4.835 $\pm$ 0.282 & -10.325 $< K_s \leq$ -9.960  \\
		-0.200 $\pm$ 0.020 & -1.207 $\pm$ 0.191 & -9.960 $< K_s \leq$ -9.595   \\
		-0.311 $\pm$ 0.015 & -2.276 $\pm$ 0.140 & -9.595 $< K_s \leq$ -9.231   \\
		-0.275 $\pm$ 0.015 & -1.942 $\pm$ 0.130 & -9.231 $< K_s \leq$ -8.866   \\
		-0.225 $\pm$ 0.015 & -1.498 $\pm$ 0.126 & -8.866 $< K_s \leq$ -8.502   \\
		-0.343 $\pm$ 0.014 & -2.502 $\pm$ 0.113 & -8.502 $< K_s \leq$ -8.137   \\
		-0.343 $\pm$ 0.014 & -2.501 $\pm$ 0.107 & -8.137 $< K_s \leq$ -7.772   \\
		-0.326 $\pm$ 0.015 & -2.370 $\pm$ 0.107 & -7.772 $< K_s \leq$ -7.408   \\
		-0.347 $\pm$ 0.043 & -2.528 $\pm$ 0.293 & $K_s >$ -7.408               \\ \doubleRule
	\end{tabular}
\end{table}

\begin{table}[]
	\centering
	\caption{Fitting equations of the relation between age and birth mass, $\log \, t = a \, \log \, M + b$.}
	\label{tab:taba2}
	\begin{tabular}{ccc}
		\doubleRule
		a                  & b                 & validity range               \\ \hline
		\multicolumn{3}{c}{Z = 0.0003}                                        \\ \hline
		-3.202 $\pm$ 0.021 & 9.772 $\pm$ 0.005 & $\log M \ \leq$ 0.128         \\
		-2.629 $\pm$ 0.019 & 9.699 $\pm$ 0.009 & 0.128 $< \ \log M \leq$ 0.370 \\
		-2.394 $\pm$ 0.020 & 9.612 $\pm$ 0.015 & 0.370 $< \ \log M \leq$ 0.612 \\
		-2.007 $\pm$ 0.022 & 9.375 $\pm$ 0.021 & 0.612 $< \ \log M \leq$ 0.854 \\
		-1.681 $\pm$ 0.024 & 9.096 $\pm$ 0.030 & 0.854 $< \ \log M \leq$ 1.096 \\
		-1.249 $\pm$ 0.028 & 8.623 $\pm$ 0.041 & 1.096 $< \ \log M \leq$ 1.338 \\
		-0.869 $\pm$ 0.033 & 8.115 $\pm$ 0.056 & 1.338 $< \ \log M \leq$ 1.579 \\
		-0.598 $\pm$ 0.040 & 7.688 $\pm$ 0.077 & $\ M \ >$ 1.579            \\ \doubleRule
		\multicolumn{3}{c}{Z = 0.0005}                                        \\ \hline
		-3.178 $\pm$ 0.023 & 9.777 $\pm$ 0.006 & $\log M \ \leq$ 0.130         \\
		-2.608 $\pm$ 0.021 & 9.703 $\pm$ 0.010 & 0.130 $< \ \log M \leq$ 0.372 \\
		-2.415 $\pm$ 0.022 & 9.631 $\pm$ 0.016 & 0.372 $< \ \log M \leq$ 0.614 \\
		-2.020 $\pm$ 0.024 & 9.389 $\pm$ 0.024 & 0.614 $< \ \log M \leq$ 0.856 \\
		-1.683 $\pm$ 0.027 & 9.101 $\pm$ 0.033 & 0.856 $< \ \log M \leq$ 1.098 \\
		-1.247 $\pm$ 0.031 & 8.622 $\pm$ 0.045 & 1.098 $< \ \log M \leq$ 1.340 \\
		-0.866 $\pm$ 0.036 & 8.111 $\pm$ 0.062 & 1.340 $< \ \log M \leq$ 1.582 \\
		-0.596 $\pm$ 0.044 & 7.685 $\pm$ 0.086 & $\log M \ >$ 1.582            \\ \doubleRule
		\multicolumn{3}{c}{Z = 0.001}                                         \\ \hline
		-3.174 $\pm$ 0.026 & 9.797 $\pm$ 0.006 & $\log M \ \leq$ 0.135         \\
		-2.579 $\pm$ 0.024 & 9.716 $\pm$ 0.012 & 0.135 $< \ \log M \leq$ 0.377 \\
		-2.466 $\pm$ 0.025 & 9.674 $\pm$ 0.019 & 0.377 $< \ \log M \leq$ 0.618 \\
		-2.020 $\pm$ 0.027 & 9.398 $\pm$ 0.027 & 0.618 $< \ \log M \leq$ 0.860 \\
		-1.692 $\pm$ 0.030 & 9.116 $\pm$ 0.037 & 0.860 $< \ \log M \leq$ 1.101 \\
		1.249 $\pm$ 0.035  & 8.628 $\pm$ 0.051 & 1.101 $< \ \log M \leq$ 1.342 \\
		0.868 $\pm$ 0.041  & 8.117 $\pm$ 0.070 & 1.342 $< \ \log M \leq$ 1.584 \\
		-0.599 $\pm$ 0.050 & 7.691 $\pm$ 0.097 & $\log M \ >$ 1.584            \\ \doubleRule
		\multicolumn{3}{c}{Z = 0.003}                                         \\ \hline
		-3.209 $\pm$ 0.032 & 9.862 $\pm$ 0.008 & $\log M \ \leq$ 0.150         \\
		-2.520 $\pm$ 0.029 & 9.759 $\pm$ 0.015 & 0.150 $< \ \log M \leq$ 0.385 \\
		-2.620 $\pm$ 0.030 & 9.797 $\pm$ 0.022 & 0.385 $< \ \log M \leq$ 0.620 \\
		-2.091 $\pm$ 0.033 & 9.469 $\pm$ 0.032 & 0.620 $< \ \log M \leq$ 0.855 \\
		-1.685 $\pm$ 0.037 & 9.122 $\pm$ 0.045 & 0.855 $< \ \log M \leq$ 1.090 \\
		-1.269 $\pm$ 0.042 & 8.669 $\pm$ 0.061 & 1.090 $< \ \log M \leq$ 1.325 \\
		-0.937 $\pm$ 0.049 & 8.228 $\pm$ 0.083 & 1.325 $< \ \log M \leq$ 1.560 \\
		-0.698 $\pm$ 0.057 & 7.856 $\pm$ 0.110 & $\log M \ >$ 1.560            \\ \doubleRule
		\multicolumn{3}{c}{Z = 0.006}                                          \\ \hline
		-3.252 $\pm$ 0.033 & 9.929 $\pm$ 0.009  & $\log M \ \leq$ 0.164         \\
		-2.479 $\pm$ 0.030 & 9.802 $\pm$ 0.015  & 0.164 $< \ \log M \leq$ 0.392 \\
		-2.753 $\pm$ 0.031 & 9.910 $\pm$ 0.023  & 0.392 $< \ \log M \leq$ 0.621 \\
		-2.194 $\pm$ 0.033 & 9.562 $\pm$ 0.032  & 0.621 $< \ \log M \leq$ 0.849 \\
		-1.800 $\pm$ 0.037 & 9.228 $\pm$ 0.044  & 0.849 $< \ \log M \leq$ 1.077 \\
		-1.338 $\pm$ 0.043 & 8.730 $\pm$ 0.061  & 1.077 $< \ \log M \leq$ 1.305 \\
		-0.898 $\pm$ 0.051 & 8.156 $\pm$ 0.085  & 1.305 $< \ \log M \leq$ 1.534 \\
		-0.756 $\pm$ 0.058 & 7.938 $\pm$ 0.110  & $\log M \ >$ 1.534            \\ \doubleRule
	\end{tabular}
\end{table}

\begin{table}[]
	\centering
	\vspace{1cm}
	\begin{tabular}{ccc}
		\doubleRule
		a                  & b                  & validity range               \\ \hline
		\multicolumn{3}{c}{Z = 0.008}                                          \\ \hline
		-3.263 $\pm$ 0.038 & 9.965 $\pm$ 0.010  & $\log M \ \leq$ 0.172         \\
		-2.485 $\pm$ 0.034 & 9.830 $\pm$ 0.017  & 0.172 $< \ \log M \leq$ 0.398 \\
		-2.817 $\pm$ 0.034 & 9.962 $\pm$ 0.025  & 0.398 $< \ \log M \leq$ 0.623 \\
		-2.226 $\pm$ 0.037 & 9.594 $\pm$ 0.036  & 0.623 $< \ \log M \leq$ 0.849 \\
		-1.831 $\pm$ 0.041 & 9.258 $\pm$ 0.049  & 0.849 $< \ \log M \leq$ 1.074 \\
		-1.354 $\pm$ 0.048 & 8.746 $\pm$ 0.068  & 1.074 $< \ \log M \leq$ 1.300 \\
		-0.908 $\pm$ 0.057 & 8.166 $\pm$ 0.095  & 1.300 $< \ \log M \leq$ 1.525 \\
		-0.801 $\pm$ 0.067 & 8.004 $\pm$ 0.125  & $\log M \ >$ 1.525            \\ \doubleRule
		\multicolumn{3}{c}{Z = 0.010}                                          \\ \hline
		-3.265 $\pm$ 0.039 & 9.993 $\pm$ 0.011  & $\log M \ \leq$ 0.179         \\
		-2.482 $\pm$ 0.035 & 9.852 $\pm$ 0.018  & 0.179 $< \ \log M \leq$ 0.403 \\
		-2.865 $\pm$ 0.035 & 10.007 $\pm$ 0.026 & 0.403 $< \ \log M \leq$ 0.626 \\
		-2.272 $\pm$ 0.038 & 9.635 $\pm$ 0.037  & 0.626 $< \ \log M \leq$ 0.850 \\
		-1.852 $\pm$ 0.042 & 9.278 $\pm$ 0.051  & 0.850 $< \ \log M \leq$ 1.073 \\
		-1.351 $\pm$ 0.049 & 8.740 $\pm$ 0.070  & 1.073 $< \ \log M \leq$ 1.297 \\
		-0.955 $\pm$ 0.057 & 8.227 $\pm$ 0.094  & 1.297 $< \ \log M \leq$ 1.520 \\
		-0.819 $\pm$ 0.069 & 8.020 $\pm$ 0.129  & $\log M \ >$ 1.520            \\ \doubleRule
		\multicolumn{3}{c}{Z = 0.020}                                          \\ \hline
		-3.272 $\pm$ 0.041 & 10.082 $\pm$ 0.012 & $\log M \ \leq$ 0.201         \\
		-2.516 $\pm$ 0.037 & 9.930 $\pm$ 0.020  & 0.201 $< \ \log M \leq$ 0.418 \\
		-2.998 $\pm$ 0.037 & 10.131 $\pm$ 0.028 & 0.418 $< \ \log M \leq$ 0.635 \\
		-2.406 $\pm$ 0.040 & 9.756 $\pm$ 0.039  & 0.635 $< \ \log M \leq$ 0.851 \\
		-1.947 $\pm$ 0.045 & 9.365 $\pm$ 0.053  & 0.851 $< \ \log M \leq$ 1.068 \\
		-1.388 $\pm$ 0.052 & 8.767 $\pm$ 0.073  & 1.068 $< \ \log M \leq$ 1.285 \\
		-1.028 $\pm$ 0.059 & 8.306 $\pm$ 0.095  & 1.285 $< \ \log M \leq$ 1.502 \\
		-0.778 $\pm$ 0.072 & 7.930 $\pm$ 0.132  & $\log M \ >$ 1.502            \\ \doubleRule
		\multicolumn{3}{c}{Z = 0.030}                                          \\ \hline
		-3.235 $\pm$ 0.042 & 10.116 $\pm$ 0.013 & $\log M \ \leq$ 0.207         \\
		-2.610 $\pm$ 0.037 & 9.987 $\pm$ 0.019  & 0.207 $< \ \log M \leq$ 0.418 \\
		-3.062 $\pm$ 0.037 & 10.176 $\pm$ 0.027 & 0.418 $< \ \log M \leq$ 0.630 \\
		-2.483 $\pm$ 0.040 & 9.811 $\pm$ 0.038  & 0.630 $< \ \log M \leq$ 0.841 \\
		-2.038 $\pm$ 0.045 & 9.437 $\pm$ 0.052  & 0.841 $< \ \log M \leq$ 1.052 \\
		-1.471 $\pm$ 0.051 & 8.841 $\pm$ 0.071  & 1.052 $< \ \log M \leq$ 1.264 \\
		-1.163 $\pm$ 0.061 & 8.451 $\pm$ 0.097  & 1.264 $< \ \log M \leq$ 1.475 \\
		-0.645 $\pm$ 0.075 & 7.688 $\pm$ 0.136  & $\log M \ >$ 1.475            \\ \doubleRule
		\multicolumn{3}{c}{Z = 0.035}                                          \\ \hline
		-3.229 $\pm$ 0.039 & 10.118 $\pm$ 0.012 & $\log M \ \leq$ 0.205         \\
		-2.638 $\pm$ 0.035 & 9.997 $\pm$ 0.018  & 0.205 $< \ \log M \leq$ 0.414 \\
		-3.080 $\pm$ 0.034 & 10.180 $\pm$ 0.025 & 0.414 $< \ \log M \leq$ 0.623 \\
		-2.537 $\pm$ 0.037 & 9.842 $\pm$ 0.035  & 0.623 $< \ \log M \leq$ 0.832 \\
		-2.074 $\pm$ 0.041 & 9.457 $\pm$ 0.047  & 0.832 $< \ \log M \leq$ 1.041 \\
		-1.487 $\pm$ 0.048 & 8.845 $\pm$ 0.065  & 1.041 $< \ \log M \leq$ 1.250 \\
		-1.166 $\pm$ 0.056 & 8.444 $\pm$ 0.088  & 1.250 $< \ \log M \leq$ 1.459 \\
		-0.695 $\pm$ 0.068 & 7.757 $\pm$ 0.122  & $\log M \ >$ 1.459            \\ \doubleRule
	\end{tabular}
\end{table}

\begin{table}[]
	\centering
	\begin{tabular}{ccc}
		\doubleRule
		a                  & b                  & validity range               \\ \hline
		\multicolumn{3}{c}{Z = 0.039}                                          \\ \hline
		-3.226 $\pm$ 0.038 & 10.120 $\pm$ 0.011 & $\log M \ \leq$ 0.203         \\
		-2.657 $\pm$ 0.034 & 10.004 $\pm$ 0.017 & 0.203 $< \ \log M \leq$ 0.411 \\
		-3.096 $\pm$ 0.033 & 10.184 $\pm$ 0.024 & 0.411 $< \ \log M \leq$ 0.618 \\
		-2.576 $\pm$ 0.036 & 9.863 $\pm$ 0.033  & 0.618 $< \ \log M \leq$ 0.825 \\
		-2.099 $\pm$ 0.040 & 9.469 $\pm$ 0.045  & 0.825 $< \ \log M \leq$ 1.033 \\
		-1.516 $\pm$ 0.046 & 8.867 $\pm$ 0.062  & 1.033 $< \ \log M \leq$ 1.240 \\
		-1.185 $\pm$ 0.054 & 8.457 $\pm$ 0.084  & 1.240 $< \ \log M \leq$ 1.447 \\
		-0.700 $\pm$ 0.066 & 7.756 $\pm$ 0.117  & $\log M \ >$ 1.447            \\ \doubleRule
	\end{tabular}
\end{table}

\begin{table}[]
	\centering
	\FloatBarrier
	\caption{Fitting equations of the relation between relative LPV phase duration ($\delta t/t$ where $t$ is the age and $\delta t$ is the LPV phase duration) and birth mass, $\log(\delta t/t) = D + \Sigma_{i=1}^4a _i\exp\left[-(\log M [{\rm M} _\odot]-b_i)^2/c_i^2\right]$.}
	\label{tab:taba3}
	\begin{tabular}{ccccc}
		\doubleRule
		D       & i & a       & b      & c     \\ \hline
		\multicolumn{5}{c}{Z = 0.0003}         \\ \hline
		-6.926  & 1 & 3.848   & 0.620  & 0.876 \\
		& 2 & 0.513   & 0.248  & 0.070 \\
		& 3 & 4.833   & 1.821  & 0.432 \\
		& 4 & 342.189 & -1.779 & 0.009 \\ \doubleRule
		\multicolumn{5}{c}{Z = 0.0005}         \\ \hline
		-5.054  & 1 & 1.810   & 0.586  & 0.211 \\
		& 2 & 1.882   & 0.248  & 0.159 \\
		& 3 & 3.526   & 1.666  & 0.271 \\
		& 4 & 1.496   & 1.169  & 0.197 \\ \doubleRule
		\multicolumn{5}{c}{Z = 0.001}          \\ \hline
		-4.570  & 1 & 1.336   & 0.589  & 0.166 \\
		& 2 & 0.949   & 1.148  & 0.099 \\
		& 3 & 3.481   & 1.513  & 0.252 \\
		& 4 & 1.610   & 0.265  & 0.160 \\ \doubleRule
		\multicolumn{5}{c}{Z = 0.003}          \\ \hline
		-5.323  & 1 & 1.531   & 1.826  & 0.151 \\
		& 2 & 3.421   & 1.337  & 0.405 \\
		& 3 & 2.566   & 0.306  & 0.290 \\
		& 4 & 1.288   & 0.647  & 0.127 \\ \doubleRule
		\multicolumn{5}{c}{Z = 0.006}          \\ \hline
		-5.809  & 1 & 3.149   & 0.336  & 0.366 \\
		& 2 & 4.238   & 1.290  & 0.444 \\
		& 3 & 0.676   & 0.662  & 0.095 \\
		& 4 & 2.149   & 1.757  & 0.116 \\ \doubleRule
	\end{tabular}
\end{table}

\begin{table}[]
	\centering
	\begin{tabular}{ccccc}
		\doubleRule
		D      & i & a      & b     & c     \\ \hline
		\multicolumn{5}{c}{Z = 0.008}          \\ \hline
		-10.000 & 1 & 7.386   & 0.375  & 0.676 \\
		& 2 & 0.637   & 0.695  & 0.088 \\
		& 3 & 7.380   & 1.538  & 0.289 \\
		& 4 & 5.073   & 1.133  & 0.261 \\ \doubleRule
		\multicolumn{5}{c}{Z = 0.010}       \\ \hline
		-8.225 & 1 & 3.056  & 1.062 & 0.209 \\
		& 2 & 0.697  & 0.711 & 0.095 \\
		& 3 & 6.275  & 1.444 & 0.333 \\
		& 4 & 5.593  & 0.385 & 0.598 \\ \doubleRule
		\multicolumn{5}{c}{Z = 0.020}       \\ \hline
		-5.542 & 1 & 4.273  & 1.216 & 0.584 \\
		& 2 & 2.255  & 0.304 & 0.354 \\
		& 3 & -1.954 & 0.881 & 0.091 \\
		& 4 & 0.225  & 0.450 & 0.044 \\ \doubleRule
		\multicolumn{5}{c}{Z = 0.030}       \\ \hline
		-6.646 & 1 & 5.349  & 1.208 & 0.580 \\
		& 2 & 0.966  & 0.992 & 0.037 \\
		& 3 & -2.628 & 0.918 & 0.094 \\
		& 4 & 3.189  & 0.324 & 0.435 \\ \doubleRule
		\multicolumn{5}{c}{Z = 0.035}       \\ \hline
		-5.497 & 1 & -3.283 & 0.887 & 0.102 \\
		& 2 & 2.571  & 0.381 & 0.343 \\
		& 3 & 1.719  & 0.851 & 0.223 \\
		& 4 & 4.088  & 1.248 & 0.433 \\ \doubleRule
		\multicolumn{5}{c}{Z = 0.039}       \\ \hline
		-5.321 & 1 & -2.402 & 0.884 & 0.052 \\
		& 2 & 2.247  & 0.373 & 0.308 \\
		& 3 & 0.604  & 0.688 & 0.092 \\
		& 4 & 4.053  & 1.192 & 0.468 \\ \doubleRule
	\end{tabular}
\end{table}

\subsection{Probability Function Details} \label{app:appB}

To determine the probability function, we use the data from \cite{Rejkuba2003a}, which are classified into three categories. They are marked with "Circle", "Cross", and "Triangle" Points corresponding to simulated variable stars which are considered as a function of amplitude regardless of their magnitudes for all the stars, simulated variable stars with $K_s > 20.5$ mag, and the ones brighter than $K_s < 20$ mag, respectively. They are all presented in section 6 of \cite{Rejkuba2003a} and Figs. 12 and 14. These simulated variable stars have been reproduced in terms of their amplitude, period, and mean magnitude based on the catalog.

The final probability function of simulated variable stars is parameterised in Eq.\,\ref{eq:eq5} whose coefficients are mentioned in table \ref{tab:taba4} for both Field\,1 and Field\,2.

\begin{table}[]
	\centering
	\caption{Coefficient of probability function of Field\,1 and Field\,2 for simulated variable stars by considering the amplitude regardless of their magnitude, Eq.\,\ref{eq:eq5}.}
	\label{tab:taba4}
	\begin{tabular}{ccc|c}
		\hline
		&            & Field\,1                   & Field\,2                   \\ \hline
		N  & Symbol     & Coefficient                & Coefficient                \\ \hline
		1  & $\alpha_1$ & -16.231                    & -6.348                     \\
		2  & $\alpha_2$ & 17.628                     & 7.278                      \\
		3  & $\alpha_3$ & -8.986 $\times$ 10$^{-2}$  & 7.588 $\times$ 10$^{-2}$   \\
		4  & $\beta$    & 1.227                      & 3.364 $\times$ 10$^{-1}$   \\
		5  & $\eta$     & 9.733 $\times$ 10$^{-1}$   & 2.991                      \\
		6  & $\gamma_1$ & 6.483 $\times$ 10$^{-1}$   & 2.412                      \\
		7  & $\gamma_2$ & -6.427 $\times$ 10$^{-3}$  & -2.507 $\times$ 10$^{-2}$  \\
		8  & $\gamma_3$ & 3.063 $\times$ 10$^{-7}$   & 1.178 $\times$ 10$^{-4}$   \\
		9  & $\gamma_4$ & -7.559 $\times$ 10$^{-8}$  & -2.820 $\times$ 10$^{-7}$  \\
		10 & $\gamma_5$ & 9.749 $\times$ 10$^{-11}$  & 3.545 $\times$ 10$^{-10}$  \\
		11 & $\gamma_6$ & -6.228 $\times$ 10$^{-14}$ & -2.230 $\times$ 10$^{-13}$ \\
		12 & $\gamma_7$ & 1.555 $\times$ 10$^{-17}$  & 5.536 $\times$ 10$^{-17}$  \\
		13 & $\delta_1$ & -2.014 $\times$ 10$^4$     & -4.106 $\times$ 10$^4$     \\
		14 & $\delta_2$ & -1.000 $\times$ 10$^6$     & 6.205 $\times$ 10$^6$      \\
		15 & $\delta_3$ & 1.000                      & 1.000                      \\
		16 & $\delta_4$ & 3.434 $\times$ 10$^3$      & 1.153 $\times$ 10$^2$      \\ \hline
	\end{tabular}
\end{table}

To accomplish this, we encountered a drop after reaching the peak for the simulated LPVs fainter than $K_s = 20.5$ mag. Two reasons were mentioned in \cite{Rejkuba2003a} which are related to increasing the photometry error and the difference in completeness limit for the $H$-, $J_s$-, and $K_s$-bands. The fainter stars have a lower completeness limit, as the fainter parts of their light curve would not have been detected.

\begin{figure*}[]
	\centerline{\hbox{
			\epsfig{figure=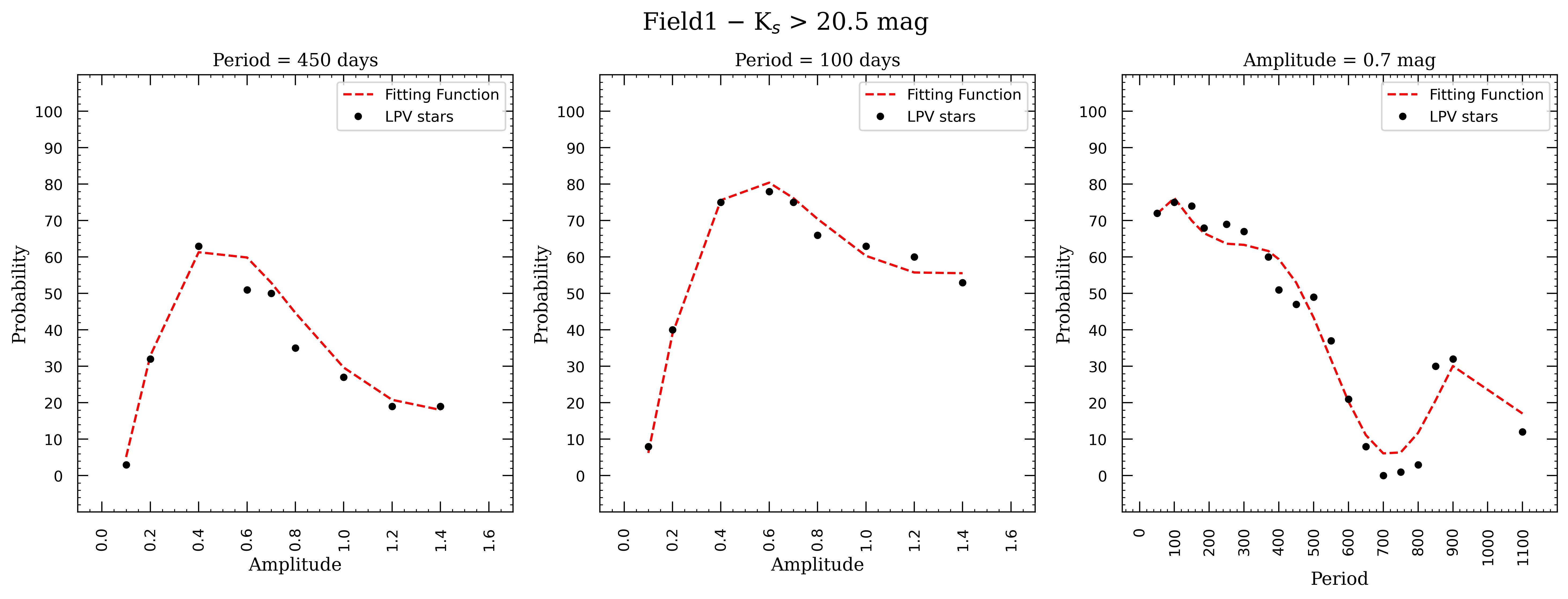,width=167mm}}}
	\caption[]{The result of fitting probability function for simulated variable stars with $K_s > 20.5$ mag in Field\,1.}
	\label{fig:fig14}
\end{figure*}

\begin{figure*}[]
	\centerline{\hbox{
			\epsfig{figure=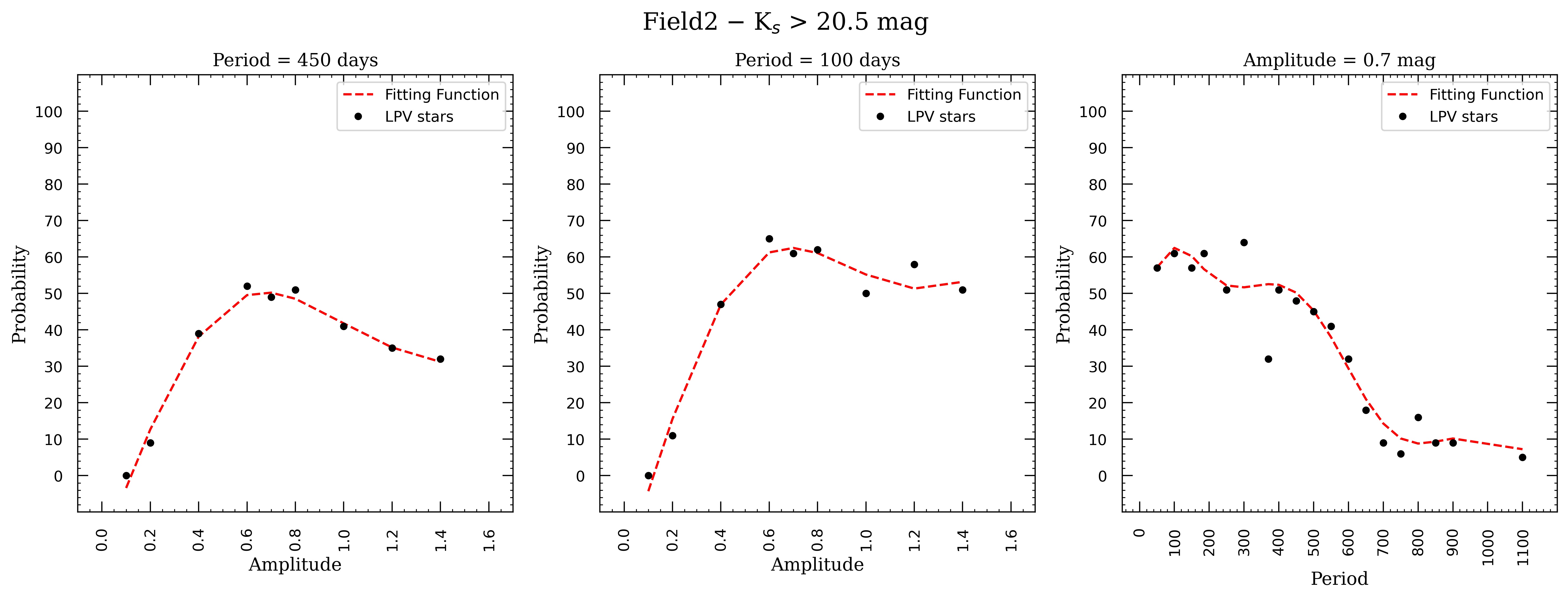,width=167mm}}}
	\caption[]{The result of fitting probability function for simulated variable stars with $K_s > 20.5$ mag in Field\,2.}
	\label{fig:fig15}
\end{figure*}

\begin{figure*}[]
	\centerline{\hbox{
			\epsfig{figure=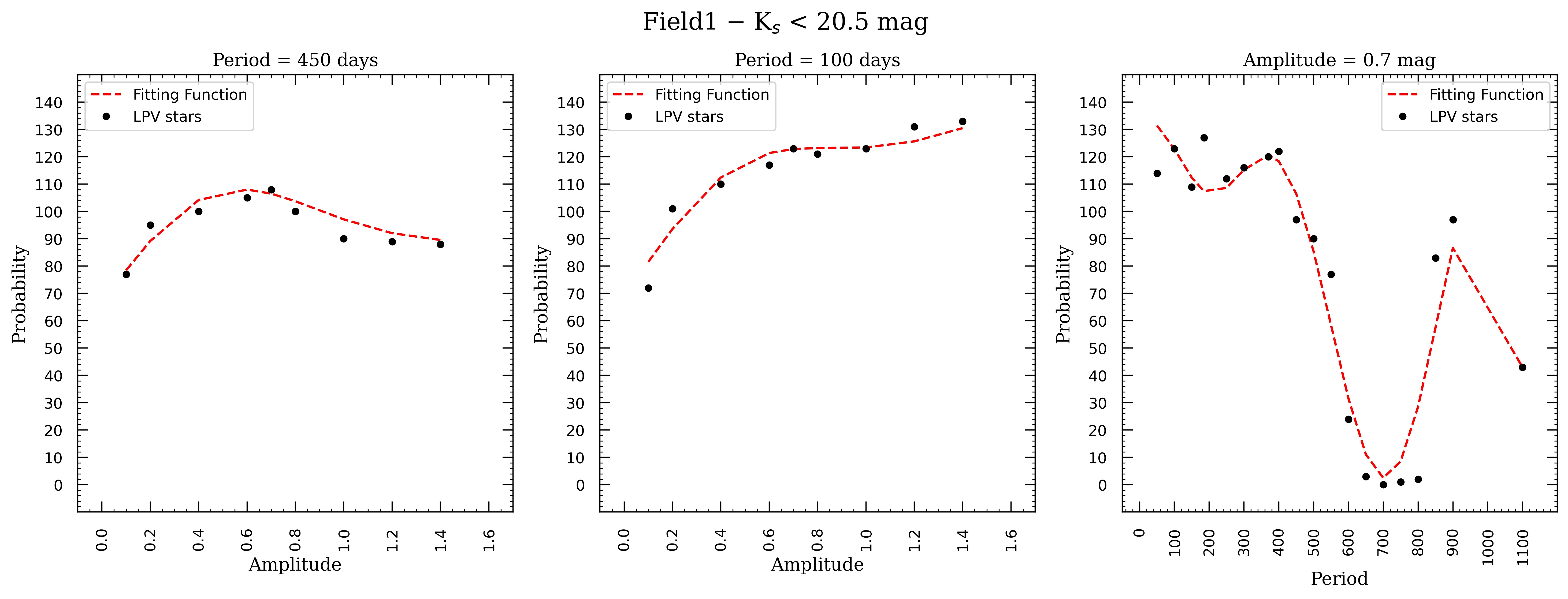,width=167mm}}}
	\caption[]{The result of fitting probability function for simulated variable stars with $K_s < 20.5$ mag in Field\,1.}
	\label{fig:fig16}
\end{figure*}

\begin{figure*}[]
	\centerline{\hbox{
			\epsfig{figure=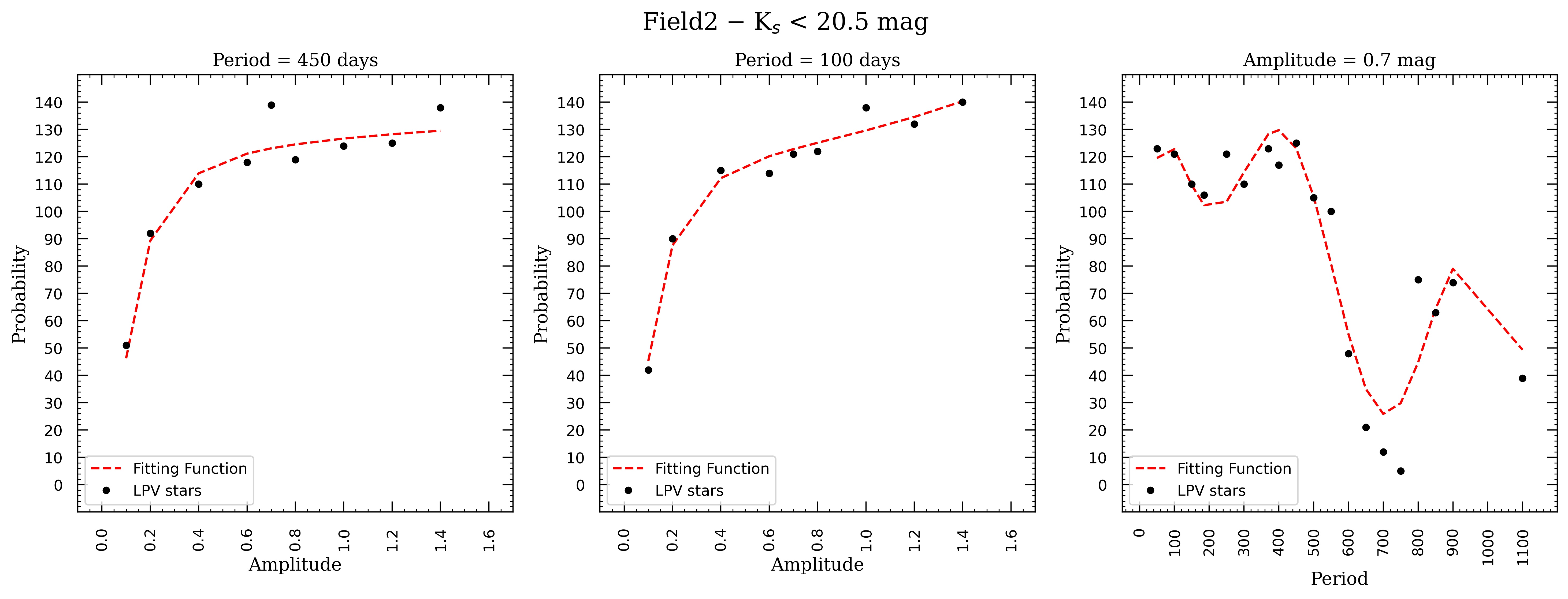,width=167mm}}}
	\caption[]{The result of fitting probability function for simulated variable stars with $K_s < 20.5$ mag in Field\,2.}
	\label{fig:fig17}
\end{figure*}

\begin{figure*}[]
	\centerline{\hbox{
			\epsfig{figure=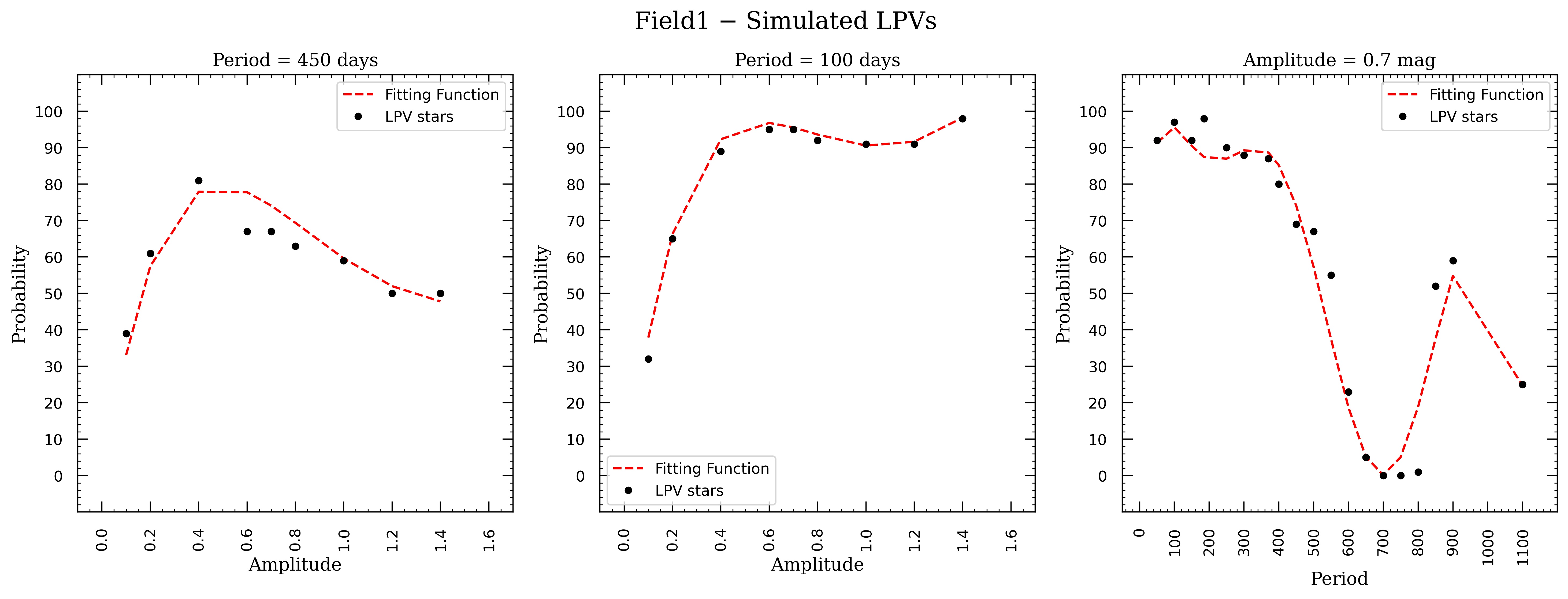,width=167mm}}}
	\caption[]{The result of fitting probability function for the simulated LPVs by considering the amplitude regardless of their magnitude of Field\,1.}
	\label{fig:fig18}
\end{figure*}

\begin{figure*}[]
	\centerline{\hbox{
			\epsfig{figure=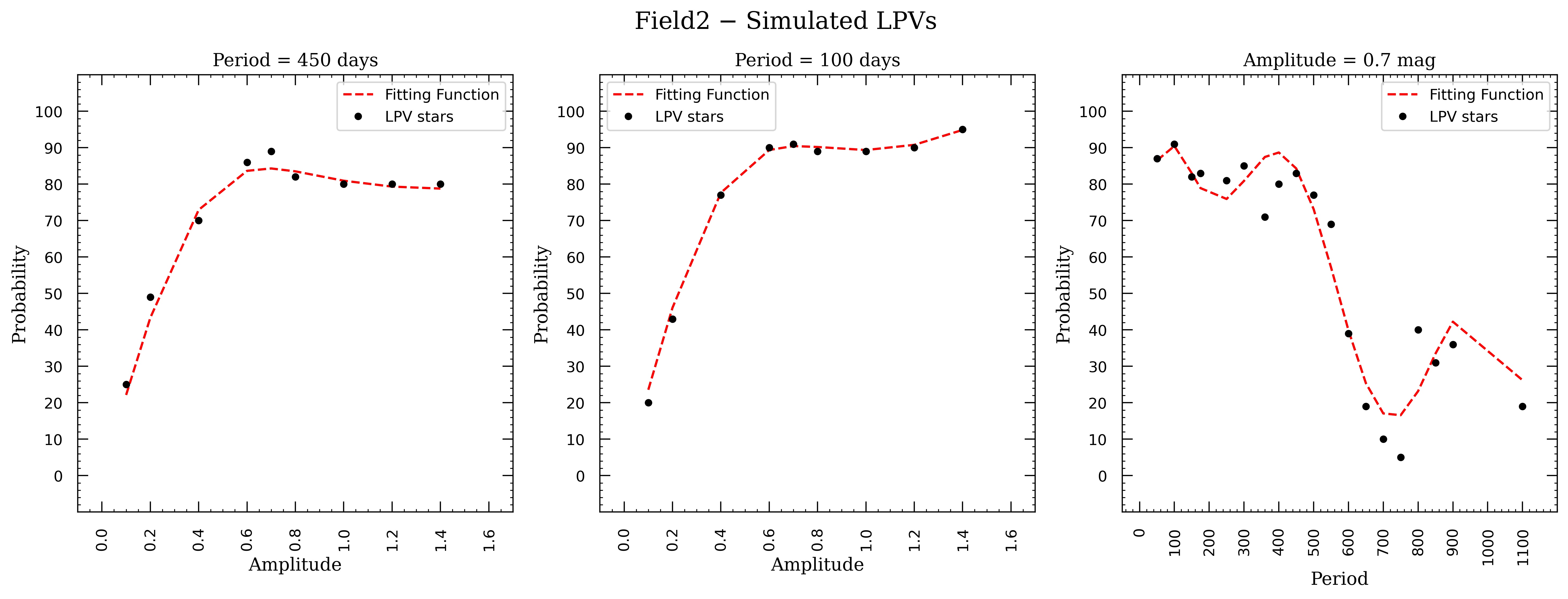,width=167mm}}}
	\caption[]{The result of fitting probability function for the simulated LPVs by considering the amplitude regardless of their magnitude of Field\,2.}
	\label{fig:fig19}
\end{figure*}

The plots which depict the accuracy of fitting probability functions are presented in Figs.\,\ref{fig:fig14} to \ref{fig:fig19} for Field\,1 and Field\,2.

\makeatletter\onecolumngrid@push\makeatother

\makeatletter\onecolumngrid@pop\makeatother

\makeatletter\onecolumngrid@push\makeatother

\subsection{Overview Plots of the SFR} \label{app:appC}

\begin{figure*}[]
	\centering{\vbox{
			\epsfig{figure=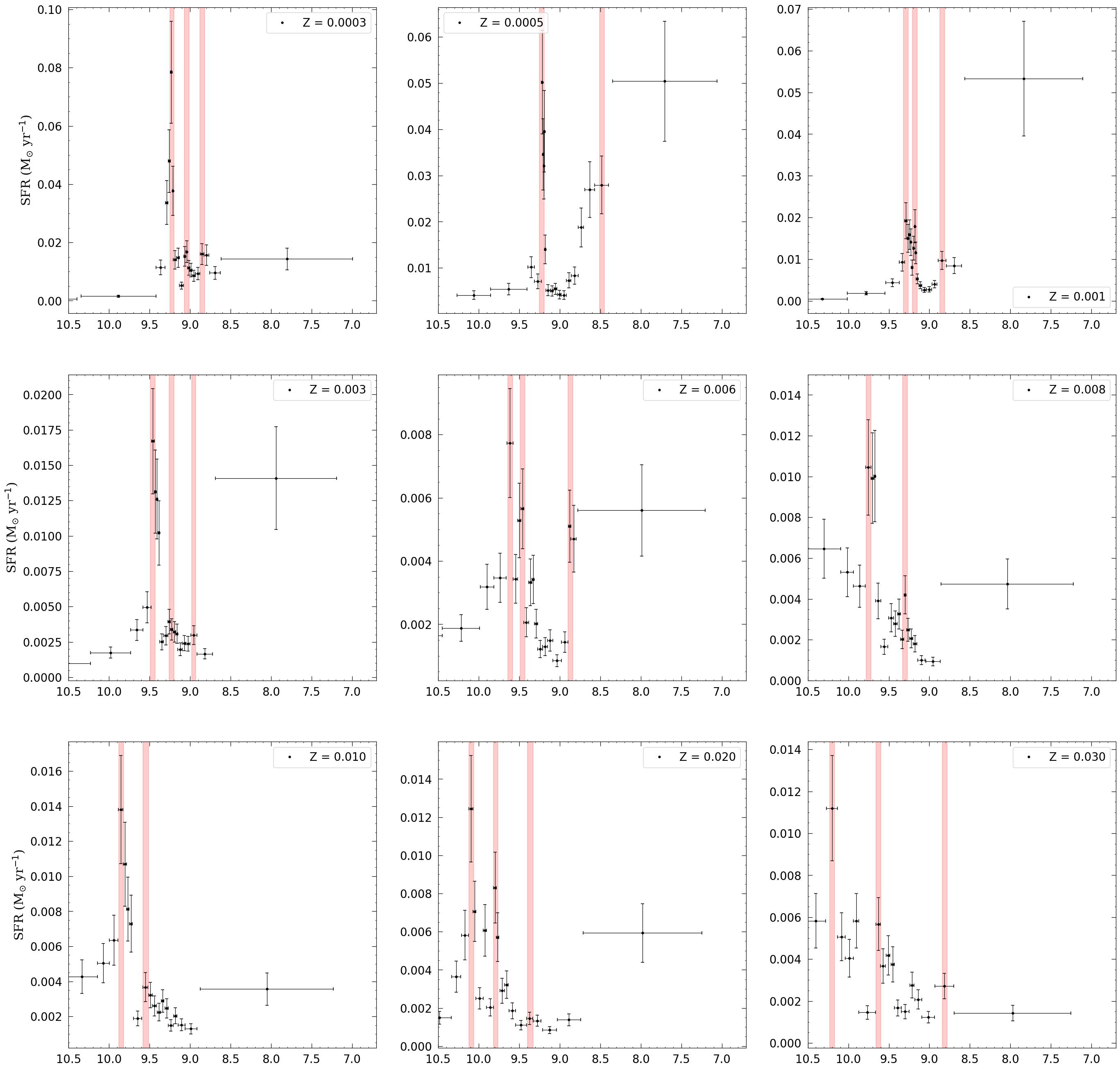,width=180mm}
			\epsfig{figure=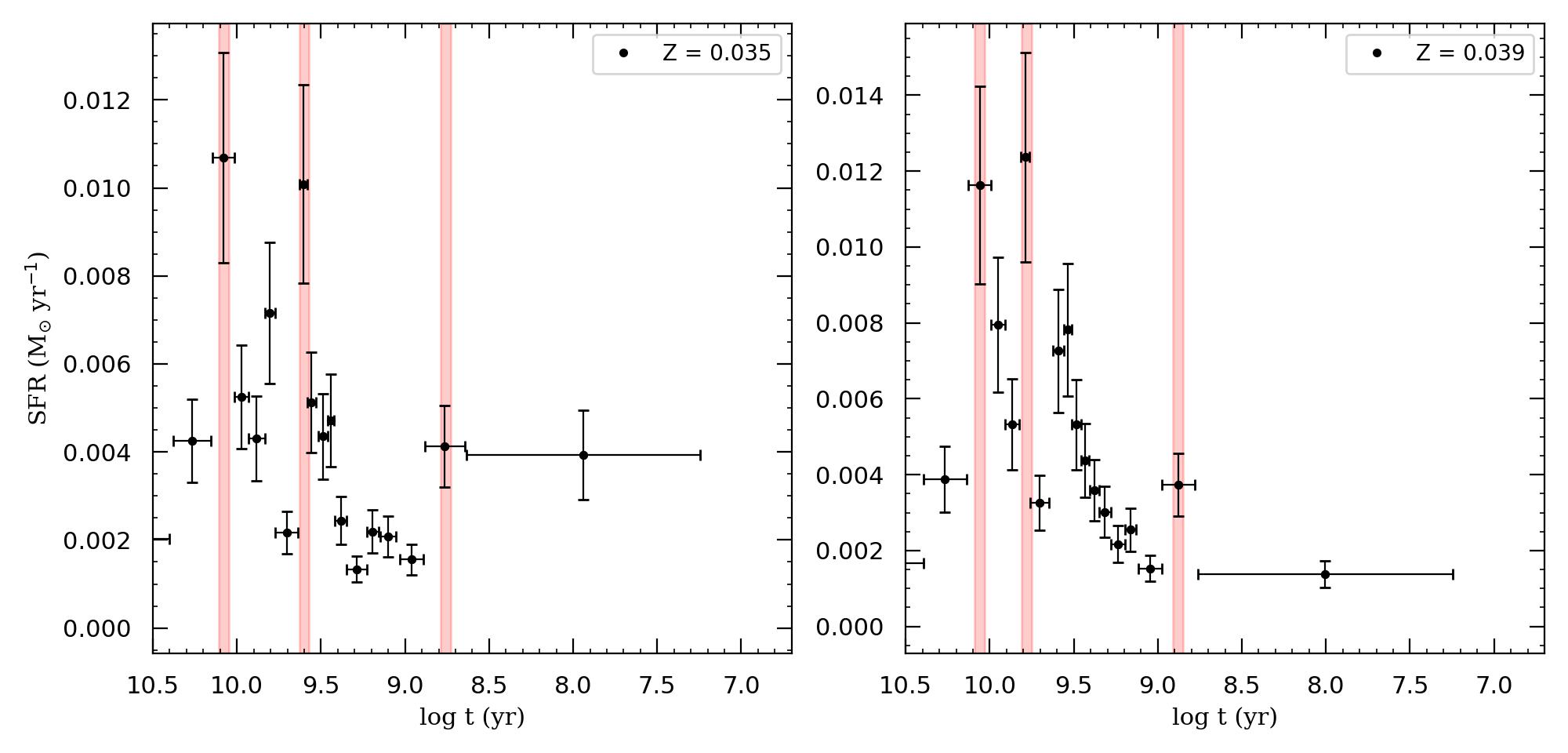,width=125mm}}}
	\caption{The SFRs of Field\,1 for different metallicities in separate panels after applying probability function. Each panel is scaled to $1$ and the red regions are the desired epochs of SFR. The The highlighted regions represent the peaks of star formation during the major epochs.}
	\label{fig:fig20}
\end{figure*}

\begin{figure*}[]
	\centering{\vbox{
			\epsfig{figure=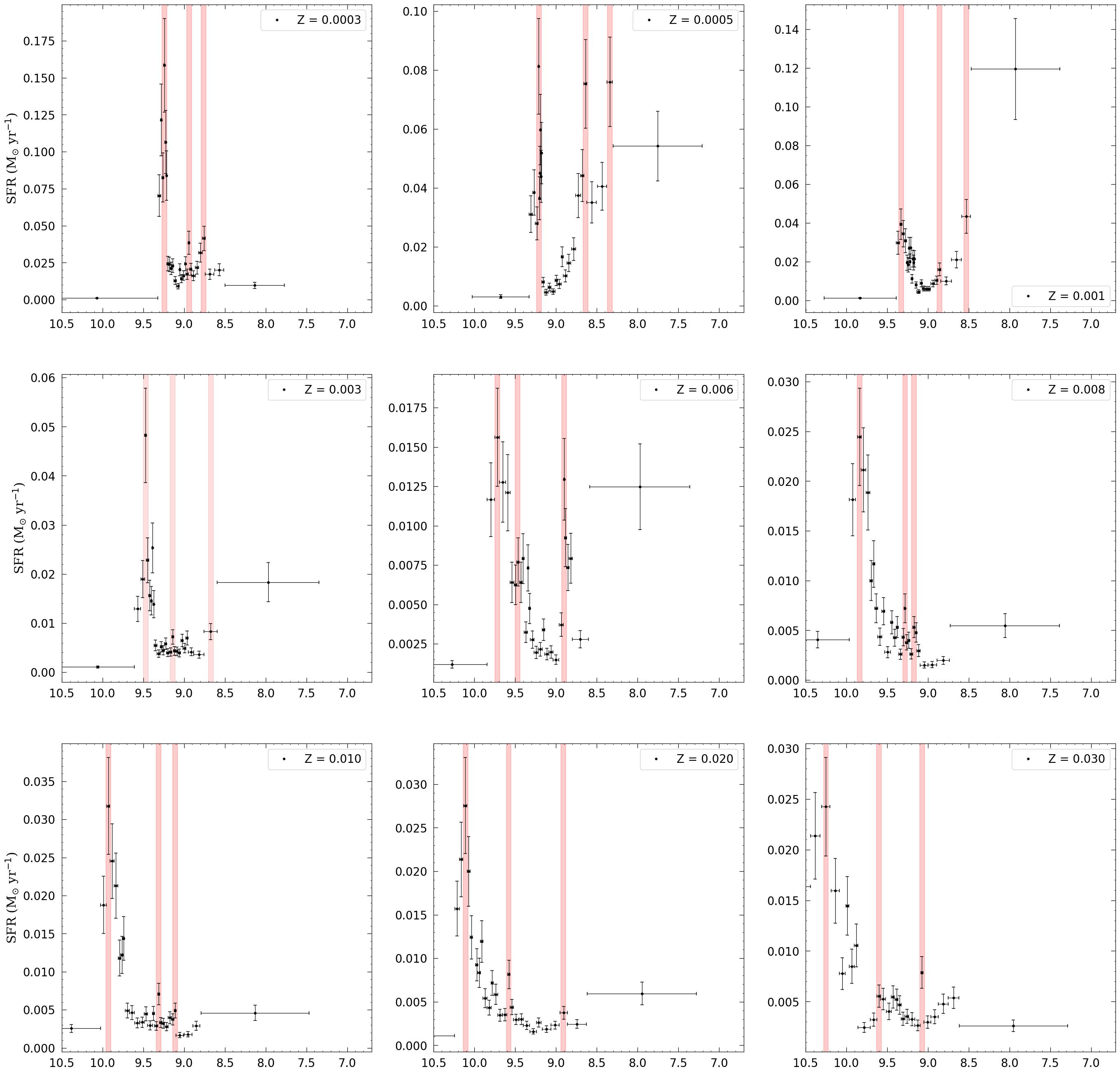,width=180mm}
			\epsfig{figure=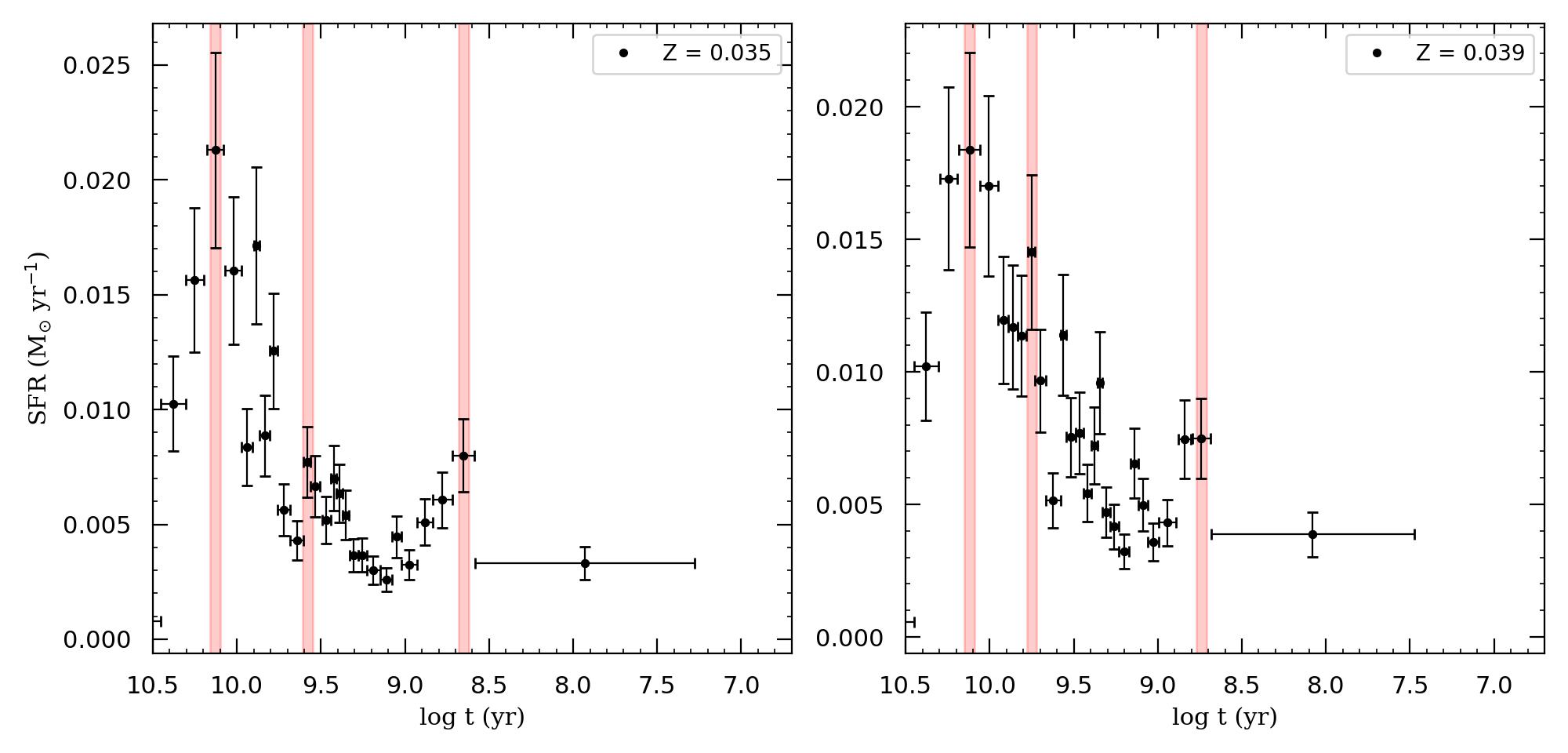,width=125mm}}}
	\caption{The SFRs of Field\,2 for different metallicities in separate panels after applying probability function. Each panel is scaled to $1$ and the red regions are the desired epochs of SFR. The highlighted regions represent the peaks of star formation during the major epochs.}
	\label{fig:fig21}
\end{figure*}

For a detailed overview of the SFR, it is plotted separately for each metallicity of Field\,1 and 2 which are Figs.\,\ref{fig:fig20} and \ref{fig:fig21}, respectively.

\end{document}